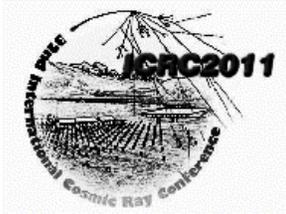
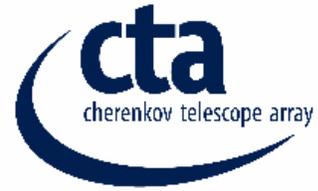

# Contributions from the

# Cherenkov Telescope Array (CTA)

# Consortium

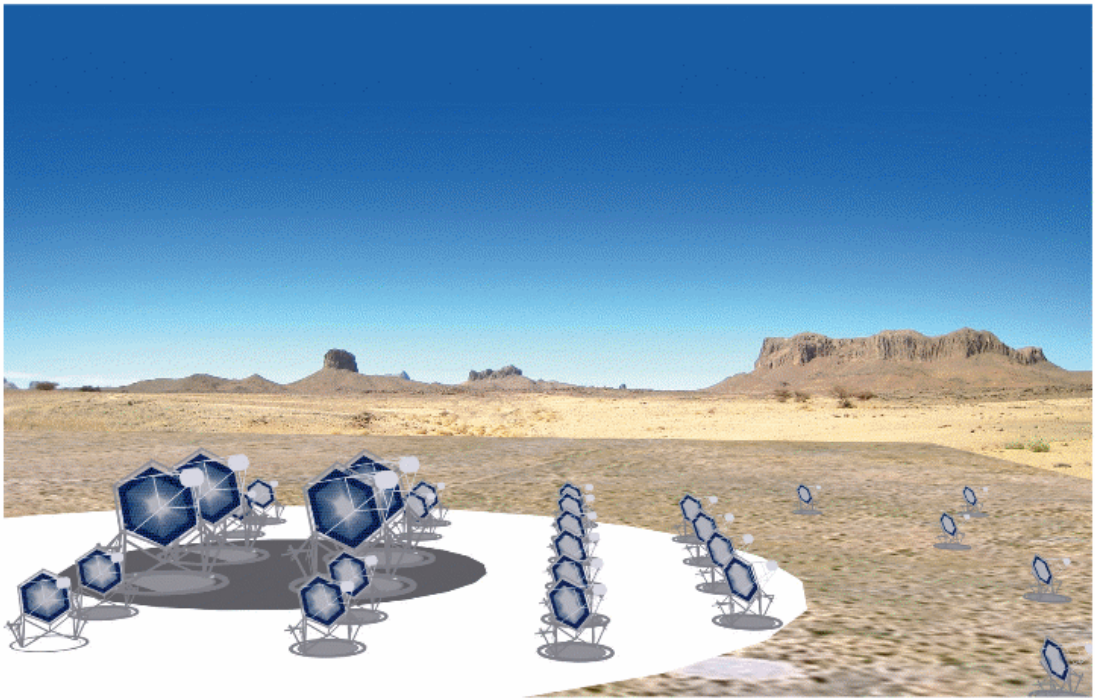

## Sessions

- OG2.2: X-ray and Gamma ray observations, Galactic Sources
- OG2.3: X-ray and Gamma ray observations, Extragal. Sources
- OG2.5: X-ray and Gamma ray observations, New experiments and instrumentation

# Contents

## Working Package Calibration

ID0408 - Development of Raman LIDARs made with former CLUE telescopes for CTA 

## Working Package Physics

ID1156 - Constraining the Extragalactic Background Light in the near-mid IR with the Cherenkov Telescope Array (CTA) 

ID1086 - On the Detectability of Dwarf Galaxies with the Cherenkov Telescope Array 

ID0883 - Search for Lorentz Invariance Violation with flaring Active Galactic Nuclei: a prospect for the Cherenkov Telescope Array 

## Authors List

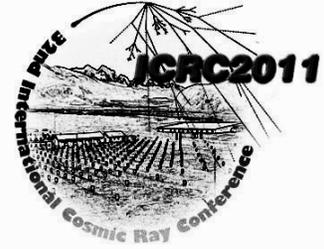

# CTA: where do we stand and where do we go ?

MANEL MARTINEZ FOR THE CTA CONSORTIUM
*IFAE, Edifici Cn, Universitat Autonoma de Barcelona, 08193 Bellaterra, Barcelona, Spain*
*martinez@ifae.es*

**Abstract:** The Cherenkov Telescope Array (CTA) is a project for the construction of a next generation VHE gamma ray observatory with full sky coverage. Its aim is improving by about one order of magnitude the sensitivity of the existing installations, covering about 5 decades in energy (from few tens of GeV to above a hundred TeV) and having enhanced angular and energy resolutions. During 2010 the project became a truly global endeavour carried out by a consortium of about 750 collaborators from Europe, Asia, Africa and the North and South Americas. Also during 2010 the CTA project completed its Design Study phase and started a Preparatory Phase that is expected to extend for three years and should lead to the starting of the construction of CTA. An overview of the CTA project will be given and the status and plans of its Preparatory Phase will be discussed.

## 1. Introduction

Very High Energy gamma-ray astronomy is consolidating as a new and very successful branch of Astrophysics and Astroparticle Physics. Exciting results have been obtained in the last few years by the current generation IACT (Imaging Atmospheric Cherenkov Telescopes) that have increased the number of detected sources from just a handful to about 150.

The present generation of Cherenkov telescopes has allowed the first detailed observations of the sky using gamma rays of energies above 100 GeV and has revealed sources with complex and resolved structures in the central band of the Milky Way and a plethora of extragalactic sources. Based on the extensive scientific knowledge gained with the Cherenkov telescopes H.E.S.S. in Namibia, MAGIC on the Canary Islands, and VERITAS in the United States, there is a high potential for lots of discoveries with a more sensitive gamma-ray observatory with full sky coverage in the energy range above some tens of GeV.

This was the motivation that, back in 2005, gathered together the scientists of the VHE gamma ray community to propose the Cherenkov Telescope Array (CTA) concept. CTA is an European-initiated worldwide project which builds upon the success and convergence of H.E.S.S., MAGIC, VERITAS and CANGAROO and gathers the current expertise in the domain of most of the worldwide community with the aim to exploit the full potential of the IACT technique during the next two or three decades.

When finally completed, CTA is expected to increase the sensitivity of actual observatories such as H.E.S.S., MAGIC or VERITAS by an order of magnitude. It will also expand the energy range coverage from some tens of GeV to above hundred TeV, opening a new window in the VHE domain never reached with such exquisite detail. Additionally, the ideal synergy between the Large Area Telescope (LAT) on board of the Fermi Gamma-ray Space Telescope and CTA will allow nearly seamless coverage from few MeV to hundreds of TeV.

This observatory shall reveal of the order of thousand sources, an order of more sources than in the current VHE catalogues, allowing for example population studies of classes such as Pulsar Wind Nebulae (PWNe) and Active Galactic Nuclei (AGN). Due to its higher sensitivity and better angular resolution it shall be able to detect new classes of objects and phenomena that have not been visible until now and will greatly improve the chances for discovery in fundamental physics issues such as Dark Matter astronomy, Lorenz invariance violation or observational cosmology.

CTA will serve as an open observatory to a wide astrophysics community and will provide a deep insight into the non-thermal high-energy universe enabling scientists to tackle a variety of fundamental open questions con-cerning supernovae, compact objects such as pulsars and black holes, the galactic centre, star forming regions, active galactic nuclei, dark matter, quantum gravity, charged cosmic rays, the physics of jets, the Extragalactic Background Light (EBL) and many more.

The main performance goals of CTA to exploit the physics potential, the conceptual design of CTA and the present status of the project are discussed in this write-up.
An overview of the CTA project will be given and the status and plans of its Preparatory Phase will be discussed.

## 2. The CTA performance goals.

The main performance goals of the CTA project are:

- A wide energy coverage: four decades, from some 10 GeV to beyond 100 TeV

- A sensitivity at least one order of magnitude better than any existing installation: about 1 miliCrab at the core energy regime around 1 TeV.

- Improved angular resolution down to a few arc-minutes and extended field of views to be able to do morphological studies.

- Two observatories, operated under a common framework, for all-sky monitoring capability: Southern, with a wide energy range covering especially the Galactic sources; and Northern, with a focus on low energies for Extragalactic objects.

The energy range extending from a few 10's of GeV to above 100 TeV shall allow new source classes to be discovered, the emission mechanisms in the known source classes to be better investigated (e.g. comparison of leptonic vs. hadronic models in SNRs, binaries) and a wide parameter space together with a larger lever-arm for signal discrimination for possible Dark Matter annihilation sources to be explored.

A jump of a factor of 10 in sensitivity over the currently-operating experiments (down to a milliCrab level), should allow about an order of magnitude increase in the number of detectable sources by providing a deeper vision of the gamma-ray sky, and allow much finer temporal resolution for variable sources (e.g. AGNs).

Additionally, an improvement in angular resolution will not only provide increased background rejection for the point-like sources (AGNs, binaries), but will allow also fine mapping of the extended Galactic sources (PWNe and SNRs) and possibly the nearest extragalactic sources (M87, Cen A). An extended field of view will allow surveying the galactic plane and other specific regions on the sky and, combined with the enhanced sensitivity, discovering hundreds of new galactic sources.

Given that the gamma-ray sky is already known to contain a rich catalogue of sources, full-sky coverage is most desirable, with Northern and Southern installations of the observatory, adapted to the sky coverage, i.e. better low-energy coverage for the Northern installation where the extragalactic sky would be the preferred target, and with a wider energy range for the Southern one for which the region around the Galactic plane will be fully accessible.

These goals could be achieved by means of two extended, mixed arrays of Cherenkov telescopes (detailed below) with the additional advantage that such large, extended arrays would have an inherent flexibility of operation, allowing both deep field investigations and surveys, in parallel with monitoring of the brightest variable sources and reactivity to alerts from other instruments.

## 3. Conceptual Design of CTA

On the basis of the experience with the current generation of IACTs, it is clear that the technique to achieve the goals of the CTA is available: extended, mixed arrays of Cherenkov telescopes, (see fig.1) with a gradation in telescope performance:

Detection at the highest energies, from 10 to 100 TeV, encounters the major difficulty of the sparseness of the signal even from strong sources. This leads to the requirement of an extended array comprising a couple of tens of telescopes over ~10 km$^2$ of small size (~5-7m diameter) but with very wide Field-of-View (FoV, 7—10 degrees) in order for air-showers to be seen in stereo by widely-separated detectors, and relatively coarse pixelization (~0.25 degrees or larger). Such an array is mainly required on the Southern site, where Galactic sources - unaffected by absorption of the extragalactic background light - are the preferred target.

In the core energy range from 100 GeV to 10 TeV, where the current generation of ACT telescopes operates most efficiently, an extended array is also required for increased collection area, providing a higher proportion of "golden events" where the shower impact parameter is contained within the array (giving better angular resolution and increased sensitivity), and which will also provide flexibility of use of sub-arrays for different targets. An array of a few tens of telescopes covering ~1 km$^2$, of mid-size (10-13m diameter), with wide FoV (5-8 degrees for mapping of extended objects) and moderate pixelization (0.2 degree or smaller) is under consideration.

At the lowest energies, a central section of a few large telescopes (20-25m diameter), with improved photo-detector performance and a FoV of 4-6 degree with finer pixelization (~0.1 degree) will provide access to the lowest energies, and therefore to the most distant AGNs and to Galactic sources with spectral cut-offs (e.g. pulsars), as well as linking up to the spectral information provided by satellite detectors (Fermi, AGILE) at lower energies.

The parameters concerning the mix of telescopes, layout, mirror area, camera pixelization, and the electronics' trigger and sampling rapidity have been part of a Design Study within the CTA consortium that was completed by mid 2010, and were optimized in a multi-dimensional parameter space with consideration given to performance for the physics goals, and to cost and reliability/durability (of particular importance for such a large project of long duration, since the installations will operate as an observatory, with a life-time of the order of 30 years).

presently it combines the worldwide experience of all relevant communities working with atmospheric Cherenkov telescopes: Europe, US, Japan and India. The groups from US (formerly AGIS collaboration) and India joined in 2010, with the US groups proposing a major upgrade to the Medium Sized Telescope subarray, possibly with a new technology consisting of dual mirror Schwartzschild-Couder Telescopes (SC-MST) still under development. At this moment (Summer 2011) CTA is a Consortium of about 800 scientists and engineers from about 140 institutions from 25 countries.

The CTA project started in fall 2010 a three-year EU-

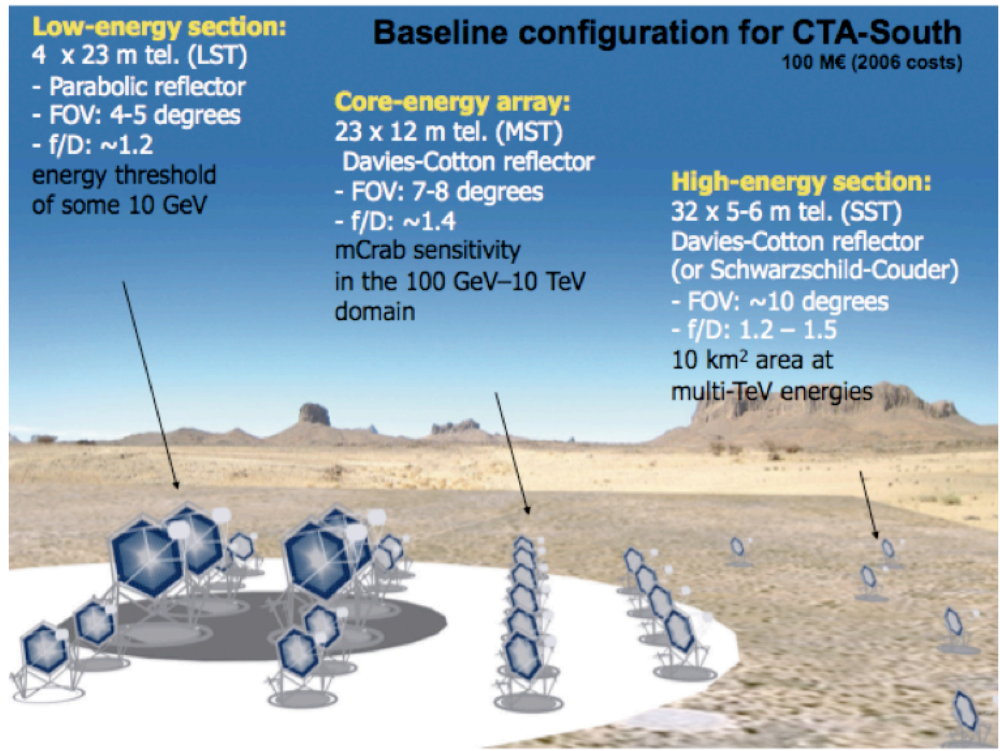

*Figure 1:* Schematic view (not to scale) of the baseline layout of the CTA south array consisting of 3 types of telescopes, with diameter sizes of approximately 6m (about 30 Small Size Telescopes -SST), 12m (about 30 Mid-Size-Telescopes -MST) and 24 m (about 4 Large Size Telescopes -LST), covering the full energy range from few tens of GeV up to 100 TeV. The details given in the sketch for each telescope type are discussed in reference 1.

The main outcome of these optimization studies is that the CTA performance goals can be achieved within the foreseen budget, as can be gleaned for the sensitivity in figure 2.

All the work performed during the Design Study phase of CTA is summarized in reference 1.

### 4. The CTA Preparatory Phase

#### *4.1 Present status and plans*

Although CTA started as an European initiative,

funded Preparatory Phase with the goal of being ready to start the construction of the Northern and Southern arrays by the end of 2013. The budget estimated in 2006 for the construction of CTA was about 100 MEuros for the Southern array and 50 MEuros for the Northern one (counting only material investment) and the expected construction time is of about 5 years. Therefore the present aim of the CTA Consortium is having the two sites of the CTA Observatory with the telescope arrays fully deployed by 2018, although partial operation may start already by 2015-2016 when the deployed telescopes shall provide already a much better performance than the existing installations.

It is worth mentioning here that, unlike the existing IACT installations, the observatory character of CTA will open this infrastructure to a much wider community, which will appear as "users" who apply for observation time, as in traditional astronomical observatories.

### 4.2 Summary of Preparatory Phase activities

The CTA Preparatory Phase will address a number of crucial prerequisites for the approval, construction and operation of CTA. It broadly splits into two themes: investigation of the science possible with CTA and technical and administrative aspects of building and operating a telescope array. The CTA Preparatory Phase activities are organized in a sizable number of work-packages that we're going to discuss briefly here.

In what concerns the science possible with CTA, a very important activity has started with the goal of establishing links (LINK work-package) with all the scientific communities (Particle Physicists, Astrophysicists, Astroparticle Physicists,...) that could take advantage from the existence of CTA and viceversa.

activities during the CTA Preparatory phase is the detailed design and prototyping of the three kinds of telescopes foreseen. The telescope activities are grouped under the task leaderships SST, MST and LST for respectively the telescope systems of small, medium and large size together with SCT for the Schwartzschild-Couder telescope of medium size proposed by CTA-USA. These leadership tasks also encompass the responsibility for the system engineering of the systems. The corresponding work packages: SST-STR; MST-STR; LST-STR and SCT-STR will provide comprehensive studies of the structure, mirrors, motorization for each telescope system. The work packages TEL (Telescope Structures) and MIR (Mirrors) were started during the Design Study phase and have provided the basis for the STR work packages. The TEL work package continues to develop a common approach regarding the types and control of telescope servo

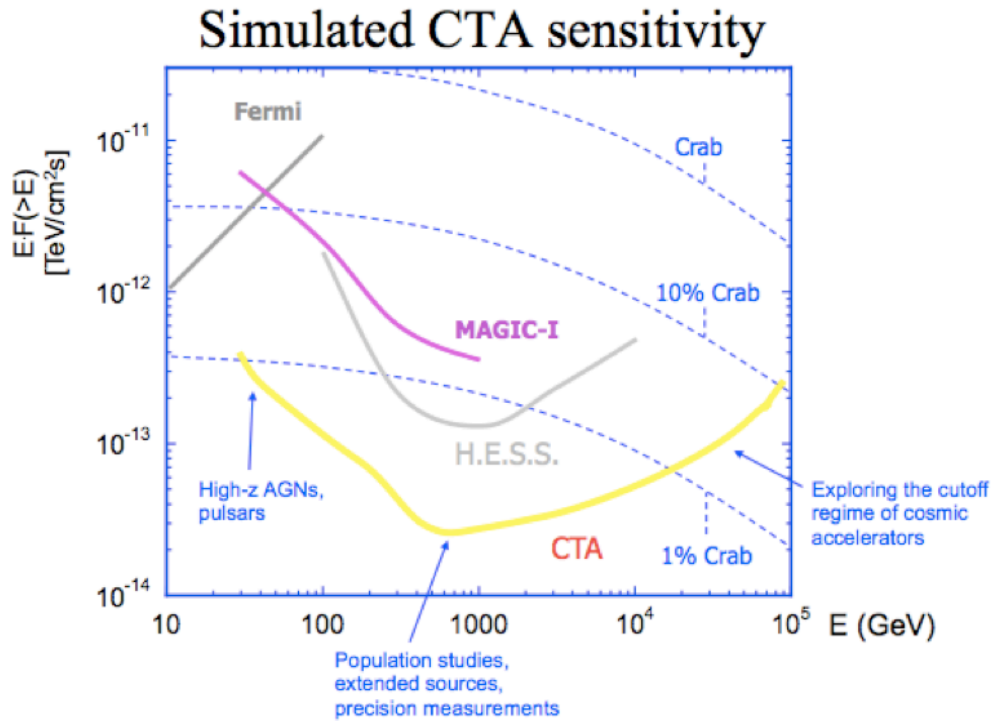

*Figure 2:* Sensitivity obtained from Monte Carlo simulations for a Southern CTA telescope array with an estimated cost within the initial budget (see text) compared with the sensitivity of some existing installations.

Besides that effort, CTA continues the exploration of the physics potential of CTA and its feedback in the optimization studies through the MC group, that performs the Monte-Carlo simulations to characterize the performance of candidate realizations of CTA. The optimization of the array layout and general telescope characteristics are achieved via quantitative assessment of candidate array performance on key science topics. This work package also provides input for the development of reconstruction and analysis tools.

In what concerns the technical aspects, one of the main systems, and the development of implementation guidelines regarding interlocks and safety systems. The MIR work package continues to investigate different mirror production technologies and reflective coatings, to develop methods to evaluate mirror facets by measuring wavelength-dependent local and global reflectivity and facet point spread function and to carry out accelerated mirror aging tests. The work packages SST-CAM; MST-CAM; LST-CAM and SCT-CAM are responsible for the technical design of the cameras for the corresponding telescopes. These activities will take over at an appropriate time from the work packages FPI (Focal Plane Instrumentation) and ELEC (Camera Electronics) that started during the Design Study phase. FPI provides common concepts for the photon capture and conversion part of the imaging cameras, including, for example, light sensors and amplifiers, light guide systems, power supplies for the light sensors, mechanics and cooling systems of the cameras. FPI is responsible for the evaluation of components from different manufacturers, especially of photon detectors and HV systems, and for developing methods of quality control for mass production. ELEC is concerned with the trigger, digitization and readout electronics of the cameras, with the aim of deriving cost effective solutions that can be used across several telescope types. ELEC will optimize readout options, which can then be adapted by the camera prototype projects and later will evaluate inter-telescope trigger options following existing concepts. In parallel, ELEC will develop and provide tools to evaluate the performance of prototype cameras, and test and verify electronics sub-assemblies during mass production.

Since the Atmosphere is an integral part of our instruments, a dedicated work-package ATAC (Atmospheric Monitoring and Calibration) is focused towards atmospheric monitoring, associated science and instrument calibration, with the goal of better understanding the atmosphere as part of the detector and the calibration of the optical properties of the telescopes.

Another fundamental Preparatory Phase activity the software development. There the work is splitted into three work-packages, namely ACTL, CEIN and DAFA. ACTL (Array Control) will design a failure-tolerant system integrating both an automatic control layer and a software layer for the control and operation of single telescopes as well as the telescope array. CEIN (Computing and E-Infrastructures) will develop and implement the infrastructures for efficient and high performance data management and data archiving. The work package will develop the science user gateway including user-configured tools for access to shared computing resources, tools for data access, data processing, and data analysis as well as the implementation plans for the offline e-infrastructure. DAFA (Data Facilities) is responsible for defining and implementing data formats, data archives, and analysis tools to guarantee the efficient analysis of data.

A crucial activity during the Preparatory Phase is the site selection and two work-packages deal with this issue: SITE (Site Characterization) identifies possible locations for the arrays of the CTA Observatory based on meteorological, geological, geographical, political and social criteria and SDEV (Site Development) provides a plan for the development of, and infrastructure at, the sites of the CTA observatory.

For what concerns the administrative aspects, the group of work packages LEGAL, GOV and FINANCE collectively studies possible legal frameworks, governance schemes and financial models for organizing, constructing and operating CTA. Recommendations will be made to the Consortium and to the Resource Board. PROC (Procurement) develops policies for procurement and industrial involvement for the implementation phase of the CTA project and to produce an industrial capability register. The related work package IRD will establish a coherent interfacing of technical work packages with companies to promote industrial R\&D where possible.

Last but not least, OUTR (Outreach) will provide the initial means of engaging the public as well as decision makers with CTA and produce an Outreach Plan for the future phases of CTA.

Summarizing, the CTA (Cherenkov Telescope Array) consortium is meeting the challenge of preparing an installation fulfilling the above goals in a Preparatory Phase work that started by fall 2010 and which aims to result in 2013 in the start of the construction of two arrays of telescopes that meet the requirements explained.

More information about the CTA project and its Preparatory Phase organization and activities can be found in http://www.cta-observatory.org/

**Summary**

The Cherenkov Telescope Array (CTA) is presently the worldwide project for the next generation of ground-based Cherenkov Telescopes for Very High Energy (VHE) Gamma ray astronomy.

The CTA project started in 2010 a three-year EU-funded Preparatory Phase with the goal of being ready to start construction by the end of 2013. For that phase the vast amount of activities needed is organized in a number of Work Packages whose work is advancing.
Prototypes of components, systems and even some telescopes are being constructed, which shall allow the final decisions to be made and construction to begin in 2013, with aim of completing the deployment of both arrays by 2018. CTA will then be the major observatory in VHE gamma-ray astronomy, combining guaranteed astrophysics and physics returns with significant discovery potential.

**Acknowledgements**

We gratefully acknowledge financial support from the agencies and organisations listed in this page: http://www.cta-observatory.org/?q=node/22

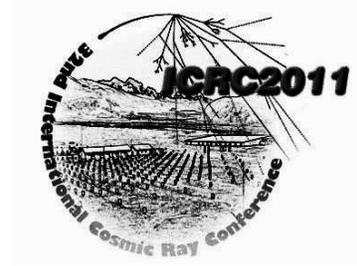

# Performance studies of the CTA observatory

F. DI PIERRO[1], K. BERNLÖHR[2], C. FARNIER[3], J.A. HINTON[4], J.-P. LENAIN[3], H. PROKOPH[5], V. STAMATESCU[6] AND P. VALLANIA[1] FOR THE CTA CONSORTIUM

[1] *Istituto di Fisica dello Spazio Interplanetario dell'Istituto Nazionale di Astrofisica, Torino, Italy*
[2] *Max-Planck-Institut für Kernphysik, P.O. Box 103980, D 69029 Heidelberg, Germany*
[3] *ISDC Data Centre for Astrophysics, Chemin d'Ecogia 16, 1290 Versoix, Switzerland*
[4] *Dept. of Physics and Astronomy, University of Leicester, Leicester, LE1 7RH, UK*
[5] *DESY, Platanenallee 6, D 15738 Zeuthen, Germany*
[6] *Institut de Fisica d'Altes Energies (IFAE), Edifici Cn. Universitat Autonoma de Barcelona, 08193 Bellaterra (Barcelona), Spain*

*dipierro@ifsi-torino.inaf.it*

**Abstract:** The Cherenkov Telescope Array (CTA), currently in the Preparatory Phase, will study gamma rays in the energy range from a few tens of GeV to >100 TeV with unprecedented sensitivity and angular resolution. In order to cover such a vast energy range, an array of three complementary telescope types is envisaged: a small number of Large Size Telescopes (LST, 23 m-diameter dish) for the low-energy range (below 200 GeV), few tens of Medium Size Telescopes (MST, 12 m dish) for the central energy range (100 GeV - 10 TeV) and a large number of Small Size Telescopes (SST, 4-7 m dish) for the high energy range (above 10 TeV). The optimization of the performance (sensitivity, angular and energy resolutions) of such a complex system requires intensive simulation activities. Some preliminary results in terms of off-axis performance, operation under moonlight conditions and high altitude performance are reported and discussed.

**Keywords:** CTA, Imaging Atmospheric Cherenkov Telescope, gamma-rays, Monte Carlo simulations

## 1 Introduction

CTA (Cherenkov Telescope Array) [1] is an international project included in the European roadmap for research infrastructures [2]. The project aims at building a large ground-based gamma-ray observatory for the study of the Universe in the Very High Energy (VHE) range of the electromagnetic spectrum, from tens of GeV to >100 TeV, through the observation of the Cherenkov light emitted by Extensive Air Showers (EAS) of particles initiated by gamma-rays. The goal of CTA is an order of magnitude improvement in sensitivity with respect to the current operating Imaging Atmospheric Cherenkov Telescope (IACT) arrays (HESS, MAGIC and VERITAS), together with expanded field-of-view and dramatically improved angular and energy resolution. The observatory will consist of ∼100 Cherenkov telescopes operating from two sites in the northern and southern hemispheres, and will be designed and built by a consortium of scientific institutes belonging in 25 countries worldwide. CTA is currently in a Preparatory Phase, which began in October 2010 and will run for three years; during this phase the project design will be defined and tested through the realization of prototypes. The Monte Carlo working group [1] plays a key role in this phase, as detailed simulations of the development of show-ers, and their detection and reconstruction, are required to characterise the performance of the instrument.

Presently operating IACTs have a field of view limited to typically 4°−5° in diameter [3, 4]: the planned increase up to ∼8° could greatly improve the full sky survey capability, the potential for serendipitous source discovery and the analysis of very extended objects. Dedicated Monte Carlo studies are underway to study the off-axis performance of CTA and are critical to the science-based optimisation of the observatory design. Another key input to this process is the duty cycle of the observatory and the array performance under non-optimal conditions. Studies of performance under high Night Sky Background (NSB) conditions are in progress: the nominal duty cycle of ∼ 1000 h/y (11%) due to solar, lunar and weather constraints may be significantly extended by operating under moonlight (and twilight) conditions.

## 2 The Monte Carlo simulations

The simulations performed to optimize CTA performances consist of three major steps: 1. the development of the extensive air shower in the atmosphere and the emission of Cherenkov light; 2. the response of the telescopes; 3. The analysis of the simulated raw data. The EAS development is calculated using the CORSIKA program [5], while the *sim_telarray* [6] program implements the ray-tracing, photon detection, and the electronics for trigger and signal processing. Finally different analysis programs perform: image cleaning, image analysis (in the standard pipeline based on second-moment Hillas parameters [7] ) and stereoscopic event reconstruction (see e.g. [8]). The aim of the simulation is to optimize the array sensitivity over the whole energy range, keeping close to optimal angular and energy resolution. In order to reach this goal about $2.5 \times 10^9$ $\gamma$-ray events and a factor 20 more background events (hadrons and electrons) have been simulated using several different array layouts. In addition to those presented here, many activities are currently in progress concerning, for instance: the optimization of the LST design, the impact of field of view, pixel size and Schwarzschild-Couder optical design for the SST, the trigger strategy, the digitization sampling rate and the dynamic range, and the optimization of analysis algorithms. Fig. 1 shows the configuration used for the sensitivity curves which are presented in the following sections.

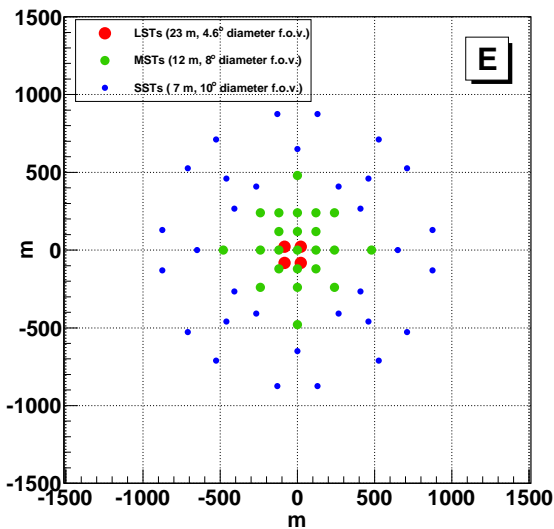

Figure 1: Layout of one of the configurations giving the best sensitivity/cost in the whole energy range (from a few tens of GeV to >100 TeV).

## 3 The off-axis performance

A key goal for the CTA observatory will be to perform surveys of a significant fraction of the sky. The southern array will be used to perform a Galactic plane survey as well as targeted observations of Galactic and extragalactic objects and will cover the full energy range, and the northern array is optimized for lower (from a few tens of GeV to a few

TeV) energies, and will be focussed on the study of extragalactic sources. For each telescope type in these arrays, a wide field of view (f.o.v.) will benefit the survey capability, both in terms of producing a relatively flat off-axis response at higher energies, and by improved determination of the residual cosmic-ray background through the selection of more off-source background-control regions.

We investigate the off-axis performance of CTA by using simulated diffuse gamma-ray showers, that have been produced at off-axis angles up to 10°. The data are analyzed for the CTA candidate array E [1], that consists of 4 LSTs (4.6° diameter f.o.v.), 23 MSTs (8° diameter f.o.v.) and 32 SSTs (10° diameter f.o.v.). Given that the analysis method used in this study is optimized for on-axis gamma rays, the current off-axis performance estimate may be considered to be conservative.

The off-axis differential sensitivity is estimated by first determining an off-axis cut on the reconstructed direction in each energy bin. This cut is obtained by scaling the corresponding optimum on-axis cut by the relative (with respect to on-axis) worsening in the off-axis angular resolution, which is shown in Fig. 2 for several off-axis angles. The on-axis performance is obtained with an analysis method that delivers marginally better differential sensitivity at the expense of worse angular resolution below ∼1 TeV, with respect to the standard analysis pipeline on-axis performance [1]. Optimized on-axis cosmic-ray background rejection and telescope multiplicity cuts are applied to the diffuse gamma, proton and electron data, and relative effective areas (with respect to on-axis) are determined as a function of the off-axis angle for each primary particle type. The off-axis gamma-ray and cosmic-ray background rates for each energy and off-axis angle bin are obtained by scaling the on-axis rates with the corresponding relative effective area and with the relative (with respect to on-axis) difference in reconstructed direction cut efficiency. The minimum detectable flux is then determined by demanding for each energy bin a minimum 5 σ detection, at least 10 gamma-ray events and a minimum gamma-ray excess of at least 5% of the residual cosmic-ray background. Fig. 3 shows the expected differential sensitivity for different off-axis angles. The plot shows, for instance, that at an energy of ∼12.6 TeV the differential sensitivity degrades by a factor of ∼1.4 at 1.25° off-axis and by a factor of ∼4.8 at 3.25° off-axis.

## 4 The moonlight operations

Traditionally Cherenkov telescope arrays have made observations only during astronomical darkness. The presence of moonlight reduces the observation time and hampers the possibility of observing the entire evolution of transient and variable phenomena. An important example of this is given by the recently discovered (by Agile and Fermi) flare activity from the Crab Nebula, The major flares in September 2010 [9] and April 2011 [10] both happened very close to full Moon. Moonlight observations are possible but the increase in NSB during moonlight and twilight conditions leads to a higher trigger and analysis thresholds and to less sensitivity (at least at low energies). Monte Carlo simulations have been carried out to assess the performances of the CTA array under these conditions.

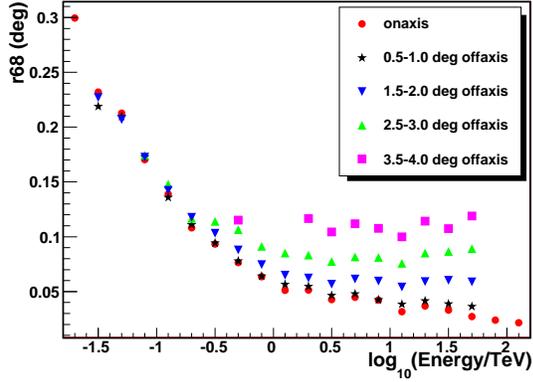

Figure 2: Angular resolution of CTA candidate array E, given as the 68% event containment radius, as a function of estimated energy and off-axis angle. The on-axis angular resolution of the array is also shown. The missing points are due to limited Monte Carlo statistics.

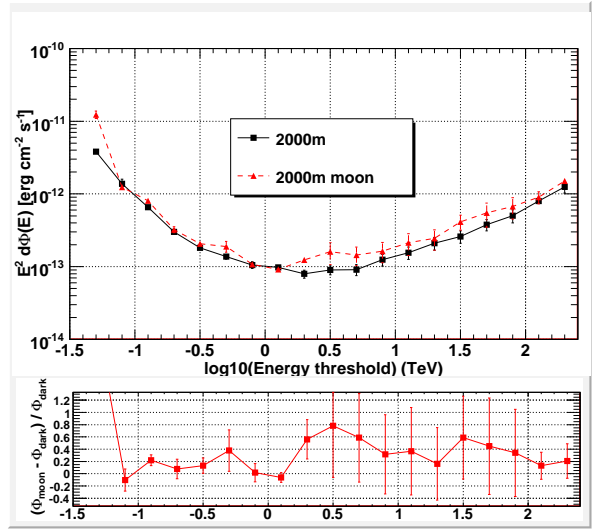

Figure 4: *Top:* Expected sensitivity of CTA candidate array E with (in red) and without moonlight (in black), for 50 h of observations. *Bottom:* Deviation of the sensitivity with moonlight with respect to the one under dark conditions.

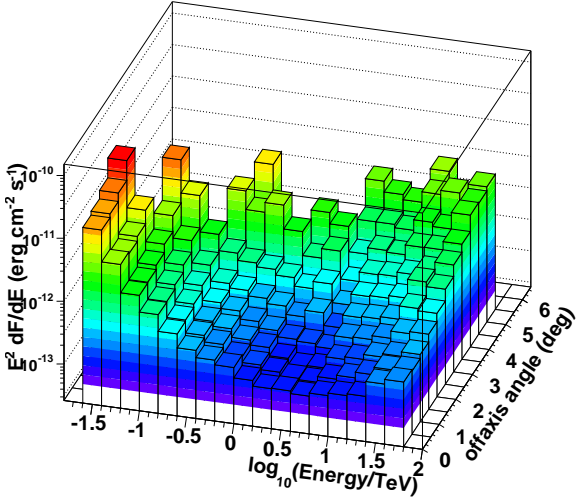

Figure 3: Differential sensitivity (see text for details) of CTA candidate array E calculated as a function of off-axis angle.

The adopted NSB level simulates a lunar phase of ~60%, with the telescopes pointing in a direction of 90° from the Moon, which corresponds to a NSB level 4.5 times higher than under standard, dark conditions. The performance is computed using the standard moment analysis for $\gamma$/hadron separation [7], for 50 h of observations, with $5\sigma$ detection per energy bin. The sensitivity under high NSB conditions is presented in Fig. 4 for the array candidate E, compared to the sensitivity without moonlight. For the used analysis algorithm, a pre-cleaning of the camera images is used, based on the amplitude of the signal contained in each pixel. In case of dark condition observations (standard NSB), only the pixels with a content at least 10 times higher than the RMS of the noise ($\mathrm{Amp_{mean}}$) of the camera surrounded by pixels with a content of 5 $\mathrm{Amp_{mean}}$ are kept, commonly referred as 5/10 image cleaning. For high NSB observations, the energy threshold implied by the higher value of $\mathrm{Amp_{mean}}$ appeared to be a too stringent condition. Therefore, a new image cleaning of 4/7 was introduced for the treatment of simulated data acquired in moonlight condition. The corresponding energy threshold is higher compared to dark conditions, at about 80 GeV, however the moonlight sensitivity almost recovers the level under low NSB at higher energies. The expected angular and energy resolutions are roughly not affected by moonlight. Same results are found for other array candidates for CTA. This good performance of CTA with high NSB indicates that it is possible to observe even when the Moon is partially present, increasing the observing time and mainly the continuity of the light curves. The expected gain for the duty cycle is about 30%, corresponding to a total observing time of ~1300 h per year.

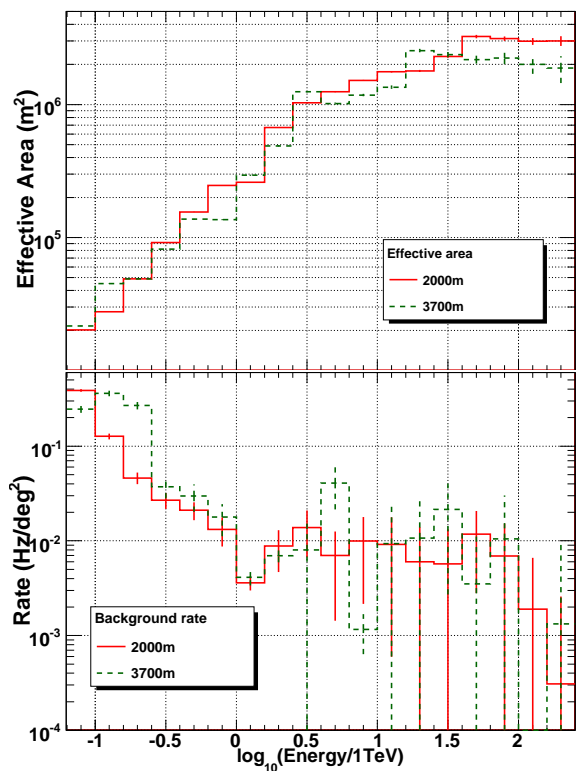

Figure 5: Effective area for the CTA candidate array E of gamma rays (*top*) and background rate (*bottom*) after cuts for standard and high altitude sites.

## 5 High altitude performance

A consequence of a high altitude site for CTA is to reduce the typical distance to the shower maximum and hence increase the density of Cherenkov light at the observation level and reduce the energy threshold somewhat. This is due to the smaller diameter of the light pool, which in general will also lead to reduced effective area and/or telescope multiplicity, which may degrade performance at high energies. To study this effect in detail and help in the selection of the CTA a set of shower simulations has been performed for a site located at 3700 m above sea level (a.s.l.). Data are analyzed using the same 5/10 image cleaning described in the previous section. For energies above few hundred GeV, the results of the high altitude simulations show that for identical array layout, the collection area and background rate follow the same trend of the simulations performed for at 2000 m a.s.l. array (Fig. 5). Consequently, the sensitivities in the energy range between few hundred GeV and few tens of TeV at the two altitudes are comparable within a factor of 2. The angular resolution does not significantly differ from the one for a site at 2000 m. For very high energy showers (E>10 TeV), the energy resolution degrades with respect to the 2000 m altitude performance for some array layouts.

## 6 Conclusions

Monte Carlo simulations are crucial to define the array and telescope configuration (including number of telescopes, spacing and layout, mirror diameter, optical psf and pixel size) necessary to get the required performance in the relevant energy range. The array design must therefore be driven by simulation results, but with constant feedback on costs and feasibility from the technical work packages. At the moment the Monte Carlo work has helped to define many of the key parameters of the telescopes to be built (in particular for the mid-sized telescope). However, many challenges remain in this area, including the comparison of dual mirror and single mirror designs for the SST and also at intermediate energies. New methods of reconstruction, based on image time gradients or on image fits from a EAS model, advanced gamma/hadron separation methods, using new discrimination variables and Multi Variate Analysis [11, 12], give already significantly better sensitivities and will be used to test of influence of the site altitude on the performance.

## Acknowledgements

We gratefully acknowledge support from the agencies and organisations listed in this page: http://www.cta-observatory.org/?q=node/22

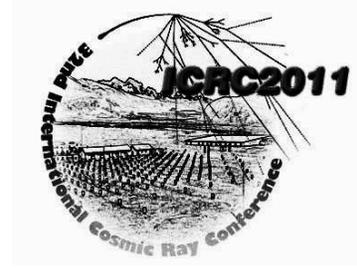

## Mirror Development for CTA

A. FÖRSTER[1], T. ARLEN[2], A. BONARDI[3], P. BRUN[4], R. CANESTRARI[5], P. H. CARTON[4], P. CHADWICK[6], G. DECOCK[4], M. DORO[7], D. DURAND[4], A. ETCHEGOYEN[8], LL. FONT[7], E. GIRO[9], J. F. GLICENSTEIN[4], M. GOMEZ BERISSO[10], S. HERMANUTZ[3], C. JEANNEY[4], A. KNAPPY[6], L. LESSIO[9], M. MARIOTTI[9], M. C. MEDINA[4], J. MICHALOWSKI[11], P. MICOLON[4], J. NIEMIEC[11], G. PARESCHI[5], B. PEYAUD[4], G. PÜHLHOFER[3], F. SANCHEZ[8], C. SCHULTZ[9], A. SCHULZ[12], K. SEWERYN[13], C. STEGMANN[12], F. STINZING[12], M. STODULSKI[11], V. VASSILIEV[2] ON BEHALF OF THE CTA CONSORTIUM, AND C. FABER[14], G. HÄUSLER[14], R. KROBOT[14], E. OLESCH[14]

[1] Max-Planck-Institut für Kernphysik, Heidelberg, Germany
[2] University of California Los Angeles, California, USA
[3] IAAT, Universitaet Tuebingen, Tuebingen, Germany
[4] IRFU, CEA, Saclay, France
[5] INAF - Osservatorio Astronomico di Brera, Merate, Italy
[6] University of Durham, Durham, United Kingdom
[7] Universitat Autonoma de Barcelona, Bellaterra, Spain
[8] Instituto de Tecnologias en Deteccion y Astroparticulas (CNEA-CONICET-UNSAM), Buenos Aires, Argentina
[9] INFN, Padova, Italy
[10] Centro Atomico Bariloche and Instituto Balseiro (CNEA-CONICET- UNCuyo), San Carlos de Bariloche, Argentina
[11] Institute of Nuclear Physics, Polish Academy of Sciences IFJ-PAN, Krakow, Poland
[12] ECAP, Universitaet Erlangen-Nuernberg, Erlangen, Germany
[13] Space Research Center - Polish Academy of Science, Warsaw, Poland
[14] Institute of Optics, Universitaet Erlangen-Nuernberg, Erlangen, Germany

andreas.foerster@mpi-hd.mpg.de

**Abstract:** CTA will be an array of Imaging Atmospheric Cherenkov Telescopes (IACTs) for VHE gamma-ray astronomy with a proposed total mirror area of approximately 10000 square meters. The challenge is to develop lightweight and cost-efficient mirrors with high production rates and good long-term durability. Several technologies are currently under rapid development: sandwich structures based on carbon/glassfibre-epoxy composite materials and monolithic carbon fibre structures, either with glass or epoxy surfaces; cold-slumped glass sheets with aluminium honeycomb or glass foam as structural material; all-Aluminium mirrors. New surface coatings are under investigation with the aim of increasing the reflectance and long-term durability. In addition, new methods for a fast and reliable testing of thousands of mirrors are being developed.

**Keywords:** CTA, Imaging Atmospheric Cherenkov Telescope, Gamma-rays, Optics

## 1 Introduction

In recent years, ground-based very-high energy gamma-ray astronomy has experienced a major breakthrough demonstrated by the impressive astrophysical results obtained with IACT arrays like H.E.S.S., MAGIC, and VERITAS [1]. The Cherenkov Telescope Array (CTA) project is being designed to provide an increase in sensitivity of at least a factor ten compared to current installations along with a significant extension of the observable energy range down to a few tens of GeV and up to > 100 TeV [2]. To reach the required sensitivity, several tens of telescopes will be needed with a combined mirror area of up to 10000 $m^2$. Current design studies investigate three telescope sizes: small-sized telescopes with a diameter of approximately 6 m, several medium-sized telescopes (12 m) and large-sized telescopes (23 m). In addition, telescopes with dual mirror optics (Schwarzschild-Couder configuration) are under investigation.

The individual telescopes will have reflectors of up to 400 $m^2$ area. The requirements for the focal point spread function (PSF) are more relaxed compared to those for optical telescopes. Typically, a PSF below a few arcmin is acceptable which makes the use of a segmented reflector consisting of small individual mirror facets (called mirrors in the following) possible. IACTs are usually not protected by domes, the mirrors are permanently exposed to the environment. The design goal is to develop low-cost, lightweight, robust and reliable mirrors of $1 - 2$ m$^2$ size with adequate reflectance and focusing qualities but demanding very little maintenance. Current IACTs mostly use polished glass or diamond-milled aluminium mirrors, entailing high cost, considerable time and labour intensive machining. The technologies currently under investigation for CTA pursue different methods such as sandwich concepts with cold-slumped surfaces made of thin float glass and different core materials like aluminium honeycomb, glass foams or aluminium foams, constructions based on carbon fiber/epoxy or glass fibre substrates, as well as sandwich structures made entirely from aluminium.

## 2 Basic specifications

The mirrors for the CTA telescopes will be hexagonal in shape, with an anticipated size between $1 - 2$ m$^2$, well beyond the common size of $0.3 - 1$ m$^2$ of the currently operational instruments. IACTs are normally placed at altitudes of $1,000 - 3,000$ m a.s.l. where significant temperature changes between day and night as well as rapid temperature drops are quite frequent. All optical properties should stay within specifications within the range $-10°$C to $+30°$C and the mirrors should resist to temperature changes from $-25°$C to $+60°$C with all possible changes of their properties being reversible.

Intrinsic aberrations in the Cherenkov light emitted by atmospheric showers limit the angular resolution to around 30 arcsec [3]. However, the final requirements for the resolution of the reflectors of future CTA telescopes, i.e. the spot size of the reflected light in the focal plane (camera), will depend on the pixel size of the camera and the final design of the telescope reflector. There is no real need to produce mirrors with a PSF well below the half of the camera pixel size, which is ordinarily not smaller than 5 arcmin. A diffuse reflected component is not critical as long as it is spread out over a large solid angle. The reflectance into the focal spot should exceed 80% for all wavelengths in the range from 300 to 600 nm, ideally close to (or even above) 90%. The Cherenkov light intensity peaks between 300 and 450 nm, therefore the reflectance of the coating should be optimized for this range.

## 3 Test facilities

The standard way to determine the PSF of such mirrors is a so-called $2f$-setup: the mirror is placed twice the focal distance $f$ away from a pointlight light-source and the return image is recorded using a CCD or photodiodes. Using waveband filters or narrowband LEDs measurements at different wavelengths are possible. Normalizing for the intensity of the light-source the total directed reflectance into the focal spot can be estimated as well. Comparable setups currently exist in several institutes involved in the development and characterization of CTA mirrors.

While being a reliable method, $2f$-measurements need a lot of space (several 10s of meters) and are rather time-intensive. An alternative approach with a compact setup especially for testing huge numbers of mirror is being pursued at the University of Erlangen: Phase Measuring Deflectometry (PMD) [4, 5]. The basic idea of PMD is to observe the distortions of a defined pattern after it has been reflected by the examined surface and from them to calculate the exact shape of the surface. For this, sinusoidal patterns are projected on a screen and cameras take pictures of the distortions of the patterns due to the reflection on the mirror surface. The primary measurement of PMD is the slope of the mirror in two perpendicular directions. A map of the mirror's curvature can be calculated by differentiating the slope data. Using a ray-tracing script in which the normal and slope data from the PMD measurements are the input parameters, it is possible to calculate the PSF at arbitrary distances from the mirror.

IACTs usually operate without domes and the mirrors are exposed to the environment for many years. Therefore, an extensive set of long-term durability tests is being defined by the University of Durham, trying to use ISO standards wherever applicable. Apart of classical temperature and humidity cycling for accelerated aging the intended test series involve corrosion tests in salt fog atmospheres, abrasion tests by sand blasting, pull tests with sticky tape to check the adhesion of the coating, or tests of the influence of bird faeces on the reflective coating.

## 4 Technologies under investigation for CTA mirrors

Several institutes within the CTA consortium are developing or improving different technologies to build mirrors, most of which are in a prototyping phase at moment:

### 4.1 All-aluminium mirrors

The entire reflector of MAGIC I and more than half of the MAGIC II mirrors are made of a sandwich of two thin aluminium layers interspaced by an aluminium honeycomb structure that ensures rigidity, high temperature conductivity and low weight, as shown in fig. 1a [6]. The assembly is then sandwiched between spherical moulds and put in an autoclave, where a cycle of high temperature and pressure cures the structural glue. The reflective surface is then generated by precision diamond milling. The final roughness of the surface is around 4 nm and the average reflectance is 85%. The aluminium surface is protected by a thin layer of quartz (with some admixture of carbon) of around 100 nm thickness. For CTA, this technology is being further developed especially by the use of either a thin coated glass sheet as the front layer or a reflective foil to reduce the cost imposed by the diamond milling of the front surface.

## 4.2 Glass replica mirrors

The basic concept of this method, originaly developed by INAF Brera, is to form a thin sheet of glass on a high precision mould to the required shape of the mirror and glue a structural material and a second glass sheet or other material to its back to form a rigid sandwich structure. This concept is being pursued by three institutes (INAF Brera, Italy, CEA Saclay, France, and Sanko, Japan). A sketch of the basic layout of these mirrors is shown in fig. 1b.

*INAF Brera, Italy*

Almost half of the reflector facets of MAGIC II are cold-slumped glass-aluminium sandwich mirrors [7, 8]. A thin sheet of glass is cold-slumped on a high precision spherical mould. This glass sheet, an aluminium honeycomb and a back sheet are then glued together with aeronautic glue. The shaped substrates are coated in the same way as traditional glass mirrors. For CTA R&D activities are going on to improve the process and to reduce the costs. While Al honeycomb is the baseline design, in addition the use of FoamGlass® as structural material is being investigated. This material has a low weight, $(0.1 - 0.165 \text{ g/cm}^3)$, a very low thermal expansion coefficient (CTE $\simeq 9 \, \mu\text{m K/m}$), is water tight, can easily be machined, has high strength and is very competitively priced.

*CEA Saclay, France*

A similar method is being pursued by the IRFU group at CEA (Saclay) [9]. Here as well a sandwich structure is formed by 2 glass sheets and an aluminium honeycomb core and the spherical shape of the front surface is created by cold-slumping the front sheet on a high-prescision mould. First hexagonal mirrors of 1.2 m flat-to-flat (the planned size for the medium-size telescopes of CTA) with 16.7 m focal length have been produced this way.

*Sanko, Japan*

The same technology is also being pursued by Sanko in Japan, concentrating on hexagonal mirrors with a size of 1.5 m flat-to-flat as planned for the large-sized telescopes of CTA. First prototypes have been produced and a closed-cell aluminium foam as alternative core material is being investigated.

## 4.3 Composite mirrors

Carbon fibre/epoxy based substrates have good mechanical properties and show the potential for fast and economical production in large quantities. The challenge is to produce mirrors with good surface qualities without labour-intensive polishing. In addition, variations of the same designs using glass fibre and/or aluminium as structural material are being studied. These types of composite mirrors are under development at CEA Saclay, France, SRC-PAS, Warsaw, Poland, and IFJ-PAS, Krakow, Poland.

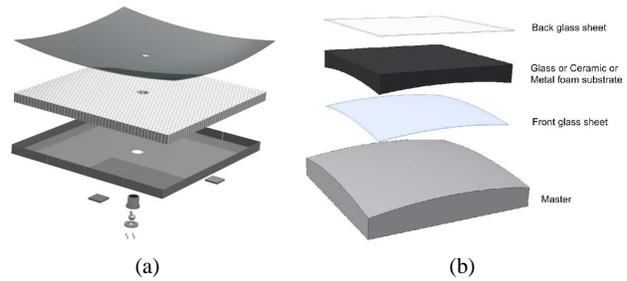

(a)            (b)

Figure 1: (a) All-aluminium mirrors (INFN Padova). (b) Cold-slumped glass mirrors (INAF Brera, CEA Saclay, Sanko)

*CEA, Saclay, France*

The CEA composite mirror design [9] has a core of rectangular strips of either carbon fibre, glass fibre or aluminium. On one side they are machined to the radius of curvature. To this core a front and a back sheet of the same material are glued, the front having been shaped on a mould with the appropriate radius of curvature. In a second step a thin glass sheet is glued to the front side, again using the mould, that is coated with a reflective coating. Several hexagonal mirrors of 1.2 m have been produced and are being tested. A principal sketch is shown in fig. 2a

*SRC-PAS, Warsaw, Poland*

The SRC is investigating the sheet moulding compound (SMC) technology, in which a composite material (Menzolit®) is formed in a spherical steel mould at high pressures (60 bar) and high temperatures (150°C). Menzolit has a carbon fibre content of 60%, a Young's modulus of $20 - 50$ GPa (depending on fibre direction) and 0% shrinkage. The moulding process takes approximately 10 min. The whole mirror structure is made as a single part and of one material, with ribs formed on the rear to increase mechanical stability. The spherical surface is formed by an in-mould coating process (IMC) during the forming process of the structure itself which is later coated or by using a reflective aluminium material called Alanod®. A sketch of such a composite mirror and the respective mould is shown in fig. 2b.

*IFJ-PAN, Krakow, Poland*

The composite structure under investigation is a rigid sandwich which consists of two flat panels of either carbon fibre, glass fibre or aluminium separated by perforated aluminium tubes of equal length. In a second step a spherical epoxy layer is formed on the front panel using a master surface. Alternatively front surfaces made of a cold-slumped glass sheet or of Alanod® are under investigation. The open sandwich structure enables good cooling and ventilation of the mirror panels and avoids trapping water inside the structure. The flatness and uniform thickness of the sandwich structure facilitates production, while the robustness of the structure ensures easy handling of the mirror. A sketch of the principal design is shown in Figure 2c.

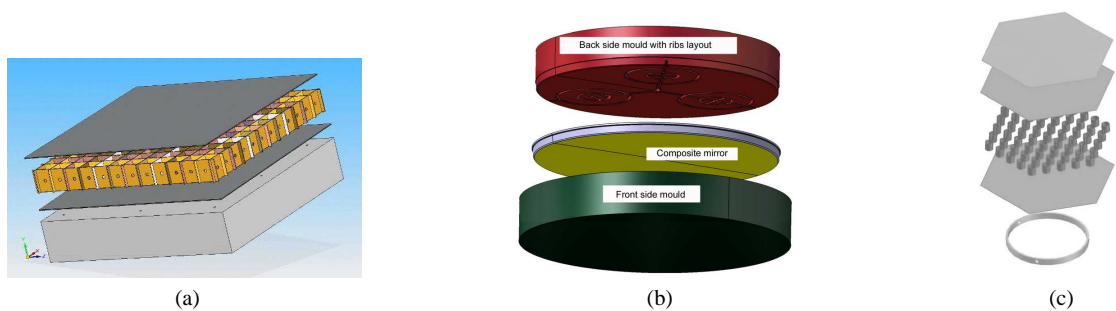

Figure 2: (a) Sandwich design with rectangular hoeneycomb (CEA Saclay). (b) Monolithic composite mirror (SRC-PAS Warsaw). (c) Open structure composite mirror (IFJ-PAN Krakow).

## 5 Reflective and protective coating

IACTs need to have a good reflectance between 300 and 600 nm wavelength which makes aluminium the natural choice as reflective material. The mirrors are exposed to the environment all year round, therefore this aluminium coating is usually protected by vacuum deposited $SiO_2$ (in the case of H.E.S.S.), $SiO_2$ with carbon admixtures (for MAGIC) or $Al_2O_3$ obtained by anodizing the reflective Al layer (in the case of VERITAS). Nevertheless, a slow but constant degradation of the reflectance is observed.

The Max-Planck-Institut für Kernphysik, Heidelberg, together with industrial partners, is performing studies to enhance both the reflectance and the long-term durability of mirror surfaces [10]. Coatings under investigation include: *a)* Multilayer dielectric coatings of alternating layers of materials with low and high refractive index (e.g. $SiO_2/HfO_2$) on top of the aluminization. Simple 3-layer designs are already able to increase the reflectance between 300 and 600 nm by 5%. *b)* Purely dielectric coatings without any metallic layer avoiding the rather low adhesion of aluminium on glass. These show a reflectance greater than 95% in the wavelength-region of interest and very low reflectance of only a few percent elsewhere. Extensive temperature and humidity cycling as well as corrosion tests in salt-fog atmospheres show a very stable long-term behaviour of these purely dielectric coatings. More extensive durability testing is ongoing at the moment. In addition, the H.E.S.S. experiment is re-coating the mirrors of its telescopes at the moment. 99 of these mirrors have been re-coated with a purely dielectric coating, several hundred with a three-layer protective coating on top of the aluminium layer, so that durability data from a real application in the field will become available.

In addition, the University of Tübingen is working on simulations to improve the design of the multi-layer coatings and operates a coating chamber for the production of small mirror samples to systematically study various coating options [11]. Furthermore, groups from Argentina and Brasil with experience in the field of mirror coating have joint the efforts recently.

## 6 Summary

The demand for a few thousand mirrors with a total reflective area of up to 10,000 $m^2$ for CTA is a challenge in quite a few aspects such as the production of large size facets of up to 2 $m^2$ in area, low weight ($\simeq$ 20 $kg/m^2$), high optical quality, easy and rapid series production and especially low costs. One of the major constraints is the requirement for a very slow degradation of the reflectance allowing at least 10 years of operation without re-coating. Currently, quite a few mirror technologies are under study with the goal to improve the performances substantially and to minimize the production and maintenance costs.

## Acknowledgements

We gratefully acknowledge support from the agencies and organisations listed in this page: http://www.cta-observatory.org/?q=node/22.

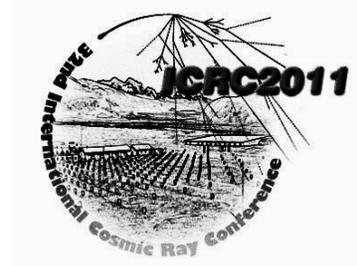

# Developments for coating, testing, and the alignment of CTA mirrors

A. BONARDI[1], J. DICK[1], A. FÖRSTER[2], E. KENDZIORRA[1], S. HERMANUTZ[1], G. PÜHLHOFER[1], S. SCHWARZBURG[1], A. SANTANGELO[1] FOR THE CTA COLLABORATION
[1] *Institut für Astronomie und Astrophysik Tübingen*
[2] *Max Planck Institut für Kernphysik*
*antonio.bonardi@uni-tuebingen.de*

**Abstract:** The telescopes of the Cherenkov Telescope Array (CTA) will have segmented mirrors, with mirror facets of ∼1-2.5 m$^2$ area. In the framework of the CTA mirror work package, the Institute for Astronomy and Astrophysics in Tübingen (IAAT) is participating in the developments for procedures to coat glass-substrate based mirror facets, is preparing a mirror facet test facility, and is prototyping Active Mirror Control (AMC) alignment mechanics and electronics. The developments are based upon the experiences the group has gained through its participation in the preparations for the 27 m dish of H.E.S.S. phase II. We will present the current status of our work and plans for future developments. The devices and procedures will be relevant for all classes of telescopes that will finally form CTA.

**Keywords:** Cherenkov Telescope, Optics

## 1 Introduction

The Cherenkov Telescope Array (CTA) consortium aims at deploying two arrays of Imaging Atmospheric Cherenkov Telescopes (IACT), in the Southern and in the Northern hemisphere, with a surface covering ∼3-5 km$^2$. At least three different telescope types are foreseen: A large size telescope (LST) type with ∼24 m dish diameter, a medium size (MST) with ∼12 m, and a small size (SST) with ∼4-7 m. The reflective surface of each telescope primary mirror will consist of segmented mirror facets of ∼1-2.5 m$^2$ area. The baseline design for the MST (a close to Davies-Cotton design with camera in the primary focus) foresees hexagonal segments with 1.2 m flat-to-flat diameter, and 84 mirrors per telescope.

The total reflective mirror surface of the entire CTA is huge (∼ $10^4$m$^2$), with hence several thousands of mirror segments. High efficiency during testing and alignment of the segments is therefore necessary. High durability of the mirror reflective surface is equally necessary to avoid the need for frequent re-coating. In general, segments should persist with sufficient reflectivity for at least 10 years. In this paper, activities by the Institute for Astronomy and Astrophysics Tübingen (IAAT) conducted in the field of mirror alignment, mirror testing, and mirror coating are reported. The work is performed in the framework of the CTA mirror work package.

## 2 Mirror coating

Current mirror facet specifications demand that the reflected light should largely be contained in a 1 mrad diameter area, the reflectance in the 300 nm ≤ λ ≤ 600 nm (hereafter WR$_{300-600}$) range should be ≥ 80%, and facets should be robust against aging for several years [1]. The goal is of course to provide coatings with the highest possible reflectivity and longest lifetime that is economically possible.

Currently, various designs for mirror facets are under study, including aluminum or carbon fiber honeycomb structures. The focus of the study presented here is on mirror types where a reflective layer on top of a glass substrate is needed. Such coating consists of a thin (100-1000 nm) single or multiple reflective layer plus, possibly, a protective overcoating.

### 2.1 Mirror coating study

The aim of our study is to obtain a mirror coating solution with reflectance > 90% (measured vertically close to the mirror surface) for the entire WR$_{300-600}$, with long term endurance. At the same time, it should be as simple as possible to be suitable for mass production. We currently use the McLeod [2] simulation software for optimizing the coating design. Coating of small glass samples is performed in the recently refurbished IAAT coating chamber (shown in Fig. 1).

The IAAT coating chamber is equipped with:

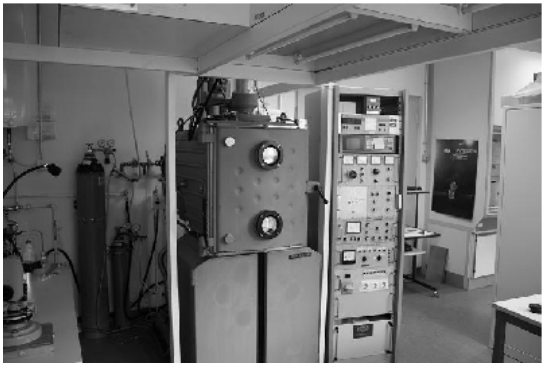

Figure 1: Coating chamber of the IAAT

- a rotative pump for vacuum production ($10^{-6}$ mbar);

- Nitrogen and Argon flooding system for humidity and impurities removal;

- water warming-cooling system (15-65 $^o$C);

- one resistor crucible for metals evaporation;

- a 4-fold electron beam crucible for dielectric materials evaporation;

- a single-sensor quartz micro-balance for measuring the deposited layer thickness.

For the coating, we focus along three main directions:

1. one metallic reflective layer plus a very robust protective overcoating;

2. one metallic reflective layer plus a multilayer interferometric overcoating for both protection and reflectance improvement;

3. a purely dielectric multilayer interferometric coating.

The first solution usually consists of an Al layer with a protective coating made of $SiO_2$ or $Al_2O_3$, widely in use for current IACT mirror facets because of its good performance, easiness, and low cost. The biggest drawback of this approach is that the long term mirror reliability achieved so far is probably not suitable for CTA scales.

The second solution is an intermediate solution enhancing a simple reflective layer with a protective coating, which will not only provide a protection to the metallic layer but also improve the reflectance in the $WR_{300-600}$.

The third solution is the most interesting and so far the least explored one. Light, passing from a material to another one with different refractive index, is reflected according to Fresnel's law

$$R = \left( \frac{n_1 - n_2}{n_1 + n_2} \right)^2$$

where $R$ is the reflection coefficient, $n_1$ and $n_2$ the refractive index of the first and second material.

For a light ray passing from one layer to another one, the reflectance depends also on its wavelength. The maximum value is for

$$\lambda = \frac{4 \cdot n_1 \cdot L_1}{\sin(\theta)}$$

where $\lambda$ is the reflected wavelength in vacuum, $L_1$ and $n_1$ the thickness and the refractive index of the first layer, and $\theta$ the impinging angle. By alternating many dielectric layers with different thickness and refraction index, it is possible to achieve very high reflectance inside the $WR_{300-600}$, with very low reflectance outside this WR. In this way, the Cherenkov light collection will be maximized, and at the same time the pollution by the Night Sky Background (NSB) – mostly at large wavelength – will be minimized. Furthermore, since no metallic layer will be used for the mirror coating, deterioration e.g. due to oxidation will not occur. On the other hand, a large number ($> 30$) of layers is required. This could result in layer deterioration in the field, because of thermally induced stress between different layers.

## 2.2 First results from mirror coating

For the first solution we focus on improving the resistance of the protective overcoating, on which the long term mirror reliability depends on. We have produced several samples in the coating chamber with different coating procedures (coating chamber and glass substrate cleaning, pumping time, deposition rate) in order to find the optimum one. Furthermore, we checked the agreement between simulation prediction and experimental measurement on the coated samples.

In Fig. 2 we report the experimental and simulated reflectance curve for almost vertical ($\theta = 83$ degrees) light of a glass sample coated with one 116 nm thick Al layer plus one 100 nm thick $SiO_2$ layer. The two lines show reasonably good agreement especially in $WR_{300-600}$, while for $\lambda < 300$ nm the agreement worsens to some extent. Whether this is a limitation of the software or due to variations in the coating procedure is still under investigation.

For a dielectric multilayer coating we found several solutions applying commonly used materials, like $HfO_2$, $ZrO_2$, $MgF_2$ and ZnS. In Fig. 3, the reflectance and transmittance for a multilayer coating as obtained by the McLeod simulation software [2] is shown. Such a design could not yet be realized in the current stage of completion of our coating chamber, since a multi-sensor quartz micro-balance is needed together with a substrate warming system, in particular for the ZnS deposition.

For the intermediate solution, we are about to realize different samples of $\sim 100$ nm thick Al coating with a protective overcoating made of $SiO_2$-$HfO_2$-$SiO_2$. This solution is not completely new, since it is long time used by industry and it has been recently applied for the the Hess I mirror re-coating.

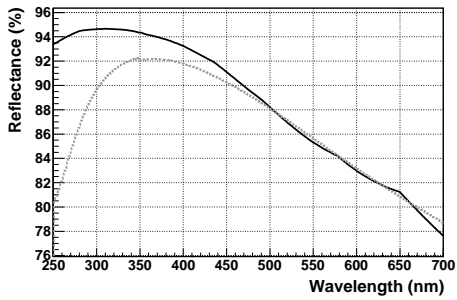

Figure 2: Reflectivity vs. wavelength for 83 degrees impinging angle light for a glass substrate coated with one Al layer 116 nm thick plus one $SiO_2$ layer 100 nm thick, simulation (black) and real data (grey dotted).

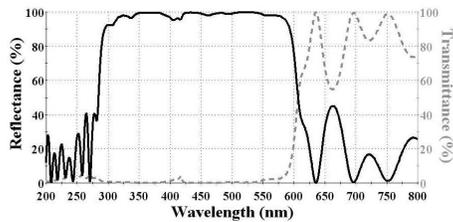

Figure 3: Simulation of reflectivity (black line) and transmittance (grey dotted line) vs. wavelength for normally impinging light for a 35 layer $ZrO_2$, $MgF_2$ and ZnS coated glass substrate.

Our aim is to put the realized samples through long term reliability tests. If the results will be positive, such a design might present a good alternative to the more complicated and expensive pure dielectric coating.

## 3 Mirror reflectivity and psf testing

Using a mirror test stand developed by the MPI-K Heidelberg, the reflectivity and spot size of all the mirror facets for the H.E.S.S. II telescope have been tested at IAAT. The setup is shown in Fig. 4 and summarized in the following:

- a halogen lamp lights up the mirror facet from a distance two times its focal distance;

- the reflected light converges on a target, placed at two times the mirror facet focal distance, producing a light spot;

- the reflected light spot is monitored with a digital camera to measure the spot size;

- the total reflectance at different wavelengths is measured with a photo-diode which is scanned across the light spot, using different optical filters.

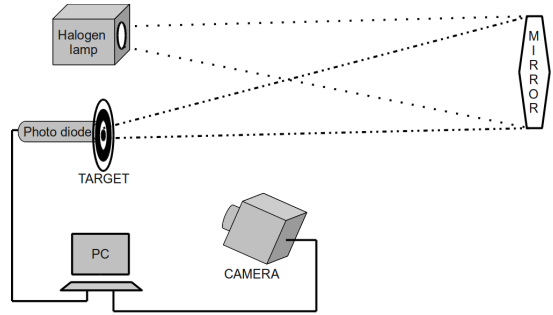

Figure 4: Current 2f mirror test setup based at IAAT.

We are working towards an improvement of such a mirror test setup by making the following changes:

- replacing the present digital camera with a new one sensitive additionally in the UV region. In this way the global reflectance can be measured avoiding the photo-diode;

- replacing the halogen lamp with four monochromatic high power LEDs with different wavelengths, covering the entire $WR_{300-600}$. The light source will be much more point-like and its time stability will be considerably improved. Furthermore, the optical filters and the lamp cooling system will not be needed anymore;

- installing an optical fiber system for monitoring the four LEDs, using the same digital camera used for observing the reflected light spot. Information about the emitted light, camera stability, and target aging will be constantly acquired, allowing a more precise measurement of the mirror reflectance.

The improved IAAT mirror test setup could be used for testing mirror prototypes as well as perhaps a good fraction of the total number of the CTA mirror facets.

## 4 Mirror alignment system

The large number of mirror facets, especially for the MST and LST types, demand motorized control of the mirror actuators, even if the mirror dish is stiff enough so that only an initial alignment after mirror mounting or mirror exchange is needed. In the current baseline design for the MST, the dish is expected to be indeed stiff enough so that an active alignment (i.e. frequent realignment during telescope observation time) is not needed. On the other hand, the LST structure will most likely require such active alignment. While the demand on mirror actuators is certainly lower for initial alignment procedures, nevertheless options which might be suited for both types of alignment are currently pursued.

The design which the IAAT is currently developing and testing is based on the actuator mechanics developed for the large, 27 m telescope of H.E.S.S. phase II [3]. Mirror facets are supported by two motor driven actuators and one freely-rotating bolt which is fixed to the telescope structure. Several of these actuators are currently being prepared to be tested in the course of the MST prototype development project led by DESY. Here, the actuators are connected to a supporting triangle, which is fixed to the structure of the telescope dish by two clamps. One actuator is tightly anchored to the supporting triangle not allowing any tilting movement; the second one is equipped with a pivot allowing a free rotation along a tilting angle. The motor of each actuator is housed in a watertight box vented by a sintered bronze valve. Based on the results of outdoor testing performed already at IAAT and also during the MST prototyping phase, any necessary improvements on the actuator mechanics will be made.

Compared to the electronics which is used to control the actuators at H.E.S.S. II [3], a new architecture based on the Controller Area Network (CAN) interface [4] will be employed. For each mirror there is a electronic control board, composed by a driver for each motor and a microcontroller (Atmel AT90CANxx), which is placed in one of the two actuator boxes. The other actuator is connected to the control electronics by an external cable. For each mirror, the two actuators plus the electronic control board form a Mirror Control Unit (MCU). In Fig. 5, an actuator with its motor and electronic control board are shown.

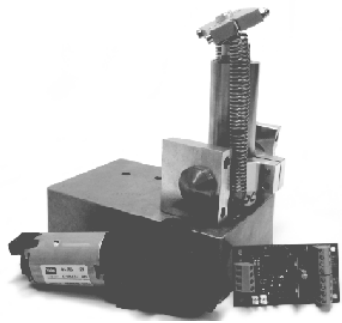

Figure 5: Mirror actuator together with its motor (normally housed in the bottom box) and the CAN control board.

MCUs are serially connected to a central unit. Each such unit serves one telescope, and can be placed in the center of the dish. The connection is realized with a four-wire shielded cable: two wires are used for transmitting the CAN signal, the other two for the power supply. A limitation of the chain length is not imposed by the CAN interface, but by the Ohmic power loss of the cables transmitting the motor current. As an alternative, more expensive cables could be used, but we found the best solution will consist of a distribution of the MCUs along different CAN chains, each of them serving 7-8 MCUs. The central unit, an embedded PC working as a TCP/IP to CAN gateway, is controlled by a PC. The control PC could be Client or Server and be placed in the telescope tower or in the central control room, depending on the architecture of the telescope control system. A power gate for the AMC DC power supply is also present in the telescope tower. The scheme of the actuator control system is shown in Fig. 6.

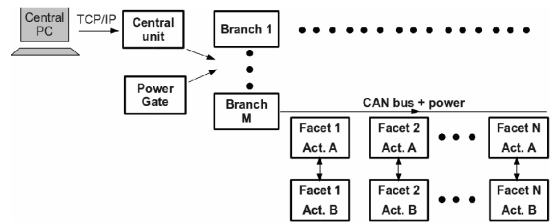

Figure 6: Scheme of the actuator control system developed for CTA telescopes.

While in the H.E.S.S. II control frame only one mirror facet can be aligned at a time[3], with our new frame it will be possible to align a large number of mirror facets at the same time. The only limitation will be given by the AMC power consumption.

Furthermore, such control frame is extremely simple and robust as evidenced by our laboratory and external tests. During different tests, we transmitted various millions of CAN commands without experiencing any transmission failure. It is also extremely flexible and can be easily adapted to any of the CTA telescope types requiring motorized actuators.

## Acknowledgement

This work has been partially funded by the BMBF/PT-DESY, grants 05A11VT1 and 05A10VTA. We gratefully acknowledge support from the agencies and organizations listed in this page: http://www.cta-observatory.org/?q=node/22 .

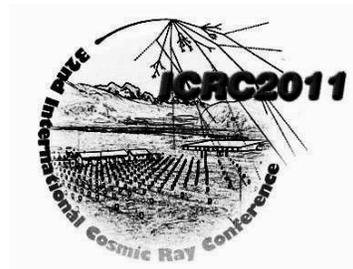

# High-reflectance, High-durability Coatings for IACT Mirrors

A. FÖRSTER[1], R. CANESTRARI[2], P. CHADWICK[3], A. KNAPPY[3] ON BEHALF OF THE CTA CONSORTIUM

[1] Max-Planck-Institut für Kernphysik, Heidelberg, Germany
[2] INAF - Osservatorio Astronomico di Brera, Merate, Italy
[3] University of Durham, Durham, United Kingdom
andreas.foerster@mpi-hd.mpg.de

**Abstract:** Increased reflectance and longterm durability are major goals in the development of mirrors for CTA, the future observatory for very-high energy gamma-ray astronomy. Multilayer protective coatings on top of an aluminium coating and especially purely dielectric coatings show an increase in reflectance by 5 to 10% and a significantly better resistance to environmental impact simulations in the laboratory than aluminium coatings with a single $SiO_2$ protective layer as are used in present day experiments.

**Keywords:** CTA, Imaging Atmospheric Cherenkov Telescope, Gamma-rays, Optics

## 1 Introduction

Imaging Atmospheric Cherenkov Telescopes (IACTs) for very-high energy (VHE) gamma-ray astronomy image Cherenkov light of particle showers in the atmosphere onto a photosensitive detector. The wavelength range of interest is roughly between 300 and 600 nm. Typically IACTs have tesselated mirror areas of the order 100 $m^2$ and larger. The current standard (e.g. in H.E.S.S., VERITAS and partially MAGIC) are mirrors with glass surfaces, coated on the front surface with aluminium (Al) and a single protective layer (e.g. $SiO_2$, $Al_2O_3$). Not being protected by a dome, the mirrors are constantly exposed to the environment and show a loss of reflectance of a few per cent per year. This requires re-coating of all mirrors after approximately 5 years. For the future CTA observatory (see [1]) with a total planned mirror area of about 10000 $m^2$ this would mean a significant maintenance effort. Coatings which increase the lifetime of the mirrors can therefore play a major role in keeping the maintenance costs of the observatory lower. If new coatings in addition show a better reflectance than the classical Al + $SiO_2$ coatings the sensitivity of the instrument will be increased.

## 2 New Coating Options

Aluminium coatings with a single $SiO_2$ layer typically show a reflectance of 80 to 90% between 300 and 600 nm. To enhance this reflectance and the durability of the coatings two commercially available options are currently under investigation:

(a) A three-layer protective coating ($SiO_2 + HfO_2 + SiO_2$) on top of an Al coating. Already this enhances the reflectance by about 5% as can be seen in Fig. 1 in comparison to the reflectance of an Al + $SiO_2$ coating.

(b) A dielectric coating, consisting of a stack of many alternating layers of two materials with different refractive indices, without any metallic layer. This allows to custom-taylor a box-shaped reflectance curve with ¿95% reflectance between 300 and 600nm, and ¡30% elsewhere. The reflectance curve of this coating is as well shown in Fig. 1. The design could be adjusted to the required wavelength range such, that e.g. a cut-off at 550nm allows to reduce the night-sky background (first emission line around 556nm). The latter might become important in combination with a possible future replacement of the current photomultiplier tubes (which are less succeptible to night sky background) by silicon detectors that have a good quantum efficiency as well for wavelengths above 600 nm.

## 3 Durability Testing

A series of durability tests have been performed with small glass sample that have been coated with the different coatings to evaluate their resistance to environmental impact in the laboratory.

**Temperature and humidity cycling:** The samples have been exposed to overlapping cycles in temperature (-10° C < T < 60° C; 5 h cycle duration) and in humidity (5% to 95%; 8h cycle duration) for a total of approximately 8000h. The different cycle duration was chosen to expose the sam-

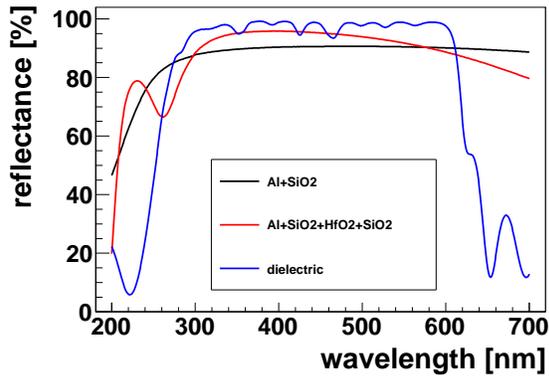

Figure 1: Comparsion of the spectral reflectance of the two newly investigated coatings (Al + SiO₂ + HfO₂ + SiO₂ and dielectric) relative to Al + SiO₂.

ples to all possible combinations of temperatures and humidity. The reflectance as a function of wavelength of the samples has been measured with a spectrophotometer (angle of incidence 7°) before and after the cycling. The results of these measurements are shown in Fig. 2

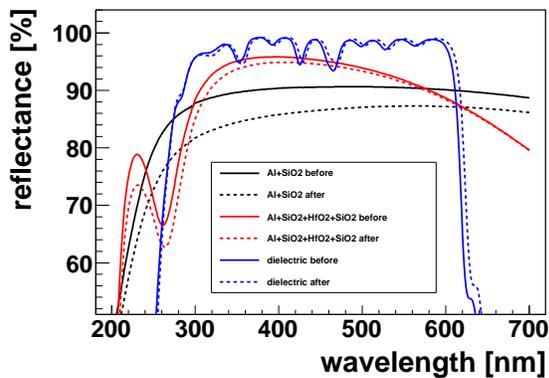

Figure 2: Comparsion of the spectral reflectance before (full lines) and after (dashed lines) temperature and humidity cycling for all three coatings.

The classical Al + SiO₂ coating shows a significant loss of refeltance. The Al coating with the three layer protective coating exhibts a much smaller change in reflectance after the cycling and the dielectric coating has not changed its reflective properties at all within the accuracy of the measurement.

**Salt-fog test:** All samples have been exposed for 72 hours to a salt-fog atmosphere at a temperature of about 20° C with a salt concentration of 5%. The samples with Al + SiO₂ showed small spots of damaged coating at the sample edges visible by eye, the other two coatings did not exhibit any damage in the visual inspection. Figure 3 shows the reflectance before and after the test, measured in the center of the samples where no obvious damage was visible.

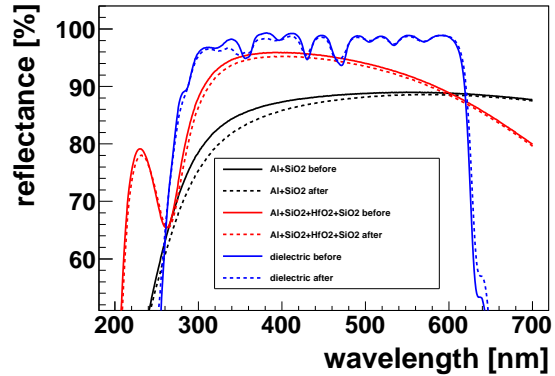

Figure 3: Comparsion of the spectral reflectance before (full lines) and after (dashed lines) a 72 h salt fog test for all three coatings.

The changes in reflectance are less pronounced as for the temperature and humidity cycling. Neverthess, again the standard Al + SiO₂ coating shows the strongest decrease.

**Abrasion tests:** Three different abrasion tests have been on samples with all three caotings:

a) A standard cheesecloth test using a force of 5 N and 50 strokes on the coated surface does not show scratches on all coatings. After increasing the force to 10 N a significant amount of fine scratches were created on the samples with the Al + SiO₂ coating. The depth of these scratches was determined to be of the order of 10 nm (with the SiO₂ coating having a thickness of 70 to 100 nm) using a Zygo profilometer. Figur 4 shows the results of the measurement of the depth of such a scratch.

The samples with the three layer overcoating and with the dielectric coating were hardly affected by this test also at a force of 10 N.

b) In a more severe test an eraser was used and 20 strokes with a force of 10 N were performed. After this test all three coatings showed scratches, the samples coated with SiO₂ significantly more than the those with the three-layer overcoat and those more than the dielectric samples. The scratches were deeper (up to 30 nm) than for the cheesecloth test, for the dielectric coatings the width was narrower than for the other two coatings.

c) Samples with all three coatings were exposed to a sand-blasting test. The abrading medium used was silicon carbide with a grade of 220 μm. The flow rate was approximately 25 g/min and the test duration 5 min. The setup was operated using an air pressure of 15 kPa and the air was fed in at a rate of 50 l/min. The sample was placed under an angle of 45° under the abrasive jet nozzle. The test results into an ellipse on the coated surface in which the coating is fully removed. The size of this ellipse is a measure how easy or not the coating is abraded. Figure 5 shows three samples after the sand-blasting test, on the left with the Al + SiO$_2$ coating, in the middle with the Al + SiO$_2$ + HfO$_2$ + SiO$_2$ coating and on the right with the dielectric coating.

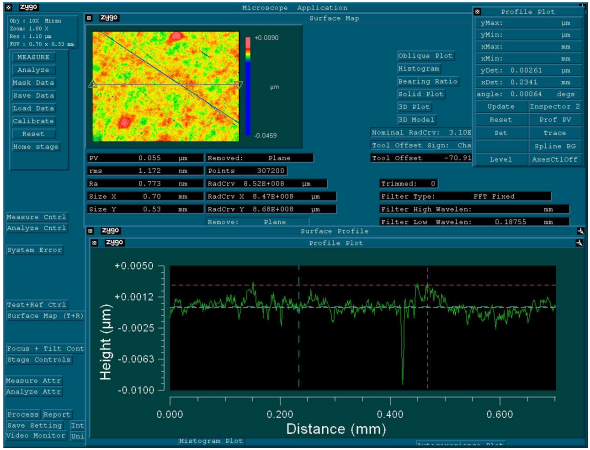

Figure 4: Measurement of the depth of a scratch in an Al + SiO$_2$ coating after the cheesecloth abrasion test.

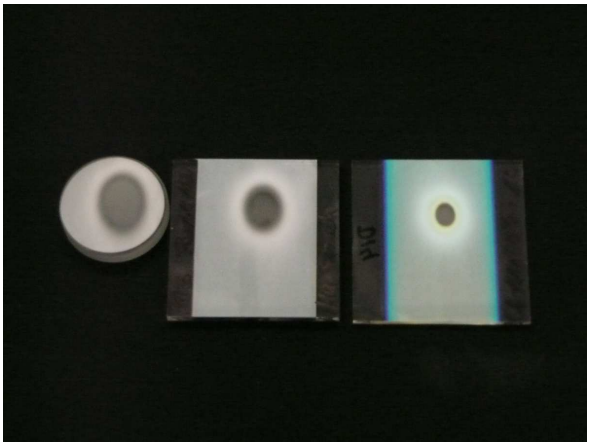

Figure 5: Coating samples after the sand-blasting test. Left:Al + SiO$_2$. Middle: Al + SiO$_2$ + HfO$_2$ + SiO$_2$. Right: dielectric coating.

The areas of the ellipses in Fig. 5 are 150 mm$^2$ for the sample with the SiO$_2$ coating, 85 mm$^2$ for the sample with the three-layer overcoating and 35 mm$^2$ for the dielectric coating. The test has been performed with similar results for 5 samples of each type of coating.

**Artificial Bird Faeces:** Samples of all coatings have been treated with pancreatin, a pancreas enzyme that is regularly used to simulate the effects of bird faeces on lacquers and other material. A 1:2 mixture of pancreatin and de-ionized water has been applied to the coated surfaces of the samples and they have been "baked" for 4 weeks at 40° C in a climate chamber to simulate the effekt of the bird faeces staying on the mirror surface for some time in a hot and dry environment as is typical for locations of Cherenkov telescope experiments. No influence on either of the the three coatings was observed after cleaning the samples.

## 4 Conclusions and Outlook

In laboratory tests like temperature and humidity cycling, exposure to salt fog atmospheres or simulated abrasion the three layer protective coating on top of an aluminium coating performs slightly better than the standard Al + SiO$_2$ coating. The dielectric coating shows even a significantly better performance. Nevertheless, the predictive power of these laboratory tests for the real outdoor performance is not clearly established. Currently the approximately 1800 mirrors of the H.E.S.S. experiment in Namibia are being re-coated. The Al + SiO$_2$ + HfO$_2$ + SiO$_2$ coating was chosen for most mirrors of this project. The first 280 mirrors with the three layer overcoatimg have been deployed in autumn 2010, another 380 mirrors in spring 2011. Further 380 mirrors will follow in autumn 2011. In addition, 100 mirrors with the dielectric coating have been installed in autumn 2010. Further, one telesope was refurbished with 380 mirrors with a classical Al + SiO$_2$ coating and can serve as a baseline for comparison. So far no negative influence on the performance of the experiment has been recorded. Detailed data on the performance of the three different coatings in a realistic outdoor application will become available soon.

Concernig the reflectance the three-layer overcoating shows an increase by approximately 5% compared to Al + SiO$_2$, the dielectric coating of the order of 10%. The latter has the additional advantage of a box-shaped reflectance curve of tunable width might help to reduce the night-sky background, especially in case of the possible future use silicon based detectors that are more sensitive at longer wavelengths as the currently used photomultipliers.

Costwise there is not significant difference between the classical Al + SiO$_2$ and the Al + three layer coating since costs for relatively simple coatings are mainly driven by the time the coating chamber is occupied rather than material costs (for the types of materials used here). The dielectric coatings are currently more expensive, but this is based on orders of much smaller quantities.

Both, the Al + SiO$_2$ and the Al + three layer coating can be applied at low substrate temperature during the coating process. This was not of importance for classical glass mirrors as used in H.E.S.S. but might become an issue for CTA, since most mirror substrate technologies currently under investigation are sandwich structures that consist of different materials that are glued together, mostly with a thin cold-slumped glass sheet as the front surface (see for example [2, 3, 4, 5] for details). The dielectric coating currently needs a substrate temperature of 150° C which is too high for most of these sandwich structures. The plan for the near future is to develop an improved coating design that keeps the performance of the currently investigated dielectric coating and can be applied at significantly lower temperatures. In parallel, the possibility of coating the glass sheets (at high temperatures) before mounting the sandwich structure should be investigated.

Both, the Al + SiO$_2$ and the Al + three layer coating are available "out of the shelf" for substrate sizes as currently envisaged for CTA (up to 2 m$^2$). For the type of dielectric coating investigated here the largest mirrors we have tested so far are the H.E.S.S. mirrors (circular, 60 cm diameter). Since for this type of coatings a very precise control of the film thickness of all layers over the full mirror area is essential, further investigations are needed to proof that mirrors up to 2 m$^2$ can be successfully coated.

To conclude, the Al + SiO$_2$ + HfO$_2$ + SiO$_2$ is a readily available alternative to the standard Al + SiO$_2$ coating that provides an about 5% better reflectance and a slightly better performance in durability tests in the laboratory at no significant extra cost. The dielectric coating provides a significantly better reflectance in the desired wavelength range, the possibility to suppress night-sky background and a significantly better performances in the durability tests, but it needs further investigation concerning the application on large mirror surfaces and a possible application at lower substrate temperatures for future sandwich mirrors.

## Acknowledgements

We gratefully acknowledge support from the agencies and organisations listed in this page: http://www.cta-observatory.org/?q=node/22.

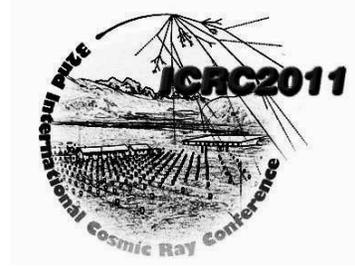

# First results on mirror design for CTA at Irfu-Saclay

M. C. MEDINA[1], P. BRUN[1], P. H CARTON[1], G. DECOCK [1], D. DURAND[1], J. F. GLICENSTEIN[1], C. JEANNEY[1], P. MICOLON[1], B. PEYAUD[1].

[1]*Institute de Recherche sur les Lois Fondamentales de l'Univers (IRFU), CEA, F-91191 Gif-sur-Yvette Cedex, France.*

*clementina.medina@cea.fr*

**Abstract:** The Cherenkov Telescope Array (CTA), as the future ground based $\gamma$-ray astronomy facility, is currently in a very advanced design phase. CTA will comprise several tens of large Imaging Atmospheric Cherenkov Telescopes of several different sizes. The total reflective surface needed is about 10,000 m$^2$ requiring unprecedented technological efforts, mainly towards cost reduction. In this paper we present a new mirror concept for CTA specially developed by the IRFU group and intended to fulfill the technical and optical specifications established by the Consortium. Lightweight, reliability and cost-effectiveness were sought with this design. The mirrors consist of a thin glass layer glued to a sandwiched honeycomb structure by pressure against a dedicated spherical mould. The first series of nominal size mirrors have been built and, with this series, the technique and different materials (carbon fiber, glass fiber, aluminum) were evaluated. The results of this evaluation are discussed in this paper, together with current and future mirror testing activities.

**Keywords:** Cherenkov Telescopes Array, mirrors.

## 1 Introduction

Because of its large size, the reflector of a Cherenkov telescope is composed of many individual mirror facets. A hexagonal shape was chosen, with a panel size of 1 - 2.5 m$^2$ area. More precisely, the baseline idea for the Middle Size Telescope (MST) is to use hexagonal mirrors of 1.2 m (flat to flat) diameter, with a spherical shape of about 32 m of radius of curvature. The criteria to evaluate the performance of the facets are equivalent to those of current instruments with regard to the spot size, to the reflectivity and to the long-term durability. 80% of the incident light to the mirror should be reflected in a 1 mrad diameter spot within the wavelength range of 300 - 600 nm, and facets must be robust against aging when exposed to the environment at the chosen site for several years [1][2]. Weight reduction is also an important goal but should not come at the expense of optical quality. About 20 - 30 kg/m$^2$ is acceptable. Also mirror deformation under gravity must be small enough to maintain the specifications for the PSF and the alignment. Is most likely that CTA will be placed on a site with high temperature amplitude implying that mirrors should resist temperature changes from -15°C to +50°C and keep their optical properties between -10°C to +30°C.

Finally, maybe the most critical point is long-term stability of the reflectivity under the expected environmental conditions. Due to the large mirror surface in CTA, it is intended to maximize the time interval between two re-coating process, demanding a better protection than is currently used on Cherenkov telescopes mirrors.

In order to account for all these requirements, the IRFU team has worked on the development of mirrors facets, particularly using the cold-slumping technique [3] (see Fig. 1). This is a two steps technique. At first, a thin sheet of glass, which is intended to receive the reflective coating at a further step, is shaped with a high precision spherical mould by applying a uniform pressure on it. In a second step a stiff back panel is fixed behind the glass to maintain its spherical shape. The back panel is required to be lightweight, cost efficient and to be assembled easily . Several material combinations have been tested and the design is now converging to its final stage.

This paper reports on the various steps that led to our current best-guess designs for MST mirrors. The test procedures both optical and mechanical for the prototypes are presented, as well as the results for the best specimens. Eventually the perspectives for mass-production and extrapolation of the technique to Large-Scale telescope (LST) mirrors and Small-Scale telescope (SST) mirrors are discussed. The paper is organized as follow: section 2 is dedicated to the evolution on the design of the first series of prototype mirrors. In section 3 the thermal behavior of the mirrors is discussed while in section 4 their optical properties are evaluated. Finally we present the next steps on the development of a final design in section 5 and the conclusions in section 6.

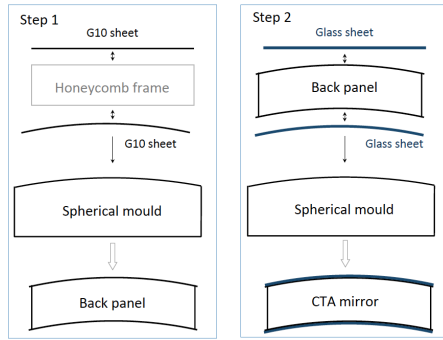

Figure 1: Basic concept of Saclay cold-slumped mirrors.

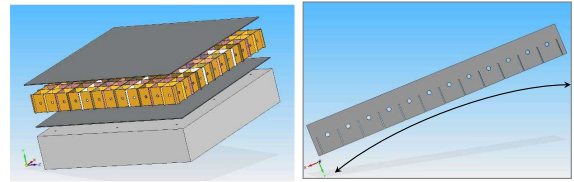

Figure 2: *Left*: Back panel principle with custom made honeycomb sandwich. *Right*: pre-shaped strip with the given radius.

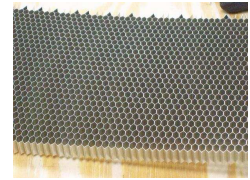

Figure 3: Commercial Aluminum honeycomb which is sandwiched by two plates to get the final mirror structure.

## 2 Prototypes description

The very first prototypes were realized with a 50cm x 50cm spherical mould. Many types of back panels have been produced with it. They include pre-formed foam panel, injected foam, and honeycomb both commercial and custom-made. These preliminary prototypes allowed us to learn about the geometries of the back panel, the different materials behavior and the properties of several types of glue.

Mirrors with foam back panels have been ruled out essentially due to their bad thermal behavior. Indeed, the average sag of the mirrors (measured with a basic mechanical sensor) can change significantly and permanently after the mirrors have been submitted to temperatures not higher than 50°C.

We then focused on honeycomb custom-made structures, made of aluminum, carbon fiber or glass fiber, as well as commercial aluminum honeycomb. The results obtained with our reduced-size mould convinced us to purchase a high precision 1.2 m flat-to-flat hexagonal mould of spherical shape. Its radius of curvature is 33.6 m, which matched previous specifications for CTA MST mirrors (32.14 m is the current demanded radius).

The custom-made structure is made of strips of a given material for which one edge is cut according the desired spherical shape. The strips are assembled by hand and the structure is sandwiched between two sheets of material. The structure is glued applying pressure on the rear face and keeping it against a dedicated mould (see Fig. 2). The commercial aluminum honeycomb is not previously milled and is also sandwiched between two sheets of material. The shape is therefore maintained by the glue, which is active on a very large surface thanks to the large number of honeycomb cells (see Fig. 3). As a second step, thin glass sheets are glued on both faces while the structure is maintained against the mould with pressure.

Note that out of the number of prototype mirrors that have been built, not all of them received high quality aluminization. Indeed almost all of them have a glass that is aluminized on its rear face. That allows getting a rough estimate of the optical quality of the mirror without proceeding to a rather expensive front-side aluminization. For some of the prototype mirrors, the mechanical measurements and the geometrical quality inferred from the optical estimate were very promising and the mirrors have been sent to industry to receive a real front-side reflecting coating. The reflectivity has been measured for those ones.

In Table 1 the main characteristics of prototypes are given and some of them are shown on Fig. 4.

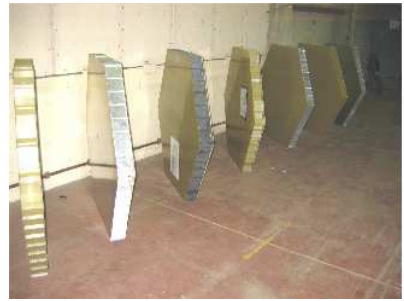

Figure 4: Some of the nominal size prototype mirrors built at Saclay. The different materials used can be distinguished.

## 3 Thermal Behavior

As already mentioned above, one important point to study is the stability and durability of the geometry of our mirrors with time and with different environmental conditions. Using a facility installed at Saclay, we exposed the prototypes to different temperature changes, from 0 °C to 36°C, and we mechanically measured the deviation from the nominal spherical shape suffered in each temperature step. As the different materials (aluminum, glass, carbon) have different thermal expansion coefficients and they are glued to each other with an epoxy resin, different stresses may appear when the temperature varies and produce a global deformation of the mirror. As CTA will most likely be installed on a desert site, the temperature difference between day and night can play an important role even if the temperature can be relatively stable during the observation time. The two important features are: the magnitude of this deformation and the capacity of the mirror to recover its original shape. In Fig. 5 the behavior with temperature of six of our mirrors is shown. We can see from these results that the amplitude of the deformation for three of them (prototypes 1, 7 and 10, mainly made of Aluminum) is quite important. A rough explanation of this behavior is the following: as the temperature drops, the mirrors tend to become flatter because the Aluminum sandwich behind the glass sheet tends to contract. On the contrary, when temperature rise, the dilatation of the Aluminum allow the structure to reach higher curvatures.

| Mirror | Sandwich plates | Inner structure | Reflective Surface | Glue type |
|--------|-----------------|-----------------|--------------------|-----------|
| 1 | Al (1mm) | Al strips (1mm) | 2mm glass sheet[2] | Araldite |
| 2 | Al (1.5mm) | Al strips (1.5mm) | 2mm glass sheet[2] | Araldite + Optical Gel |
| 3 | G10 (1.5mm) | G10 strips (1.5mm) | 2mm glass sheet[2] + front coating[3] | Araldite + F50 |
| 4 | G10 (1mm) | G10 strips (1mm) | 2mm glass sheet[2] + front coating[3] | Araldite + F50 |
| 5[1] | G10 (1.5mm) | G10 strips (1.5mm) | 2mm glass sheet[2] | Araldite AW 106 + F50 |
| 6 | G10 (1.5mm) | commercial Al honeycomb [4] | 2mm glass sheet[2] | Araldite AW 106 + F50 |
| 7 | G10 (1.5mm) | commercial Al honeycomb | 2mm glass sheet[2] | Araldite AW 106 + F50 |
| 8[1] | G10 (1.5mm) | G10 strips (1.5mm) | 2mm glass sheet[2] | Araldite AW 106 + F50 |
| 9 | G10 (1.5mm) | Carbon fiber strips (1.6mm) | 2mm glass sheet[2] | Araldite AW 106 + F50 |
| 10 | Al (1mm) | commercial Al honeycomb | 2mm glass sheet[2] | Araldite AW 106 + F50 |
| 11 | G10 (1.5mm) | commercial Al honeycomb[4] | 2mm glass sheet[2] + front coating[3] | Araldite AW 106 + F50 |
| 12 | 3mm glass sheet | commercial Al honeycomb[4] | none | Araldite AW 106 |
| 13 | 2mm glass sheet | commercial Al honeycomb[4] | front coating[3] | Araldite AW 106 |

Table 1: Prototype mirrors main characteristics. 1) 1.5 mm thick G10 side walls. 2) Rear metalized. 3) Aluminization + 100 nm of SiO₂. 4) 19 mm cell and 8 cm height.

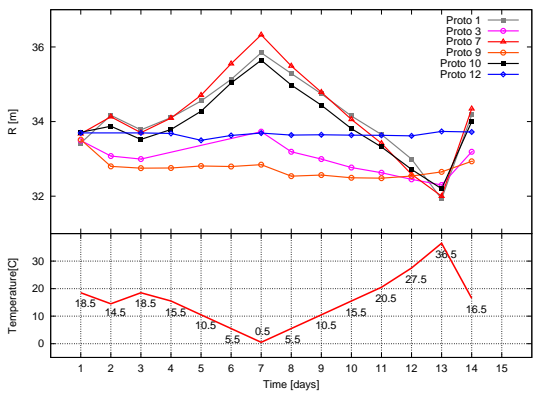

Figure 5: Thermal behavior of some of the prototypes. See table 1 for details on the mirrors.

Concerning the carbon fiber structure (prototype 9), its behavior is quite stable with temperature variation even though the original curvature was not completely recovered. The glass fiber mirror tends to follow the same trend as the Aluminum ones but the amplitude of the deforma-tion is smaller. Finally, the one which seems to experiment no transformation with temperature is the one representative of the final concept of our mirrors (symmetric mirror: same material at two faces). The older prototypes were built without the rear glass sheet, which correspond to the previous asymmetrical design.

## 4 Optical properties

In order to characterize the prototypes and evaluate them according to the CTA Consortium specifications, we have built a test facility dedicated to measure three important parameters of the mirrors: the Point Spread Function (PSF), the effective reflectivity and the focal distance. Using the 2f (twice the focal distance) test bench sketched on Fig.6 we determined the optical properties of the first series of mirrors. The results for the front-side coated mirrors are presented in Table 2. The mirror is uniformly illuminated by a LED type light source with a strong blue component. As a first approximation, we use two filters (V and B) to get the reflectivity at the relevant wavelengths for Cherenkov detectors. The light flux arriving to the mirror and that reflected in the spot at 2f are measured by planar diffused Silicon photodiodes with 611 mm² active area. The PSF is the angular size of the spot produced on a screen (Fig. 6). Images of this screen are taken with a CCD camera (ATIK_4000M) and analyzed offline to get the size of the region that contains 80% and 90% of the reflected light.

The improvement on the reflectivity and the PSF from the first prototypes to the last one is due, basically, to the progress on the control of the technique. Problems related to the deviation from nominal curvature at the edges of the mirrors were found analyzing the images at 2f. As the mirrors have no flexural constrains on the sides, the radius at the edges tended to be larger, generating a large directional dispersion of the light as one can see on the *top/left* panel of Fig. 7. In order minimize this dispersive effect, thick side walls were added to the mirrors to prevent the edges to bend as is shown on the *bottom/right* image of Fig. 7. The final result is a clean concentration of the light as the *top/right* panel of Fig. 7 shows.

| Prototype | $\rho$ (B)* | $\rho$ (V)* | PSF size (mrad) |
|-----------|------------|------------|-----------------|
| 3 | $77 \pm 2$ % | $75 \pm 2$ % | $\sim 1.1$ |
| 4 | $78 \pm 2$ % | $78 \pm 2$ % | $\sim 1.1$ |
| 11 | $84 \pm 2$ % | $81 \pm 2$ % | $\sim 1$ |

Table 2: Prototype mirrors optical performance.*$\rho$ is the % of incident light reflected on the 611 mm$^2$ area photodiode placed at 2f.

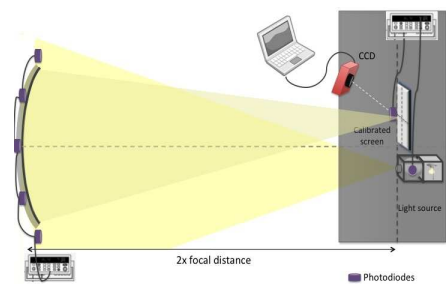

Figure 6: Schematics of Saclay Test Bench setup.

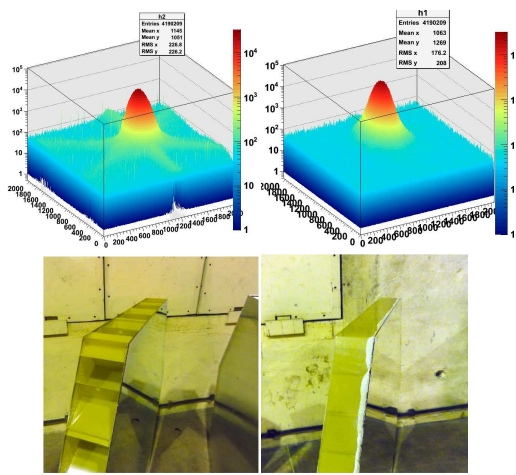

Figure 7: *Left*: image obtained with prototype N° 3 (*Bottom/Left*) at twice the focal distance and a point like source. The light dispersion on six directions perpendicular to the flat sides can be seen on logarithmic scale. *Right*: image obtained with a G10 mirror with a side wall added (*Bottom/Right*).

## 5    Next steps

With the assumption that the construction of the CTA array will start as early as 2013, completing the design during the next year and launching the production subsequently are required. Our current effort focuses on finishing all the necessary analysis to validate what we believe to be a competitive design for the MST mirrors with the aim to establish an industrial process for manufacturing them. Future tests such as shape measurements, the influence of temperature and humidity and aging of coating will be carried out in collaboration with other parties of the Consortium.

The next step will be the production of a mini-series of about 20 facets and will be done in partnership with local industry. These mirrors will be installed on the MST prototype that is already under construction in Berlin by the Consortium. Basically, the stability and durability of different back panel combinations will be monitored under "real" environmental conditions. The integration and installation of the facets should also be tested and improved from this prototype.

## 6    Conclusions

The first series of nominal size mirrors facets for CTA telescopes have been built by the IRFU team. The results of the first technical and optical tests are encouraging. We have studied their behavior under thermal variations finding that some of the prototypes are not good enough for CTA requirements while others keep being competitive in this sense. However, this is a preliminary test and the results should be confirmed by more precise measurements in a controlled (temperature and humidity) environment. The most promising prototypes were aluminized and protected by a layer of $Si0_2$. The optical response of these mirrors fulfill the specifications of the CTA consortium for the first prototype mirrors. Based on the experience gained with these mirrors, the production of a small series of 20 mirrors will start soon. The final stage of characterization and evaluation of our mirrors will be finished after their installation on the Middle Size Telescope prototype.

**Acknowledgements** We gratefully acknowledge support from the agencies and organizations listed in this page: http://www.cta-observatory.org/?q=node/22.

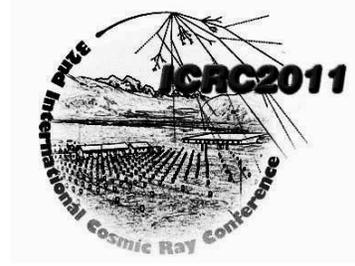

# Methods for the characterization of mirror facets for Imaging Atmospheric Cherenkov Telescopes

ANNELI SCHULZ[1], ROMAN KROBOT[2], EVELYN OLESCH[2], CHRISTIAN FABER[2], FRIEDRICH STINZING[1], CHRISTIAN STEGMANN[1], GERD HÄUSLER [2] ON BEHALF OF THE CTA CONSORTIUM

[1]Erlangen Centre for Astroparticle Physics, University Erlangen-Nuremberg
[2]Institute of Optics, Information and Photonics, Group OSMIN, University Erlangen-Nuremberg
anneli.schulz@physik.uni-erlangen.de

**Abstract:** To achieve a total mirror surface of about 10000 m$^2$ for the upcoming Cherenkov Telescope Array (CTA), individual mirror facets with a size of 1 to 2.5 m$^2$ each will be used. The production and testing of these mirror facets bears a particular organizational and logistical challenge. Here we compare two methods to determine the optical quality of mirror facets, namely the commonly used $2f$-method and a new approach, called Phase Measuring Deflectometry (PMD). PMD was developed by the OSMIN group in Erlangen. It offers a variety of advantages compared to the $2f$-method: the result of the PMD measurement is a precise map of the shape of the mirror (on a micrometer level) and of its curvature (with an accuracy up to 0.001 D). Hence it yields detailed information of the mirror surface, which is not possible with the standard $2f$-method. The PMD setup is compact, the characterization of a mirror facet is fast and the method is applicable for a variety of mirror sizes. We will present PMD, describe the different setups, and show first results for CTA mirror prototypes.

**Keywords:** CTA, Imaging Atmospheric Cherenkov Telescope, Gamma-rays, Optics

## 1 Introduction

The great potential of $\gamma$-ray astronomy has been shown by the existing Imaging Atmospheric Cherenkov Telecopes (IACTs), leading to an initiative to build a next generation instrument, namely the Cherenkov Telescope Array (CTA). A detailed description can be found in the CTA design report [1]. The projected sensitivity of CTA exceeds that of any existing IACT by one order of magnitude. CTA aims to extend the photon energy range from some tens of GeV to beyond 100 TeV. Studies of the morphology of TeV-sources will benefit from the array's enhanced angular resolution. The increased detection area will boost the detection rates and open possibilities to investigate transient phenomena. CTA aims to enhance the all sky survey capability, the monitoring capability, and the flexibility of operation.

To achieve this performance 50 – 70 telescopes of three different types will be integrated in the array. The inner part of the array will consist of few Large Size Telescopes (LSTs), with a diameter of 23 m mainly dedicated to the detection of low energy photons. These will be surrounded by several tens of Medium Size Telescopes (MSTs) covering ∼ 1 km$^2$. These telescopes will have a diameter of 9 – 12 m and be optimized for the energy range from 100 GeV to 10 TeV. A sparse array of Small Size Telescopes (SSTs) will cover the large detection area needed for the highest energies. The mirror area of the telescopes, in total about

10000 m$^2$, will be composed of mirror facets with a size of 1 – 2.5 m$^2$. The current baseline for the mirrors for the MSTs is to use hexagonal mirrors with a flat-to-flat size of 1.20 m. The total number of mirrors will be around 10000.

Currently two different types of mirrors are used in Cherenkov telescopes: glass mirrors, which are heavy and not durable and aluminium machined mirrors, which are lighter and more robust, but twice as expensive as glass mirrors. Various solutions are under study to obtain lightweight, robust, and cost-effective mirrors with the required reflectivity and focussing quality. The latter is described by the point spread function (PSF), characterized here by $d_{80}$, the diameter containing 80% of the reflected light. The PSF of the mirrors has to be < 1 mrad as stated in the design report [1]. Currently, different technologies to construct mirrors are under investigation at different institutes, ranging from cold slumped glass mirrors to diamond milled aluminium mirrors (see [1], [2], and [3] for details). Most methods are sandwich techniques, based on a honeycomb core structure to ensure rigidity and sheets made of different materials on top. To ensure the spherical shape of the mirror, all groups use moulds with the required radius of curvature ($R = 2f$) to form the shape of the sheet. The sheet is in most cases coated afterwards, and different coatings are under investigation.

Mirror development and mass production of CTA mirrors requires a fast and reliable test procedure. Commonly,

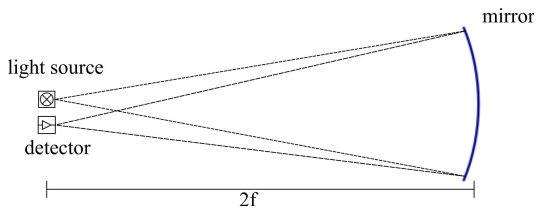

Figure 1: Sketch of the $2f$ measurement setup. The distance between the light source and the detector is extremely exaggerated to demonstrate the light rays more clearly.

the so-called $2f$-method is used to characterize mirrors of IACTs. Phase measuring deflectometry is a new method to characterize mirrors yielding superior information compared to the $2f$-method, making this a good candidate for the standard test method for IACT mirrors as will be shown in the following.

## 2 Measurement at $2f$

For a $2f$ measurement a light source is placed at a distance of $2f = R$ from the mirror and the reflected picture of the light source is detected at the same distance. The size and shape of the PSF can be tested with this kind of setup. A sketch of the $2f$-setup can be seen in Fig. 1.

This test setup is rather simple to implement, but has several disadvantages: the need of a sufficiently large room ($R = 34$ m for the MST mirrors), as well as the rather difficult alignment of the mirror, and the absence of information about the surface parameters of the mirror. The surface parameters are crucial to understand possible problems concerning the focussing quality of a mirror. Another disadvantage of the $2f$-method is that the testing conditions differ from the operation conditions on a telescope: the testing is done with an on-axis light source at $\sim 30$ m while the mirrors on the telescope are mainly operated off-axis and the light source, in this case the air-shower, is at a height of $\sim 10$ km.

## 3 Phase measuring deflectometry (PMD)

A method avoiding the disadvantages of the $2f$-method is Phase Measuring Deflectometry (PMD). This metrology is used to measure the properties of specular surfaces. It has been developed by the OSMIN group at the University of Erlangen[1] and is described in detail in [5]. PMD is, among others, used to check the quality of progressive eyeglass lenses, windscreens, and painted car bodies. A very successful project is the precise measurement of the local refractive power of progressive power eyeglasses. Solutions based on this development are now being used by all major European eyeglass manufacturing companies. The basic idea of PMD is to observe the distortions of a defined pattern after it has been reflected by the examined surface (see

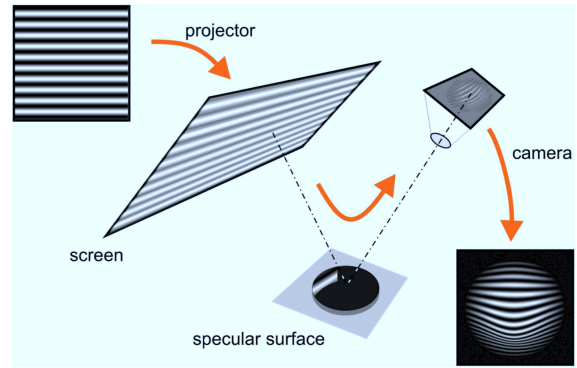

Figure 2: Sketch of the measurement principle of PMD. The sinusoidal pattern is projected on a screen or a ground glass. The camera takes pictures of the distortions of the pattern after the reflection on the object; picture from [5].

Fig. 2). From these distortions one can calculate the exact local slope and by numerical differentiation and integration the corresponding curvature and shape of the surface, respectively.

Specular Reflection is guided by the normal of the surface, that is the reason why the slope of the object is the primary quantity measured by of deflectometric systems. Since specular surfaces reflect light monodirectionally, care has to be taken that the reflected rays are visible to the camera. Therefore, the light source has to cover a large solid angle. One way to ensure this is to use a big screen, which is diffusely emitting. An alternative for concave mirrors is to place the screen and the observing cameras at a distance of $2f$, trading in the advantage of a short working distance for a smaller size of the required screen. An important part of the technique is to code the position on the screen by the projected pattern. PMD uses a sinusoidal pattern, which has the advantage that the phase of the sinus does not change if the pattern is observed out of focus [4]. It is not possible to observe the object and the pattern in focus at the same time. In order to obtain a high lateral resolution we focus on the object, so that the structured illumination pattern has to be placed out of focus, resulting in a decreased contrast. To maximize angular sensitivity, the period of the sinusoidal pattern should be chosen as small as possible. On the other hand, a smaller fringe period leads to a higher loss of contrast, scaled by the aperture of the observation system. Therefore, the lateral and angular resolution are coupled and a limit for the achievable accuracy can be determined. Using the sinusoidal pattern, it is possible to assign the phase in each pixel of the camera independently. To determine the phase, known phase shift techniques like the Bruning four-shift-algorithm [6] are used, in which four sinusoidal patterns with phase shifts of $\pi/2$ are projected onto the screen. The patterns are projected

---

1. http://www.optik.uni-erlangen.de/osmin/

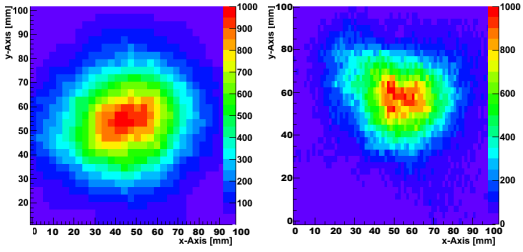

Figure 3: PSF comparison for the mirror from IRFU/CEA (50 cm), the left panel shows the $2f$-result the right panel the raytracing PSF. The color scale is identical for both panels.

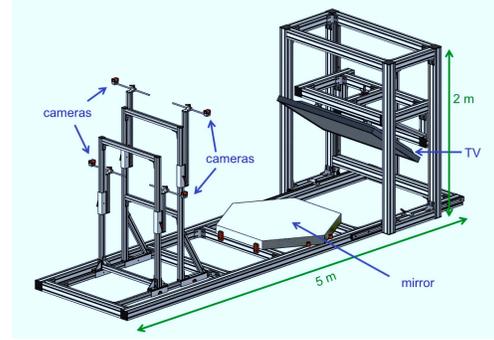

Figure 4: Sketch of the newly designed PMD setup, including four cameras and a screen.

both horizontally and vertically to determine the slope in both directions.

The primary measurand of PMD is the slope of the mirror in two perpendicular directions. The height of the mirror is obtained by integrating the slope data. A map of the mirror's curvature can be calculated by differentiating the slope data. To compare PMD results to the $2f$-data, one needs to determine the PSF of the mirror from the PMD results. We developed a ray-tracing script using the shape and slope data from PMD measurements as input parameters. This script can provide the PSF for arbitrary imaging distances and incidence angles, which is an advantage compared to the $2f$-method. The $2f$-method results in an on-axis PSF from a light source at $\sim 30$ m distance while the mirrors on the telescope are reflecting light from $\sim 10$ km distance and are used off-axis for most mirror facets.

To compare both methods the PSF calculated by the raytracing script is determined under the $2f$-conditions, meaning on-axis and the light source at a distance of $2f$. Fig. 3 shows a comparison of the PSF-results from both methods. This is the result for a prototype from IRFU/CEA in Saclay with a size of 50 cm. We found reasonably good agreement in size and shape of both PSFs. The size of the PSF of this mirror with the $2f$-method is 50 mm $\hat{=}$ 1.55 mrad and thus nearly fulfils the requirements. The value for the raytracing is smaller than for the $2f$ measurement, which is due to a perfectly pointlike light source simulated for the ray tracing, while the light source in the $2f$ measurement has an extension of a few mm.

The comparison of the PSFs obtained by both methods has also been done with mirrors from H.E.S.S. and MAGIC, resulting in detailed maps of the surface demonstrating the good performance of the method.

It is possible to measure a mirror with a diameter of 1.20 m with PMD using two different approaches: the so-called Long Working Distance setup (LWD) and Short Working Distance (SWD) setup. The LWD uses a $19''$ monitor as screen and two cameras to observe the pattern after the reflection, and the mirror is placed at a distance of $2f$. This setup has the same disadvantage of the large space needed like the $2f$-method, but reveals the surface parameters of the mirror in a quick measurement. The SWD is a more compact solution using a $60''$ screen; a sketch is shown in Fig. 4. The four cameras are needed to cover the mirror completely; each camera observes a separate quarter of the specular surface while all see the mirror's center. The pictures of the cameras are then combined to get a complete image of the mirror. We have built such a setup and are currently in the commissioning phase.

## 4 Results of a first CTA prototype

The CTA group at IRFU, CEA in Saclay has produced first mirror prototypes for the MST using a honeycomb technique, see [7]. These mirrors are hexagonal, have a flat-to-flat size of 1.20 m, and a radius of curvature of R= 33.4 m. We here present first results of a PMD measurement of this mirror with the LWD setup. Fig. 5 shows the slope of the mirror in the $y$-direction after subtracting the best fit plane, i.e. after subtracting the basic spherical shape of the mirror approximately. Hence, in Fig. 5 the deviations from the basic shape are shown. The underlying honeycomb structure can easily be seen in this measurement. Shape and curvature can be obtained from integrating or differentiating the slope, respectively see Figs. 6 and 7. The curvature map (Fig. 7) also shows the underlying honeycomb structure. The curvature of a mirror is the inverse of the radius of curvature, the units for the curvature are dioptre $[D] = 1/$m. The average value for the mean curvature is $c_{mean} = -0.032 \pm 0.001$ D, where the nominal value is $-0.030$ D.

This prototype is close to fulfilling the specifications for the MST for CTA; ray-tracing and $2f$-measurements also confirm the PSF to be smaller than 2 mrad. We point out that this special mirror has gone through several ageing tests and is therefore not the mirror with the best possible performance from IRFU/CEA in Saclay.

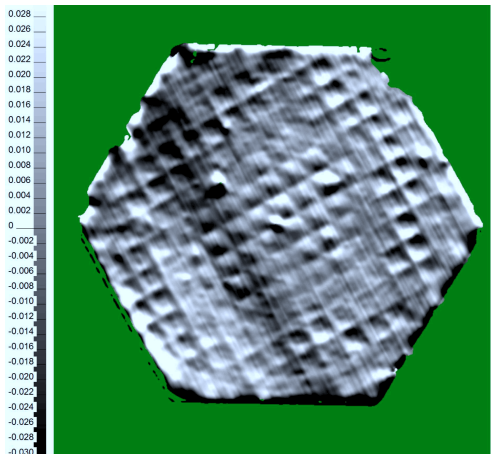

Figure 5: Slope deviation in $y$ direction of the mirror from IRFU/CEA.

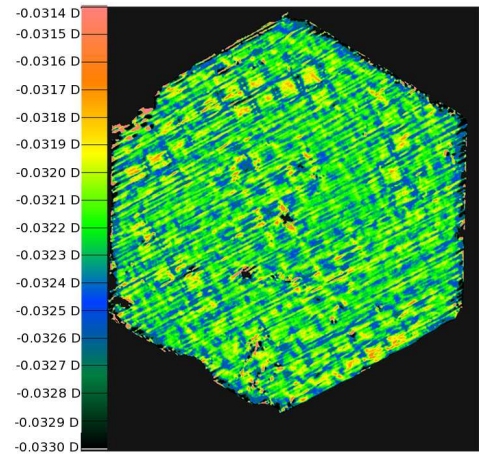

Figure 7: Mean curvature of the mirror from IRFU/CEA.

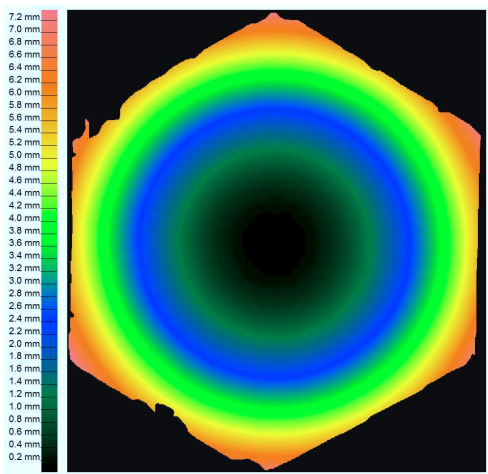

Figure 6: Surface of the mirror from IRFU/CEA.

## 5 Summary and outlook

We have presented PMD which is a new method to measure mirrors for IACTs. It has proven to give valuable informations about the tested mirrors and the results for the PSFs are in agreement with the commonly used $2f$-method.

Using PMD to characterize mirrors will give very detailed information about the performance of the different mirrors. PMD not only allows us to gather more information on the mirror surface, but it is also quicker and more compact than the $2f$-setup. The time for a measurement is shorter since the mirror does not have to be aligned. Therefore, PMD can be used efficiently for quality control.

Following the specifications in the CTA design report [1] the mirrors have to resist temperatures between $-20°$ and $+40°$ C. Particularly, for the composite mirror technologies this can be critical due to varying thermal expansion co-efficients of the different materials. One way to test the temperature behaviour of the mirrors is to heat or cool the mirror and measure the changes afterwards, but since the temperature on the site will not be stable a measurement at different temperatures is required. A climate chamber provides the needed temperature range, but it is not possible to realize a $2f$-setup inside a climate chamber. The best solution is to use the SWD PMD setup. We plan to do measurements inside a climate chamber this summer.

**Acknowledgement** We gratefully acknowledge support from the agencies and organisations listed in this page: http://www.cta-observatory.org/?q=node/22.

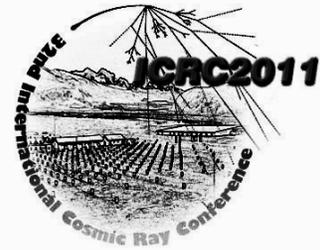

# Development of the Readout System for CTA Using the DRS4 Waveform Digitizing Chip

H. KUBO[1], R. PAOLETTI[2], M. AONO[1], Y. AWANE[1], M. BITOSSI[2], R. ENOMOTO[3], S. GUNJI[4], R. HAGIWARA[4], N. HIDAKA[5], M. IKENO[6], S. KABUKI[7], H. KATAGIRI[8], Y. KONNO[1], T. NAKAMORI[9], H. OHOKA[3], A. OKUMURA[10], R. ORITO[11], M. SASAKI[8], A. SHIBUYA[5], H. TAJIMA[5], M. TANAKA[6], M. TESHIMA[3,12], T. UCHIDA[6], K. UMEHARA[8], M. YONETANI[13], ON BEHALF OF THE CTA CONSORTIUM

[1]Department of Physics, Graduate School of Science, Kyoto University, Sakyo, Kyoto 606-8502, Japan
[2]Università di Siena, and INFN Pisa, I-53100 Siena, Italy
[3]Institute for Cosmic Ray Research, The University of Tokyo, Kashiwa, Chiba 277-8582, Japan
[4]Department of Physics, Faculty of Science, Yamagata University, Yamagata, Yamagata 990-8560, Japan
[5]Solar-Terrestrial Environment Laboratory, Nagoya University, Chikusa, Nagoya 464-8601, Japan
[6]Institute of Particle and Nuclear Studies, KEK, Tsukuba, Ibaraki 305-0801, Japan
[7]Tokai University Hospital, Isehara-shi, Kanagawa 259-1193, Japan
[8]College of Science, Ibaraki University, Mito, Ibaraki 310-8512, Japan
[9]Faculty of Science and Engineering, Waseda University, Shinjuku, Tokyo 169-8555, Japan
[10]Institute of Space and Astronautical Science, JAXA, Sagamihara, Kanagawa 252-5210, Japan
[11]Faculty of Integrated Arts and Sciences, The University of Tokushima, Tokushima 770-8502, Japan
[12]Max-Planck-Institut für Physik, Föhringer Ring 6, D 80805 München, Germany
[13]Department of Physical Science, Hiroshima University, Higashi-Hiroshima, Hiroshima 739-8526, Japan
kubo@cr.scphys.kyoto-u.ac.jp

**Abstract:** We have developed a prototype of the photomultiplier tube (PMT) readout system for the next generation VHE gamma-ray observatory, the Cherenkov Telescope Array (CTA). Several thousand PMTs along with their readout systems are arranged on the focal plane of each telescope, with one readout system per 7-PMT cluster. The signal from a PMT detecting Cherenkov light from an air shower is amplified, and the waveform is then digitized at a sampling rate of the order of GHz using an analog memory ASIC developed at Paul Scherrer Institute (PSI), called the Domino Ring Sampler (DRS4). The sampler has 1,024 capacitors per channel and the ability to cascade channels for increased depth. After a trigger is generated in the system, the charges stored in the capacitors are digitized by an external slow sampling ADC and then transmitted via Gigabit Ethernet. An onboard FPGA controls the DRS4, trigger threshold, and Ethernet transfer. In addition, the control and monitoring of the Cockcroft–Walton circuit that provides high voltage for the 7-PMT cluster is done by the same FPGA. Using a prototype named *Dragon* we successfully obtained a pulse shape of a PMT signal at a sampling rate of 2 GS/s and a single photoelectron spectrum.

**Keywords:** Imaging Atmospheric Cherenkov Telescope, Gamma-rays, Electronics.

## 1 Introduction

A ground-based imaging atmospheric Cherenkov telescope (IACT) measures Cherenkov light from an extended air shower (EAS) generated by the interaction between very high energy (VHE) gamma rays and the upper atmosphere. Night sky background (NSB) also enters a pixel photon sensor of the focal plane of the IACT with a rate of the order of 10–100 MHz, depending on the mirror size and the pixel size. The NSB therefore becomes noise that affects the sensitivity and the energy threshold. Given this NSB pollution, and the fact that the duration of Cherenkov light from EAS is a few nanoseconds, a fast digitization speed of the readout system coupled to a fast photosensor like a photomultiplier tube (PMT) is benefic to increase the pixels' signal-to-noise ratio. In addition, this system should be compact and have low cost and low power consumption because each IACT possesses several thousand photon sensor pixels and the readout system attached to the sensors is in a camera container located at the focal position. Furthermore, a wide dynamic range of more than 8–10 bits is required to resolve a single photoelectron and have a wider energy range.

A commercial flash analog-to-digital converter (ADC) satisfies the requirement of a wide dynamic range. However, it is costly and consumes relatively high power of a few watts per channel. On the other hand, an analog memory application specific integrated circuit (ASIC), that consists of several hundred to several thousand switched capacitor arrays (SCA) per channel, can sample a signal at the order of GHz, with a wide dynamic range and lower power consumption. Several types of analog memory ASICs have been developed for applications in particle physics and cosmic ray physics. With respect to IACTs, a modified version of ARS0 [1], Swift Analog Memory (SAM) [2], and Domino Ring Sampler (DRS) [3] chips are used in the H.E.S.S.-I, H.E.S.S.-II, and MAGIC experiments, respectively.

Cherenkov Telescope Array (CTA) [4] is the next generation VHE gamma-ray observatory, which improves the sensitivity by a factor of 10 in the range 100 GeV–10 TeV and an extension to energies well below 100 GeV and above 100 TeV. CTA consists of telescopes having mirrors with size 20–30 m, 10–12 m, and 3.5–7 m, which are called large size telescope (LST), medium size telescope (MST), and small size telescope (SST), respectively. Several types of readout systems are developed for CTA with different analog memories for the requirements of LST, MST, and SST. At the same time, there has been progress in the development of photon sensors such as PMT and a Geiger-mode avalanche photodiode (or silicon photomultiplier) for CTA. At this time, the primary candidate is a PMT.

Using the analog memory DRS version 4 (hereafter DRS4), we have so far developed two versions of prototypes for the PMT readout system for CTA, named *Dragon*. Using the first version of the prototype, we demonstrated that the waveform of a PMT signal can be well digitized with the DRS4 chip. The second version of the prototype was developed based on the first version, with improvements made to the sampling depth of DRS4 and a trigger. In this paper, we report the design and performance of the second version of the prototype.

## 2  Design of Readout System

### 2.1  Overview

Several thousand PMTs and their readout systems are arranged on the focal plane of each telescope, with one readout system per 7-PMT cluster. We have developed a prototype of the PMT readout system. Figures 1 and 2 show a photograph and block diagram of the prototype, respectively. The prototype consists of a 7-PMT cluster, a slow control board and a DRS4 readout board. The total size is 14 cm × 60 cm. The 7-PMT cluster and the slow control board are described in detail in reference [5]. In this paper, a brief description is provided.

Our 7-PMT cluster consists of seven head-on type PMTs with a super-bialkali photocathode and 8-stage dynodes (Hamamatsu Photonics, R11920-100 with a diameter of 38 mm), a Cockcroft–Walton (CW) circuit for high voltage supply to the PMTs (designed by Hamamatsu Photonics), and a preamplifier board. A signal from the PMT is amplified by the preamplifier (Mini-circuits LEE-39+), and fed to the DRS4 readout board.

On the DRS4 readout board the preamplified signal is divided into three lines: a high gain channel, a low gain channel, and a trigger channel. The high and low gain channels are connected to DRS4 chips. The signal is sampled at a rate of the order of GHz and the waveform is stored in a SCA in DRS4. When a trigger is generated in the trigger circuit, the voltages stored in the capacitor array are sequentially output and then digitized by an external slow sampling (~30 MHz) ADC. The digitized data is sent to a field programmable gate array (FPGA) and then transmitted to a gigabit ethernet transceiver and a backplane via a data input/output (I/O) connector. The FPGA controls a static random access memory (SRAM) that stores large amounts of data before transmission and a digital-to-analog convertor (DAC) used for thresholding in the trigger circuit.

The slow control board is equipped with a generator for generating test pulses that are fed to the preamplifier, a temperature and humidity sensor with I²C interface, a DAC for setting the voltage of the CW high voltage circuit, and an ADC for monitoring both the CW circuit and the DC anode current. These devices on the slow control board are controlled by a complex programmable logic device (CPLD). Since the CPLD communicates with the FPGA on the DRS4 readout board, the data to and from the CPLD is sent via the Ethernet.

The power supply to the DRS4 readout board and the slow control board is ±3.3V and +5V. The total power consumption is ~2W per channel.

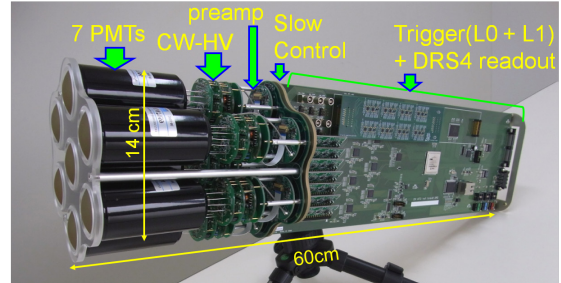

Figure 1. Photograph of the 7-PMT cluster and readout system (ver. 2).

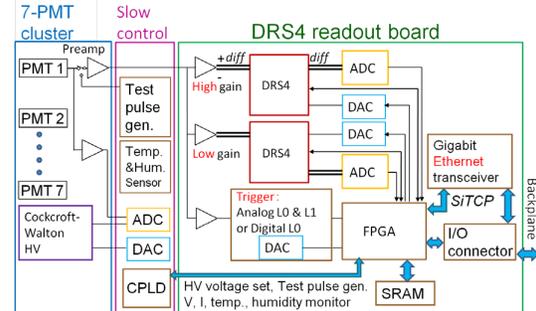

Figure 2. Block diagram of the 7-PMT cluster and readout system (ver. 2).

### 2.2  DRS4 Readout Board

Figure 3 shows a photograph of the DRS4 readout board with a size of 14 cm × 41.5 cm. The slow control board attached to the 7-PMT cluster is connected to the DRS4 readout board from the right side via two card-edge connectors. The DRS4 board has eight DRS4 chips, ADCs for digitizing a signal stored in the capacitor array in DRS4 at a sampling frequency of 25 MHz, a DAC to control the DRS4, a FPGA (Xilinx Virtex-4), a 18Mbit SRAM, a Gigabit Ethernet transceiver, and a data I/O connector to the backplane. In addition, seven pieces of main amplifier mezzanines, analog level 0 (L0) and 1 (L1) trigger mezzanines, and a digital level 0 (L0) trigger mezzanine are mounted on the DRS4 board. Details of these parts are described in the following subsections.

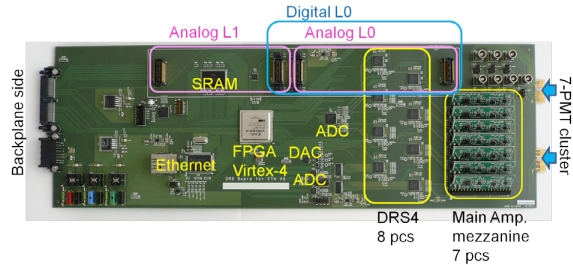

Figure 3. Photograph of the DRS4 readout board (ver. 2).

## 2.3 Main Amplifier

Figure 4 shows a photograph of the main amplifier mezzanine with a size of $1.8$ cm $\times$ 5 cm. One part of the mezzanine receives a signal from one PMT. The block diagram is shown in Fig. 5, and is designed to have a bandwidth greater than 350 MHz, lower power consumption, and a dynamic range of 0.2–3000 photoelectrons for LST. A preamplified signal from a PMT with a typical gain of $4 \times 10^4$ is fed to the main amplifier mezzanine. The signal is amplified using two differential amplifiers to meet the requirement for bandwidth and power consumption. For a high gain channel, it is amplified by a gain of 9 using Analog Devices, ADA4927. For a trigger, it is amplified by a gain of 4 using Analog Devices, ADA4927 and National Semiconductor, LMH6551. For a low gain channel, the preamplified signal is attenuated by 1/4 and then buffered with a differential amplifier using Analog Devices, ADA4950.

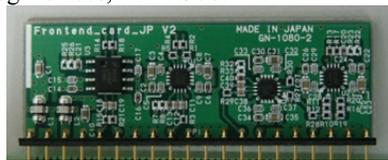

Figure 4. Photograph of the main amplifier mezzanine.

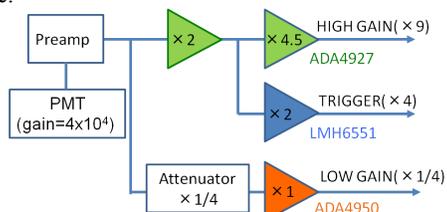

Figure 5. Block diagram of the main amplifier mezzanine.

## 2.4 Trigger

Figure 6 shows a photograph of a 4.1 cm $\times$ 16.5 cm digital L0 trigger mezzanine mounted on the DRS4 readout board. Its block diagram is shown in Fig. 7. A differential signal from the trigger channel of the main amplifier mezzanine (Figure 5) is fed to the digital trigger mezzanine through the DRS4 readout board. The signal is amplified with Analog Devices, ADA4187, and then fed to a comparator (Analog Devices, ADCMP604). A low voltage differential signaling (LVDS) output from the comparator is sent to the FPGA on the DRS4 readout board. This FPGA controls a DAC (Linear Technology, LTC2634) for thresholding of the comparator.

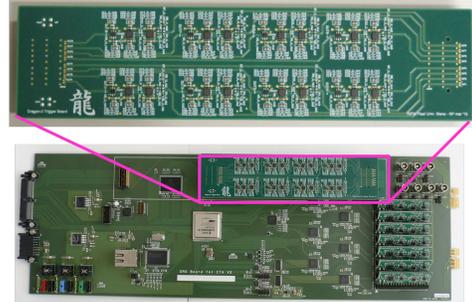

Figure 6. Photograph of the digital trigger mezzanine.

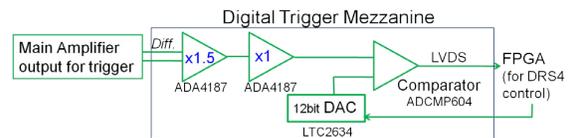

Figure 7. Block diagram of digital trigger mezzanine.

In the current phase of the CTA project, two methods of trigger generation are being developed: the digital trigger and the analog trigger. In the first version of our readout system, only the digital trigger mezzanine can be mounted. However, in the second version, the analog trigger mezzanines with level 0 or 1 can also be mounted as shown in Fig. 3. Performance evaluation of the DRS4 readout board with the analog trigger mezzanines is not yet complete. Therefore, in this paper, we report the performance of the readout system with the digital trigger mezzanine. The generated trigger is distributed via the backplane to FPGAs on backplanes of surrounding PMT clusters to generate a L1 trigger by coincidence of L0 triggers.

## 2.5 DRS4 chip

The DRS chip is being developed at the Paul Scherrer Institute (PSI), Switzerland, for the MEG experiment [3]. Four versions of the DRS chip have been designed so far by improving the functionality and performance in each version. The second version, DRS2, is used in the MAGIC-II experiment [6], and will be replaced with the current version, DRS4. The DRS4 chip includes nine differential input channels at a sampling speed of 700 MS/s–5 GS/s, with a bandwidth of 950 MHz, and a low noise of 0.35 mV after offset correction. The analog waveform is stored in 1,024 sampling capacitors per channel and the waveform can be read out after sampling via a shift register that is clocked at a maximum of 33 MHz for digitization using an external ADC. A write signal is generated for the sampling capacitors by a chain of inverters in the chip, and is stabilized by a phase locked loop (PLL). The power consumption of the DRS4 chip is 17.5 mW per channel at 2 GS/s sampling rate. The chip is fabricated using the 0.25 μm CMOS technology and is available in a 76-pin QFN package with a size of 9 mm × 9 mm.

It is possible to cascade two or more channels to obtain deeper sampling depth. In the first version of our readout system, the DRS4 chip was not cascaded. In the second version, four channels were cascaded, leading to a sampling depth of 4,096 for each PMT signal, which corresponds to a depth of 4 μs at 1 GS/s sampling rate. It enables continuous sampling without waiting the decision of trigger coincidence between multiple telescopes when a trigger of the local telescope is generated.

### 2.6 Self Calibration Function for DRS4

In the second version of the readout board, we implemented a self-calibration function for controlling the gain, offset, and timing of the SCA in the DRS4 chip. With respect to the gain and offset calibrations, the reference voltage is generated by a DAC and then fed to the DRS4 chip. With respect to the timing, a clock generated by the FPGA is fed to the DRS4 chip. After calibration, the systematic timing jitter for a given chip is below 40 ps.

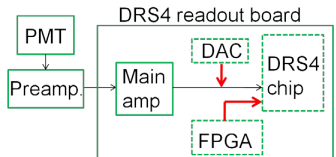

Figure 8. Self-calibration function for the DRS4 chip.

### 2.7 FPGA-based Gigabit Ethernet (SiTCP)

As described in section 2.1, the digitized waveform data and the monitor/control data are transmitted via Ethernet with only two devices: FPGA and Gigabit Ethernet transceiver (PHY). This simple composition is available on a hardware-based TCP processor, *SiTCP* [7]. The circuit size of SiTCP in the FPGA is ~3000 slices, which is enough small to allow implementation with user circuits on a single FPGA. In addition, the SiTCP has an advantage in that the throughputs in both directions can simultaneously reach the theoretical upper limits of Gigabit Ethernet.

## 3  Performance of Readout System

We have measured the performance of the DRS4 readout system (ver. 2) attached to a PMT. The preliminary results are shown here. Figure 9 shows a pulse shape of the high gain channel of the PMT signal with a gain of $2.4 \times 10^5$, which was measured with an LED light and the DRS4 readout system at a sampling rate of 2 GS/s. The PMT signal having a width of 5 ns and a height corresponding to 7 photoelectrons was successfully digitized. Figure 10 shows a single photoelectron spectrum of the high gain channel of the PMT signal with a gain of $9.6 \times 10^4$, which was measured with an LED light and the DRS4 readout system at a sampling rate of 2 GS/s. In the figure, a single photoelectron peak is clearly seen. The measurement of other performance parameters, including the case for a PMT gain of ~$5 \times 10^4$, is ongoing and will be presented at the conference.

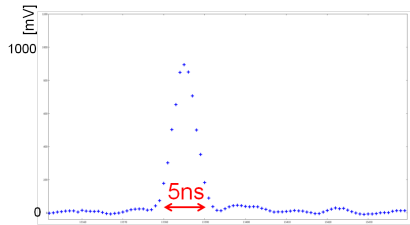

Figure 9. Pulse shape of the PMT signal measured with the DRS4 readout system at a sampling rate of 2 GS/s.

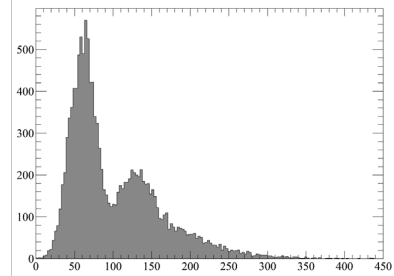

Figure 10. Single photoelectron spectrum of the PMT signal measured with the DRS4 readout system. The horizontal axis is in units of mV × ns.

## 4  Conclusion

We have developed a prototype 7-PMT cluster readout system for the next generation VHE gamma ray observatory, CTA. In the readout system named *Dragon*, a PMT signal is amplified, and its waveform is then digitized at a sampling rate of the order of GHz using an analog memory ASIC DRS4 that has 1,024 capacitors per channel. In the second prototype, four channels of the DRS4 chip are cascaded to obtain a sampling depth of 4,096. After a trigger is generated, the charges stored in the capacitors are digitized by an external slow sampling ADC and then transmitted via Gigabit Ethernet using the FPGA-based processor SiTCP. Using the prototype system attached to a PMT with a Cockcroft–Walton circuit, we successfully obtained a pulse shape of the signal of the PMT detecting an LED light at a sampling rate of 2 GS/s, and also a single photoelectron spectrum. Evaluation of the readout system's performance is ongoing, and a prototype with several PMT clusters and their readout boards will be constructed as the next step for the development of a full telescope camera.

### Acknowledgement

We gratefully acknowledge financial support from the agencies and organisations listed in this page: http://www.cta-observatory.org/?q=node/22

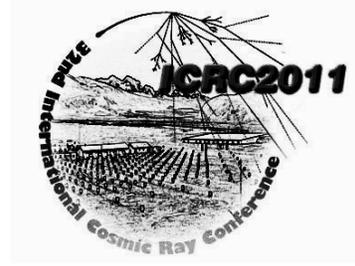

# Development of PMT Clusters for CTA-LST Camera

R. ORITO[1], H. OHOKA[2], M. AONO[3], Y. AWANE[3],A. BAMBA[4], M. CHIKAWA[5], R. ENOMOTO[2], Y. FUKAZAWA[6], D. FINK[7], S. GUNJI[8], R. HAGIWARA[8], M. HAYASHIDA[9], N. HIDATA[10], M. IKENO[11], S. KABUKI[12], H. KATAGIRI[13], K. KODANI[14], Y. KONNO[3], S. KOYAMA[15], H. KUBO[3], K. KURODA[13], J. KUSHIDA[14], R. MIRZOYAN[7], T. MIZUNO[6], T. NAKAMORI[16], K. NISHIJIMA[14], A. OKUMURA[17], O. REIMANN[7], T. SHCWEIZER[7], M. SASAKI[13], A. SHIBUYA[10], R. SUGAWARA[1], H. TAJIMA[10], H. TAKAHASHI[18], M. TANAKA[11], Y. TERADA[15], M. TESHIMA[2,7], F. TOKANAI[8], T. UCHIDA[11], K. UMEHARA[13], T. YAMAMOTO[19], K. YAMAOKA[4], M. YONETANI[6], A. YOSHIDA[4], FOR THE CTA CONSORTIUM

[1]*Institute of Socio-Arts and Sciences, The University of Tokushima, Tokushima 770-8502, Japan*
[2]*Institute for Cosmic Ray Research, The University of Tokyo, Kashiwa, Chiba 277-8582, Japan*
[3]*Department of Physics, Graduate School of Science, Kyoto University, Sakyo-ku, Kyoto, 606-8502, Japan*
[4]*Department of Physics and Mathematics, Aoyama Gakuin University, Sagamihara, Kanagawa, 252-5258, Japan*
[5]*Dept. of Physics, Kinki University, Kowakae, Higashi-Osaka 577-8502, Japan*
[6]*Department of Physical Science, Hiroshima University, Higashi-Hiroshima, Hiroshima 739-8526, Japan*
[7]*Max-Planck-Institut für Physik, Föhringer Ring 6, D 80805 München, Germany*
[8]*Yamagata University, Yamagata, Yamagata 990-8560, Japan*
[9]*Department of Astronomy, Graduate School of Science, Kyoto University, Sakyo-ku, Kyoto 606-8502, Japan*
[10]*Department of Physics and Astrophysics, Nagoya University, Chikusa-ku, Nagoya, 464-8602, Japan*
[11]*Institute of Particle and Nuclear Studies, KEK, Tsukuba, Ibaraki 305-0801, Japan*
[12]*Tokai University Hospital, Isehara-shi, Kanagawa, 259-1193 Japan*
[13]*College of Science, Ibaraki University, Mito, Ibaraki, 310-8512, Japan*
[14]*Department of Physics, Tokai University, Hiratsuka, Kanagawa 259-1292, Japan*
[15]*Graduate School of Science and Engineering, Saitama University, Sakura-ku, Saitama city, Saitama 338-8570, Japan*
[16]*Faculty of Science and Engineering, Waseda University, Shinjuku, Tokyo 169-8555, Japan*
[17]*Institute of Space and Astronautical Science, JAXA, Sagamihara, Kanagawa 252-5210, Japan*
[18]*Hiroshima Astrophysical Science Center, Hiroshima University, Higashi-hiroshima, 739-8526 Japan*
[19]*Department of Physics, Konan University, Kobe, Hyogo, 658-8501 Japan*

*orito@ias.tokushima-u.ac.jp*

**Abstract:** Following the great success of the current generation Imaging Atmospheric Cherenkov Telescopes (IACTs), a preparation of the next generation VHE gamma-ray observatory Cherenkov Telescope Array (CTA) is in progress. Here we report on the prototype photodetector module for the focal plane instrument of CTA Large Size Telescope (LST). In LST, Cherenkov photons are collected by 23 m reflective mirror and Cherenkov images are measured by $\sim$2500 photodetector pixels installed in the focal plane camera with a diameter of $\sim$2.5 m. For the LST camera, we have developed the compact PMT cluster module consisting of seven Hamamatsu R11920-100 PMTs with super bialkali photocathodes, Cockcroft-Walton type high voltage power supplies, preamplifier boards, a slow control board and fast readout electronics with low power consumption. Using this PMT cluster module, high performance Cherenkov camera for CTA-LST can be realized.

**Keywords:** Imaging Atmospheric Cherenkov Telescopes, Gamma-rays, Focal Plane Instrument, Photodetector

## 1 Introduction

The Cherenkov Telescope Array (CTA [1]) is a new generation VHE gamma-ray observatory which can achieve higher sensitivity with an order of magnitude compared to the current existing Imaging Atmospheric Cherenkov Tele-

scopes(IACTs). The CTA is planned to be constructed from an array of telescopes with different sized mirrors, large ($\sim$23 m), medium ($\sim$12 m), small ($\sim$4-7 m), which are called LST, MST, SST and SCT, respectively. Among these telescopes, LST (Large Size Telescope) is arranged in the center of the array and plays an important role for low energy gamma-ray observation below 100 GeV, which is the important energy range for studies of pulsars, high-redshift AGNs, GRBs, etc. In the current CTA preparatory phase, detailed studies of LST hardware components are ongoing[2] and some of them technically quite advanced. The focal plane camera[3] is one of the component which has much more advanced structure compared to the current IACTs. In the CTA-LST, ∼2500 photodetector pixels are planned to be installed in a sealed focal plane camera with the temperature controlled by a cooling system to avoid the deterioration of the photodetector modules. The diameter of the camera is about ∼2.5 m and each pixel has ∼0.1 deg field of view. The readout electronics is also planned to be installed inside the camera to avoid the problems caused by the long cabling between the telescope and counting house. To realize a high performance CTA-LST camera and manage the large number of photodetector channels, we are studying a compact photodetector cluster module with high performance and low power consumption. In this paper, we report on the status of the photodetector cluster development for the CTA-LST camera.

## 2 Development of PMT Cluster

The photodetector module has to be sufficiently high performance to satisfy the requirements from CTA-LST. Compactness and low power consumption are also important due to the limit of the weight and cooling ability. In addition, robustness and low cost for mass production are also necessary. After the research of last years, the current baseline of the photodetector for the CTA-LST camera is a photomultiplier tube (PMT) with a possibility of future upgrade to SiPMs[4]. We have developed the first PMT cluster module consisting of seven PMTs, Cockroft-Walton type high voltage power supplies, preamplifier boards, a slow control board and fast readout electronics. The compactness of the cluster is sufficiently enough to manage installation of several thousands channels. In addition, the cluster has a structure which the PMT can be replaced to an alternative photodetector in a future upgrade. We describe each of the components of this PMT cluster below.

### 2.1 PMT

A first candidate of the photodetecor for CTA-LST camera is Hamamatsu R11920-100 PMT which consists of 1.5 inch super bialkali photocathode with a concave-convex shape window and 8-stage dynodes. The requirements of the photodetector for CTA-LST camera are as follows: (1) peak quantum efficiency (QE) higher than 35% , (2) lifetime more than 10 years, (3) after pulse probability of less than $2\times10^{-4}$ over 4 ph.e., (4) pulse width of 2.5∼3 ns (FWHM), (5) dynamic range of ∼3000 ph.e., (6) timing resolution faster than 1.3 ns (TTS, 1 ph.e.). The Cherenkov telescope is operated under the night sky background (NSB) therefore points (2)–(4) are crucial. After some improvements of the PMT structure in last years, currently Hamamatsu R11920-100 PMT is satisfying these requirements. Here we have developed a prototype PMT module with the Hamamamtsu R11920-100 PMT. Figure 1 shows developed one PMT module consisting of R11920-100 PMT, a magnetic shield, a high voltage power supply and a preamplifier board. Figure 2 represents the QE of Hamamatsu R11920-100 PMT as a function of the wavelength.

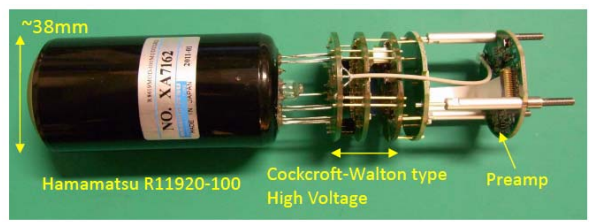

Figure 1: Photograph of the PMT module consisting of the Hamamatsu R11920-100 PMT, a high voltage power supply and a preamplifier board.

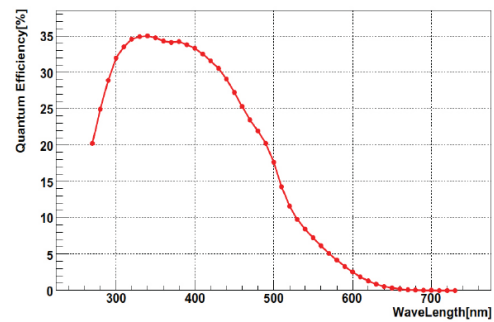

Figure 2: QE of the Hamamatsu R11920-100 PMT as a function of the wavelength.

### 2.2 High Voltage Power Supply

The Cockroft-Walton(CW) type high voltage (HV) power supply for the Hamamatsu R11920-100 PMT has been developed by Hamamatsu Photonics K.K.. The structure of the CW-HV is based on MAGIC-II CW-HV [5]. The CW-HV consists of three PCB boards and the power supply voltage is 5V. The HV from 0 ∼ -1.5 kV is controlled by an input range from 0 to 1.5 V and monitored by an output range from 0 to -1.5 V. The voltage between the photocathode and the first dynode is fixed to 300 V by low power Zener diodes for constant single photoelectron distribution. The voltage division for photocathode and 8 dynodes is 300V:1:2:1:1:1:1:2:1. Figure 3 shows the current of CW-HV as a function of HV under dark conditions. The power consumption is ∼ 40 mW maximum. Figure 4 shows the current of CW-HV under the NSB condition simulated by the LED at the PMT operation gain ∼5×10⁴.

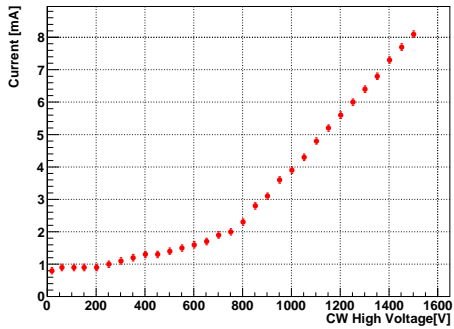

Figure 3: Current of CW-HV as a function of CW voltage under dark condition.

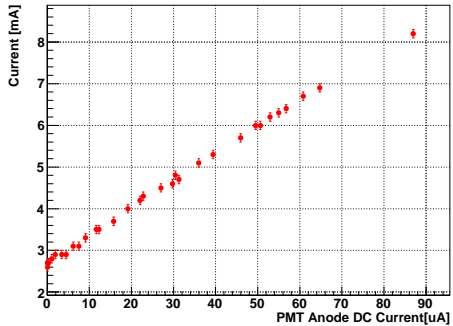

Figure 4: Current of CW-HV as a function of PMT anode DC current.

## 2.3 Preamplifier

In the CTA-LST camera, the PMT is planned to be operated with a gain of $4 \times 10^4$ considering the lifetime under NSB, moonlight or twilight conditions. To match the dynamic range with the readout electronics, the preamplifier board has been developed using the commercial chip (Mini-Circuit, LEE-39+). The circuit design is based on MAGIC-II preamp (SIRENZA, SGA-5586) [5] but here a new chip with lower power consumption is selected to reduce the power consumption in the CTA-LST camera. Figure 5 shows the photograph of the preamplifier board. The circuit for measuring the PMT anode DC current and injecting a test pulse for the electronics calibration are also implemented. The connector with a pitch of 0.5 mm is used for the connection to the following slow control board. Figure 6 shows an example of the pulse shape of R11920-100 PMT at the output point of the preamplifier board. The total power consumption of the preamplifier board is 183 mW. The PMT signal is amplified by a factor of 10.

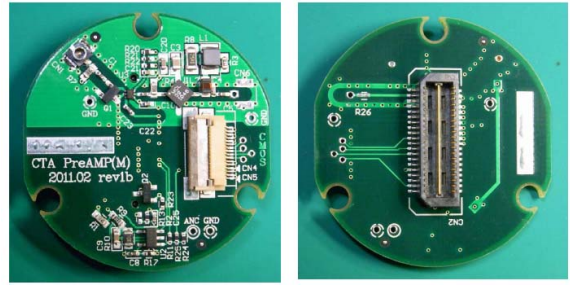

Figure 5: Photograph of the preamplifier board.

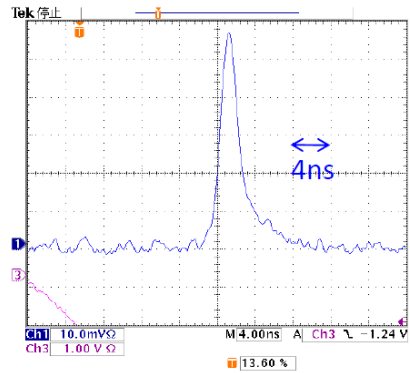

Figure 6: Pulse shape of the PMT signal after the preamplifier board.

## 2.4 Slow Control Board

The slow control board (SCB) has been developed mainly to control and monitor each PMT. The SCB controls the CW-HV with a 12-bits DAC, monitors both CW-HV and PMT anode DC current ($0 \sim 200 \ \mu A$) with a 14-bit ADC. Furthermore it generates the fast test pulse with different attenuations to be injected to the preamplifier board. A temperature sensor, humidity sensor, and a circuit for monitoring both the power voltage and current are also implemented. The power for CW-HV can be switched on or off individually. Figure 7 shows a photograph of the SCB. Seven PMT modules are attached to one side and the readout electronics board is attached to the other side. The PMT signal is fed to the readout board through SCB, therefore the pattern of PMT signal lines is carefully designed with the same layout length in SCB. To control the above mentioned functions, two Complex Programmable Logic Devices (CPLDs, Xilinx XC2C64A) are implemented in SCB. These CPLDs communicate with the Field Programmable Gate Array (FPGA) in the readout electronics board and the data can be sent via Ethernet. The power of SCB is ±3.3 V and 5 V which are common with the readout board. The power consumption of SCB is ~17 mW in standby condition.

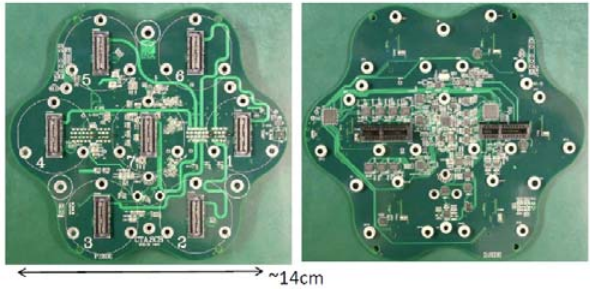

Figure 7: Photographs of the SCB. The PMT side (left) and readout board side (right), respectively.

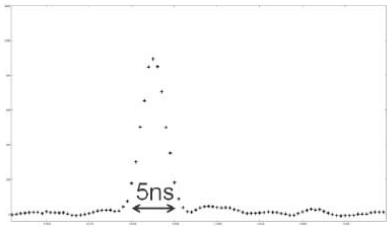

Figure 8: Pulse shape of the PMT signal taken with the DRS4 readout board at 2GS/s sampling.

### 2.5  Readout Electronics

The fast readout board has been developed using the Domino Ring Sampler (DRS [6]) capacitor array chip. This board consists of eight DRS-version4 (DRS4) chips, ADCs for digitizing a signal stored in the capacitor array in DRS4 at a sampling frequency of 25 MHz, a DAC for control of DRS4, a FPGA (Xilinx Virtex-4), an 18Mbit-SRAM, a Gigabit Ethernet transceiver, and a data I/O connector to the backplane on the left side. In this board, the PMT signal is digitized at a sampling rate of 2GS/s and then transmitted via Gigabit Ethernet. Details are described in separated paper [7]. The PMT signal has been successfully read and transmitted via Gigabit Ethernet. Figure 8 shows the PMT signal recorded by this board.

### 2.6  Mechanical Structure

Figure 9 shows a photograph of the assembled PMT cluster. Seven PMT modules are attached to the SCB with a pitch of 48 mm and the DRS4 readout board is attached to the backside of SCB. The surface level of seven PMTs is adjusted by spring adjusters installed between the CW-HV and the preamplifier board. A Winston Cone is planned to be installed in front of PMTs to collect Cherenkov light efficiently. The total power consumption and weight of this module are 14 W and 1.3 kg respectively. To be able to operate ∼350 PMT clusters in a sealed camera body with constant temperature, we are now designing a cooling system using special heat pipes and a cooling plate.

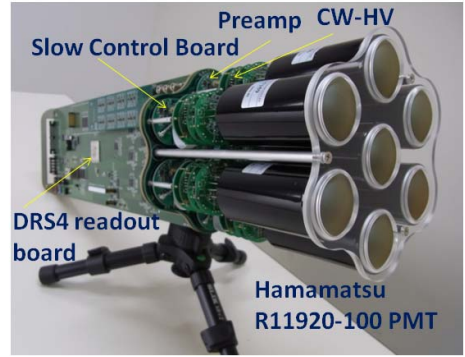

Figure 9: Photograph of the prototype PMT cluster.

## 3  Summary

The preparation of the next generation VHE gamma-ray observatory CTA is in progress. Here we reported on the development of the photodetector module for the CTA-LST camera. We have developed the prototype PMT cluster which consists of seven Hamamatsu R11920-100 PMTs, Cockcroft-Walton type high voltage power supplies, preamplifier boards, the slow control board and fast readout electronics. This is the first complete prototype of the photodetector cluster module which is optimized for CTA-LST camera. Several clusters and a small prototype camera will soon be produced to further validate the feasibility to construct the CTA-LST camera by using this cluster.

## Acknowledgement

We gratefully acknowledge support from the agencies and organisations listed in this page: http://www.cta-observatory.org/?q=node/22.

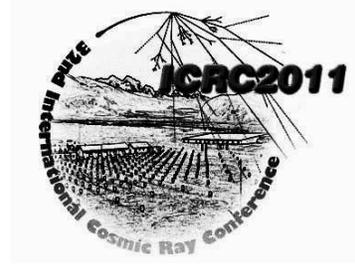

# FlashCam: A camera concept and design for the Cherenkov Telescope Array CTA

G. PÜHLHOFER[5], C. BAUER[1], A. BILAND[2], D. FLORIN[3], C. FÖHR[1], A. GADOLA[3], R. GREDIG[3], G. HERMANN[1], J. HINTON[4], B. HUBER[3], C. KALKUHL[5], J. KASPEREK[6], E. KENDZIORRA[5], T. KIHM[1], J. KOZIOL[7], A. MANALAYSAY[3], P.J. RAJDA[6], T. SCHANZ[5], S. STEINER[3], U. STRAUMANN[3], C. TENZER[5], P. VOGLER[2], A. VOLLHARDT[3], Q. WEITZEL[2], R. WHITE[4], K. ZIETARA[8], FOR THE CTA CONSORTIUM[8]

[1] *Max-Planck-Institut für Kernphysik, P.O. Box 103980, D 69029 Heidelberg, Germany*
[2] *ETH Zürich, Inst. for Particle Physics, Schafmattstr. 20, CH-8093 Zürich, Switzerland*
[3] *Physik-Institut, Universität Zürich, Winterthurerstrasse 190, 8057 Zürich, Switzerland*
[4] *Dept. of Physics and Astronomy, University of Leicester, Leicester, LE1 7RH, UK*
[5] *Institut für Astronomie und Astrophysik, Kepler Center for Astro and Particle Physics, Eberhard-Karls-Universität, Sand 1, D 72076 Tübingen, Germany*
[6] *Faculty of Electrical Engineering, Automatics, Computer Science and Electronics, AGH University of Science and Technology, Al. Mickiewicza 30, 30-059 Cracow, Poland*
[7] *Jagiellonian University, ul. Orla 171, 30-244 Cracow, Poland*
[8] *See http://www.cta-observatory.org/?q=node/342 for full author & affiliation list*
*Gerd.Puehlhofer@astro.uni-tuebingen.de*

**Abstract:** The future Cherenkov Telescope Array (CTA) will consist of several tens of telescopes of different mirror areas. CTA will provide next generation sensitivity to very high energy photons over a wide energy range from few tens of GeV to >100 TeV.

The signals of the photon detectors in the focal plane will be read out with custom-designed, fast digitization and triggering electronics. Within CTA, several camera electronics options are being evaluated. The FlashCam approach is unique here since data processing inside the camera is fully digital. Signal digitization and trigger processing are jointly performed in one single readout chain per detector pixel. For a group of pixels, such a chain consists of Flash ADCs and a Field-Programmable Gate Array (FPGA) module, both commercially available and reasonably priced. A camera-wide event trigger is subsequently computed in separate trigger units. The camera data is transferred from the cameras over ethernet to a central computer farm using a custom network protocol. Such a fully digital approach using state-of-the-art components provides accurate triggering and an easily scalable architecture in a cost-effective way.

The FlashCam team is also evaluating the concept of a horizontal camera integration. Here, the photon detector plane is sustained by a monolithic carrier holding photon detectors, high voltage supplies, and preamplifiers only. The signal digitization and triggering electronics are organized in boards and crates which are kept at the rear side of the camera body. Such an approach allows different photon detectors to be adopted and might result in cost saving.

**Keywords:** CTA, Imaging Atmospheric Cherenkov Telescopes, gamma-rays, electronics

## 1 The FlashCam camera architecture

FlashCam is a simple and straightforward concept of an electronics system for use in CTA cameras. It is based purely on commercially available microchips, in essence FADCs for signal digitization and FPGAs for further signal and trigger processing. The key point of the concept is to perform all pixel signal processing purely with digitized information. In particular, the trigger decision is computed in the camera solely from the digitized signal, a separate trigger signal path can therefore be saved. Such a trigger system is fully programmable and therefore highly flexible.

For such a camera, low power (<0.5 W/channel), 12-bit FADCs are currently only available with sampling speeds of up to ∼250 MS/s. But extensive simulations, including realistic time jitter between pixels and night sky background conditions, have shown that the trigger performance of such a system applying digital trigger options is fully competitive with higher (e.g. 2 GS/s) sampling speed systems. Moreover, the data rate of such a system (∼600 MB/s) allows the full pixel event information to be transmitted over standard gigabit ethernet infrastructure, including commercial switches, without data reduction.

Another unique feature of the FlashCam approach is the spatial separation of the photon detector plane (PDP) at the camera front from the signal digitization electronics (FADC boards) which is kept at the rear side of the camera body. Inbetween, the preamplified, analogue pixel signals are transmitted over standard CAT 5 cables. Such a layout can easily be adapted to different photon detectors and pitches, be it standard photomultiplier tubes (PMTs) or, e.g., Silicon PMT (SiPM). It also reduces the weight at the focal plane compared to drawer-based approaches. Ultimately, such "horizontal" integration could reduce camera costs.

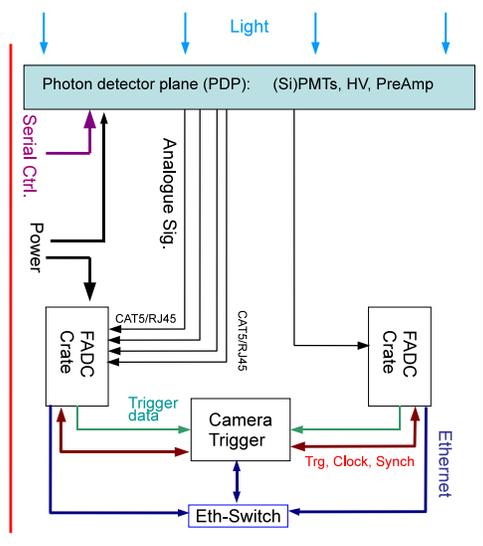

Figure 1: Schematics of the FlashCam camera architecture

Fig. 1 illustrates the global FlashCam camera layout.

## 2 Mechanical design of the photon detector plane

The current mechanical concept is based on PMTs as photon detectors and foresees modules of 12 pixels (see Fig. 2 left panel). Each module contains PMTs, pre-amplifiers, and high voltage (HV) supplies, and weighs about 1.2 kg. Two of such modules can be served by one FADC electronics board. Current drawings and numbers are for a 1764 pixel camera. The entire PDP structure weighs 81 kg, with a maximum sagging of 0.68 mm (180 kg load including electronics, camera at 90°). Such a PDP can be passively cooled. Fig. 2 shows corresponding PDP drawings and a 36 pixel mechanical prototype.

## 3 Non-linear preamplifiers at the photon detector plane

To achieve the currently anticipated dynamic range of up to 3000 photoelectrons (pe) per pixel, the FlashCam concept foresees the use of preamplifiers with non-linear gain characteristics, instead of splitting the signal into two separate

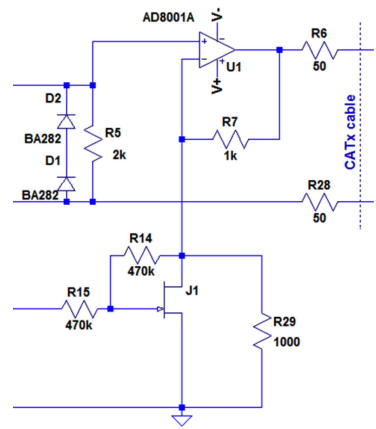

(a) Non-linear ("saturation") circuit.

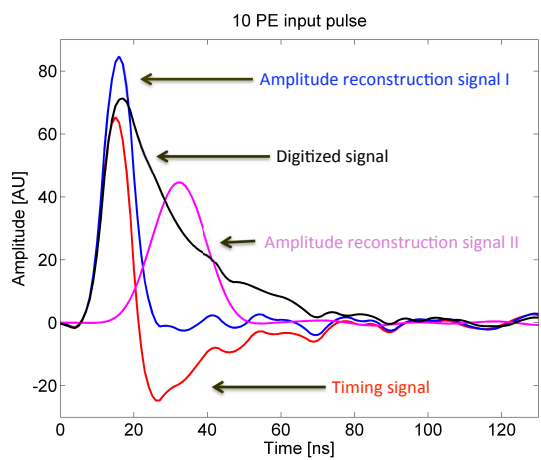

(b) Signal from one event as measured with a FlashCam demo board, sketching the signal reconstruction in the low pe regime.

Figure 3: Illustration of the non-linear preamplifier.

digitization channels with different (low and high) gain. This allows cost to be reduced (the low gain channel can be saved), and permits full flexibility should future simulations result in a reduced dynamic range requirement.

The baselign design is to run an operational amplifier with a high input resistor (1-2 kΩ instead of 50 Ω, see Fig. 3a). High amplitude signals (≳ 200 pe) will saturate the amplifier, but the input signal can be restored since the output signal broadens with a defined recovery time. The analysis of the digitized signal (performed at the FPGA directly after the FADC) will therefore run in two regimes: Below 200 pe, the full time-resolved signal can be stored (with a window of ∼60 ns), and digital filtering performed at the FPGA allows the signal amplitude to be directly computed (Fig. 3b). Above 200 pe, the signal is reconstructed from an integration over 200 ns. The available bandwidth is sufficient to transmit signals from both regimes for all pixels simultaneously, allowing e.g. for cross-calibration in the transition regime.

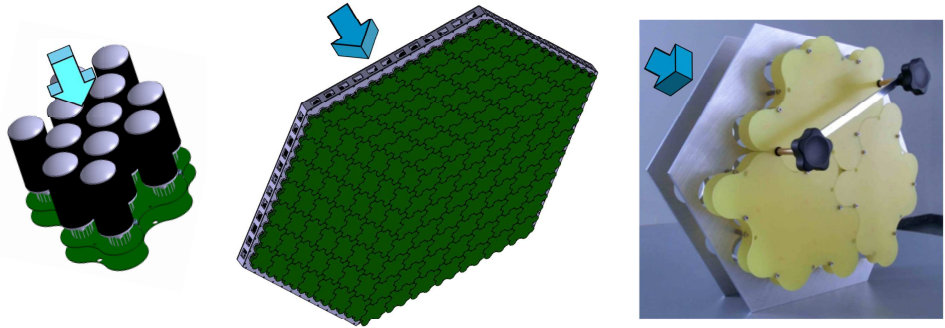

Figure 2: Left: drawing of a 12 pixel module. Middle: drawing of the entire 1764 pixel structure. Right: picture of a 36 pixel prototype. Arrows indicate the direction of the incoming light.

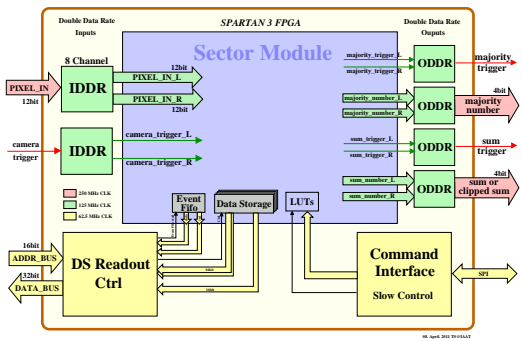

Figure 5: Schematics of the trigger and data processing unit on the Spartan 3/6 FPGA as used on the "parallel" demonstrator board shown in Fig. 4, indicating how the limited performance of that FPGA type can be optimally exploited using e.g. different clock speeds in separate chip domains.

## 4  Two electronics boards to demonstrate the performance and evaluate FPGA options

The electronics boards holding FADCs for signal digitization and FPGAs for further signal and pixel (patch) trigger processing come in large quantities (75 boards serving 24 pixel each, for a 1764 pixel camera). They are therefore pronounced in the overall cost estimate and need to be designed as cost-efficient as possible.

The FlashCam group has extensively evaluated two design options with dedicated demonstrator boards. One option is based on a Spartan 3/6 FPGA, connected with parallel data lines to the FADC chips (see Figs. 4 and 5). The second option employs a Virtex 6 FPGA, connected serially to the FADCs (Fig. 6). This second option is more pin-economical, but current FPGA price estimates strongly favour the first option. The planning towards a camera-scale prototype is therefore based on a design around a Spartan 6 FPGA.

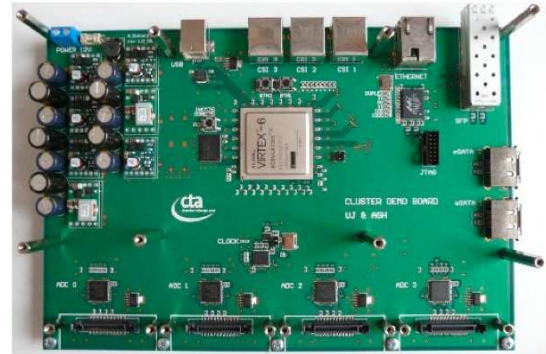

Figure 6: "Serial board": Fully functional demonstrator board for 16 channels, using a Virtex 6 FPGA.

## 5  Triggering: From pixels and pixel patches to a camera-wide event trigger decision

In the FlashCam design, the event trigger is derived from the digitized pixel signals processed in FPGAs. Various trigger algorithms (see e.g. Fig. 7) can therefore be exploited and improved even during run-time of the experiment. Such algorithms are only limited by the FPGA processing power and memory capacity, and by the trigger signal bandwidth between FADC electronics boards and a central camera trigger unit. Simulations show that a pre-grouping of pixel trigger into pixel "patch" information can be used to reduce bandwidth requirements (number of pins) without significant loss of sensitivity. A patch size of three pixels (see Fig. 8) provides excellent spatial trigger homogeneity and sensitivity combined with high versatility and is therefore used for the planning of a camera-scale FlashCam prototype.

Pin count and processing power limitations of currently available high-end FPGAs do not permit collection of the trigger information from all electronics boards into one single camera trigger unit. The camera is therefore logically separated into three sectors (see Fig. 9).

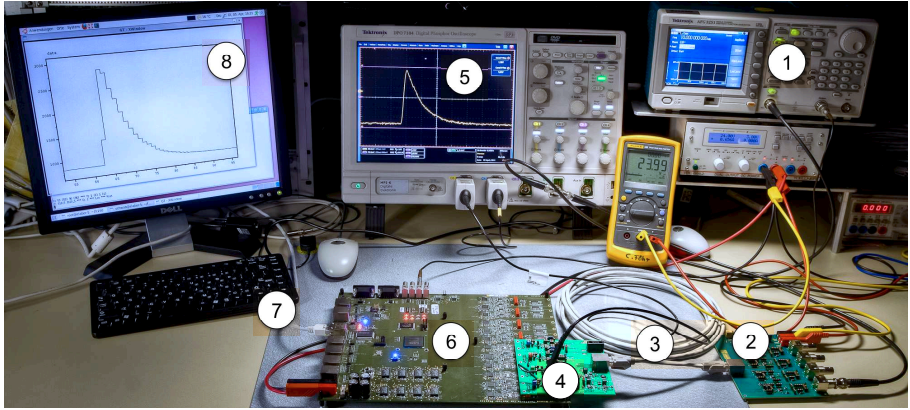

Figure 4: Test setup for the "parallel board", one of the two FlashCam demonstrator boards. 1: PMT pulse generator. 2: Preamplifier board. 3: Analogue signal transmission (CAT 5). 4: ADC driver board. 5: Analogue pulse before ADC. 6: Demo board with 8 parallel FADCs and FPGA. 7: Event transmission via LAN. 8: Digitized pulse (4 ns / step).

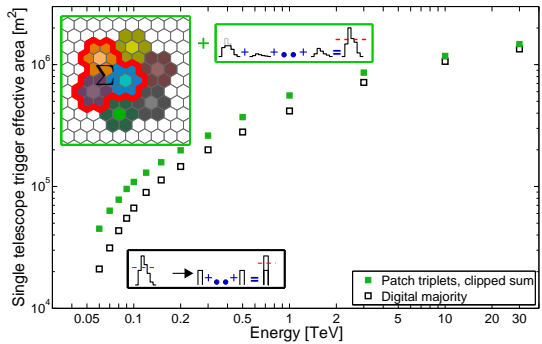

Figure 7: Simulations of the single telescope trigger efficiency to γ-ray events for two different trigger algorithms. For a fair comparison, thresholds are adapted both times to result in an accidental trigger rate (from night sky noise) of 500 Hz. Therefore, a more sophisticated triggering scheme (clipping to suppress PMT afterpulse noise, patch summation before trigger signal transmission) effectively increases the γ-ray trigger efficiency, compared to a simple pixel majority trigger.

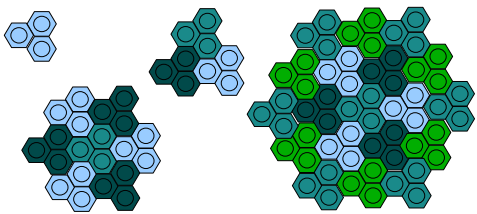

Figure 8: In the central trigger units, patches of three pixels (top left) are combined in sliding windows to derive a final trigger decison (e.g. sum of 3, 7, 19 patches as shown).

## 6 Further extensive component testing

In summary, all critical components required to construct a FlashCam camera have been evaluated and tested. In addi-tion to the components discussed in the previous sections, these comprise: (1) a 10 MHz heat-stabilized oscillator synchronized by a GPS receiver for inter-telescope event synchronisation, providing timing accuracy of better than 10 ns (max. error); (2) the infrastructure for raw ethernet data transmission from the camera to a processor farm, the tests of which resulted in sustained loss-free data transfer of 700 MB/s; (3) the use of CAT 5/6 cables for analogue signal transmission between preamplifier and FADC boards, with tests that also included the influence of RJ 45 sockets; (4) cabling for digital trigger signal transmission, tested e.g. through extensive monitoring of bit error rates.

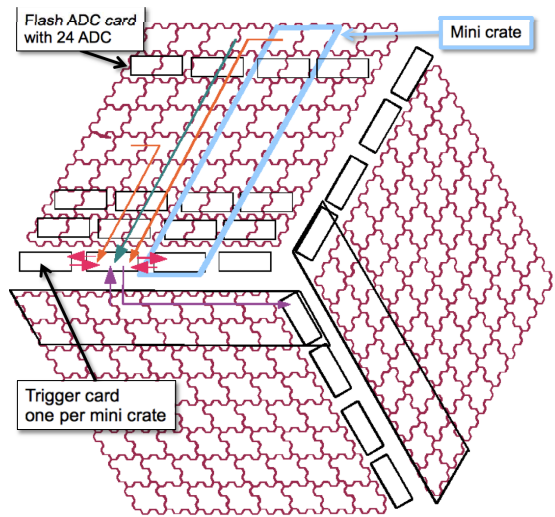

Figure 9: Logical grouping of the FADC boards in a full-scale camera. Bandwidth is provided (purple arrows) for a fully seamless trigger across sector boundaries with trigger areas of up to 19 patches (largest area in Fig. 8). The total data traffic to the trigger cards is 12*140Gb/s = 1.68Tb/s.

**Acknowledgements:** We gratefully acknowledge support from the agencies and organisations listed in this page: http://www.cta-observatory.org/?q=node/22

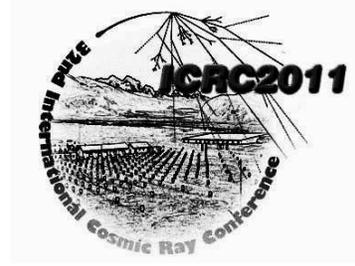

# Performance Study of a Digital Camera Trigger for CTA

R. WISCHNEWSKI[1], U. SCHWANKE[2], M. SHAYDUK[1], K. SULANKE[1], ON BEHALF OF THE CTA-CONSORTIUM [3]

[1]*DESY, Zeuthen, Germany*
[2]*Humboldt University, Berlin, Germany*
[3]*see http://www.cta-observatory.org/?q=node/22*
*ralf.wischnewski@desy.de*

**Abstract:** We present a design study for a trigger system for the imaging cameras of the "Cherenkov Telescope Array" (CTA). The proposed FPGA-based camera trigger operates on 400MHz time-sliced 1-bit-camera images. It allows to search for topological majority patterns, and is capable to search for complex space- or space-time patterns; it also allows for an image-driven readout control. A MC comparison of this and alternative trigger concepts is done.

**Keywords:** Gamma Ray, Imaging Atmospheric Cherenkov Telescope, CTA.

## 1 Triggering of CTA-Cameras

The next generation VHE gamma-ray observatory "Cherenkov Telescope Array" (CTA) is currently in its preparatory phase[?]. It will consist of 50-80 Imaging Air-Cherenkov Telescopes. The low and mid energy range of 0.010 - 10 TeV will be covered by Large Size (LST) and Medium Size Telescopes (MST) with 23m and 12m dishes. In the baseline design, the LST and MST cameras have 2841/1765 pixels (PMTs), arranged in hexagonal clusters of 7 pixels, as shown in fig.**??**. LST and MST are made of 400 and 250 cluster, respectively.

A typical gamma-ray shower generates a light flash of few nsec duration in spatially neighbored pixels (for ≫1 TeV with duration up to several tens of ns). The background is dominated by random night-sky-background (NSB) photons, causing at least 120 MHz pixel countrates (at single photoelectron (pe) threshold), and by large amplitude PMT afterpulses (APs) faking large pixel signals.

Trigger strategies used in current generation telescopes are based on either the topological distribution of pixel hits (ie. PMTs with signals above a discriminator threshold), or on the analog sum of the PMT signals : (1) Majority trigger : At least $N_{maj}$ pixels are above threshold (a few pe) for a coincidence gate of few ns
(2) SumTrigger: The analog sum of all pixel amplitudes must be $>A_{Sum}$ (20pe typical). To suppress afterpulse triggers, each pixel signal is clipped (at ∼6pe).

A camera trigger in this approach occurs, if the trigger condition is fulfilled for at least one out of a number of pre-defined static search regions (typically made up of tens of pixels each; and covering the full camera or part of it).

For this study, we use 7 pixels as base unit to build various search regions (see fig.**??**): Cluster-Doublets (pairs of neighbouring camera clusters), Cluster-Triplets (compact triplets of neighbouring clusters), and Cluster-Singlets ("scanning" singlets, build around every pixel). By construction, these search regions have a high degree of overlap, to avoid trigger inefficiencies due to edge effects.

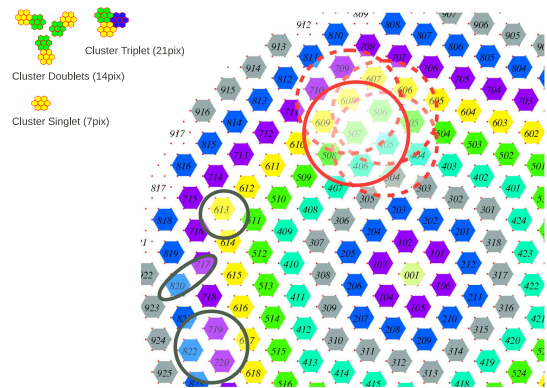

Figure 1: The cluster structure of a MST-Camera (part): each hexagon represents a 7-pixel-cluster. The grey upper circles indicate 3 neighboured supercluster (each made of 7 next neighbour clusters, corresponding to 49 pixels, ie. the region available for each Cluster-FPGA algorithm). Indicated also baseline search geometries: Cluster-Singlets, Doublets and Triplets (grey lines); also given in the insert.

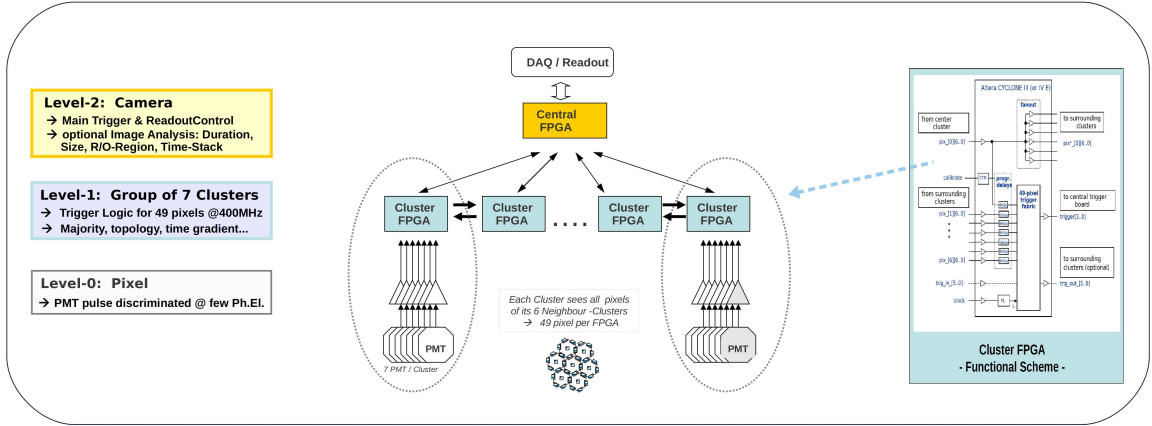

Figure 2: Block scheme of the FPGA-based Camera Trigger. Each *Cluster-FPGA* operates on 49 pixels/PMTs (7 direct and 6x7 neighbouring cluster pixels). *Central FPGA* at camera level: allows for optional Level-2 trigger / image classification. *Right insert:* Functional scheme for each Cluster-FPGA: input (7+42 pixel), digital delay lines, trigger factory, 7-pixel fanout to neighbouring cluster and trigger output (3 bits).

## 2  The FPGA Camera Trigger: Concept

The proposed digital FPGA camera trigger analyzes the discriminated PMT signals of all camera pixels, as illustrated in fig.**??**.

At *Cluster-level* (level-1 in fig.**??**), each cluster (7 pixels) is connected to its cluster-FPGA board (a single, cheap 400MHz FPGA), together with all 42 pixels of the surrounding six clusters. This 49-pixel super-cluster region is analyzed by the baseline trigger algorithms running on the 400MHz FPGA. Thus, the main trigger is essentially based on 400MHz time-sliced 1 bit-camera images. Additional advantages of the FPGA design is a simple implementation of internal programmable digital delays (auto-calibration), and of asynchronous pixel coincidences (i.e. without synchronizing the input signals to the FPGA clock) to avoid image splitting between time slices. Beyond the basic majority algorithms, also complicated topological algorithms are possible, including simple (1-bit) image-cleaning. In addition, if needed, the full pixel time history is available at super-cluster level; which can be used to find e.g. large-impact showers, but also allows to extract information after an array-level trigger for non-triggered telescopes.

At *Camera-level* (level-2 in fig.**??**), a simple centralized solution using a single FPGA is possible, collecting results of the various parallel trigger algorithms running on the Cluster-FPGAs. This provides a classification of images wrt. their duration, essential to adapt the camera readout time window to prevent signal loss.

The Cluster FPGA functional scheme is given in fig.**??**(right). First boards, with algorithms implemented, are being successfully tested (see fig.**??**).

## 3  Detailed Camera-Trigger MC Simulations

The generation of photoelectrons in the camera plane by $\gamma$-showers is done with the default CTA-MC package *SimTelArray*[**?**]. Photoelectrons are then processed by a dedicated trigger package *TrigSim* (scanning a large trigger parameter space).

The subset of investigated trigger scenarios shown here is (all for asynchronous FPGA-coincidences):
MajorityTrigger: ScanningCluster-Singlets/Majority_3 (*ScSinglMaj3*, i.e. $N_{maj}$=3), Cluster-Doublets/Majority_4 (*DoublMaj4*), Cluster-Triplets/Majority_5/7/9 (*TriplMaj5/7/9*), CompactThreeNextNeighbours (*3NN*);
SumTrigger: ScanningCluster-Singlets (*SumSingl*), Cluster-Doublets (*SumDoubl*), with clipping at 6pe (broad minimum - not optimized).

The parameters varied are : NSB-rate (MHz/pixel) = 125/250; PMT afterpulse rate (>4 pe) = 0.013% (*E4*) / 0.0434% (*E3*); analog bandwidth - the 1pe pulse-fwhm is the relevant input variable = 2.6 / 5 / 10ns; coincidence window (majority) = 1.7 / 5 / 8.3 ns.

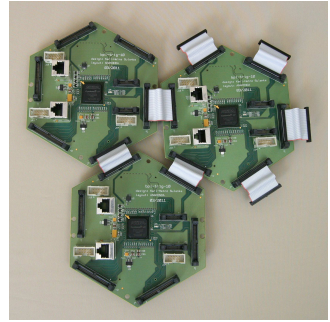

Figure 3: Prototype Cluster FPGA trigger boards, interconnected for inter-cluster data transfer.

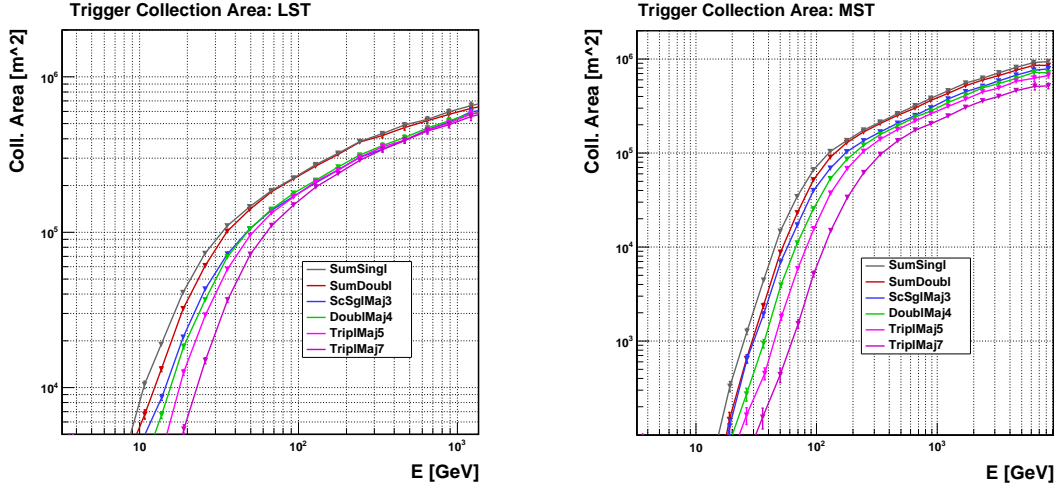

Figure 5: Trigger collection area for LST and MST for SumTrigger (SumScSingl, SumDoubl) and Majority trigger scenarios (ScSglMaj3, DoublMaj4, TriplMaj5/7).

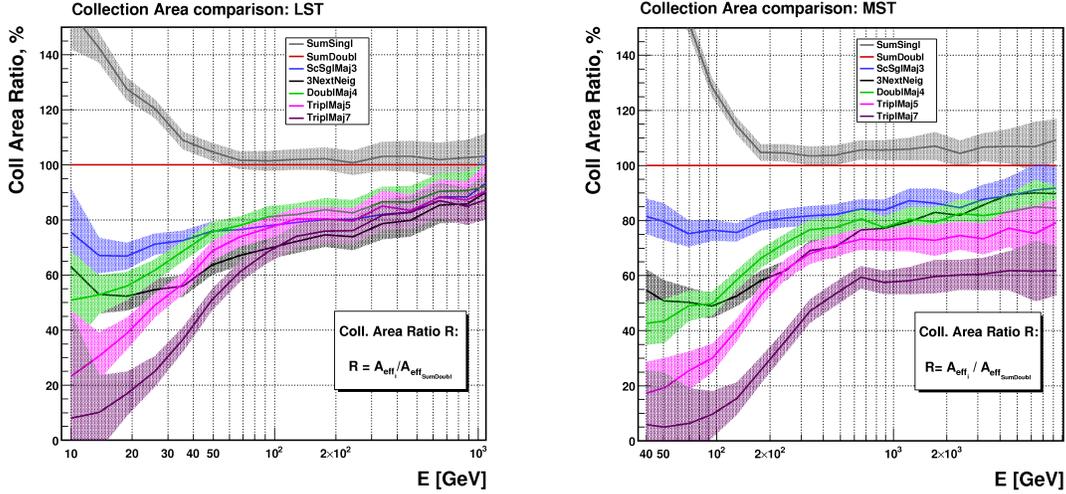

Figure 6: Trigger collection area for LST and MST from fig.**??**, normalized to Sumtrigger 'SumDoubl'.

The MC procedure for each chosen parameter set and trigger scenario is: First, obtain the working points, ie. find those pixel- / analogsum-thresholds that keep the camera trigger rate for NSB-only events at <100 Hz, see fig.**??**. For these, the trigger collection area for $\gamma$'s is determined. An optional image cleaning roughly verifies the triggered image quality.

## 4 Results: Comparing trigger Scenarios

The trigger collection areas for various algorithms for LST and MST are shown in fig.**??**. The parameter space for trigger optimization is large, results here are limited to optimal conditions (125MHz NSB, low AP-rate *E4*, high bandwith (fwhm 2.6ns) and coincidence window 1.7ns. The two SumTrigger scenarios are performing best over the entire energy range. Fig.**??** gives the trigger collection areas normalized to that of the DoubletSumtrigger (a realistic analog sumtrigger). The so far tested (simple) majority schemes perform for MST 20% (lowest) to 10% worse (highest energy) than sumtrigger; lower majority multiplicities $N_{maj}$ (ScSglMaj3, DoublMaj4) seem to work better than larger (TriplMaj5/7, with lower pixel amplitude threshold). At high energy, though, better performance is found for larger coincidence windows (5.0/8.3ns), see fig.**??**; large multiplicities (TriplMaj5/7) and low thresholds alone already reach >90% of the SumTrigger collection area. Note, that low thresholds imply less severe PMT AP-requirements.

The dependence of collection area on the bandwith (pulse fwhm of 2.6/5.0/10.0ns) is given in fig.**??** for the default majority scheme (ScSglMaj3) and coincidence window of 1.7ns. For LST, due to higher pixel thresholds, larger pulse-

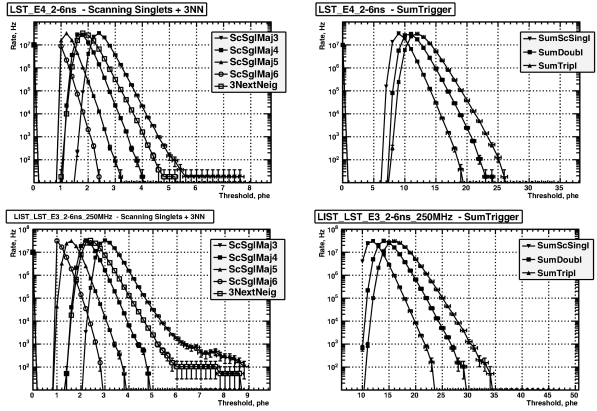

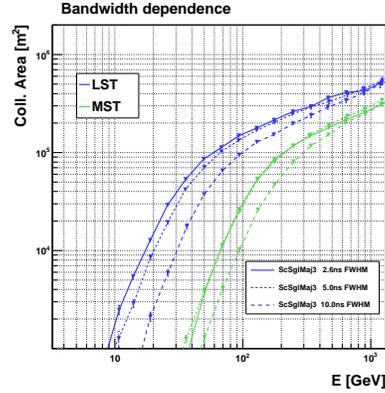

Figure 4: Pure NSB simulation. Camera trigger rate versus pixel-/analogsum-threshold. Upper plots: NSB 125, AP *E4*; lower plots: NSB 250, AP *E3*. Left: Majority trigger for ScSinglMaj3-6, 3NN; right: SumTrigger for SumScSingl, SumDoubl and SumTripl. (Rate drop at >10MHz/low thresholds is a MC-artefact due to counter saturation (10ns update window)).

Figure 8: Bandwidth dependence of collection area for LST and MST for 250MHz NSB, AP *E4*.

- an easy image-type classification at trigger level, thus for high energy large shower impacts an extended camera readout window prevents signal loss.

From a quantitative comparison of alternative trigger algorithms for the Large/MediumSize CTA telescopes, we find:

- AnalogSum and Digital-Majority scenarios show a similar performance. The AnalogSum yields lower threshold for LST, and o(15%) area improvement for MST/LST at medium energy scales;

- Best AnalogSum-trigger performance is for a scanning-single-cluster algorithm (difficult to realize in hardware, except for a digital FADC readout);

- Digital trigger scenarios require PMTs with low afterpulse-rate PMTs (unless also using AP-clipping, or working with low threshold/high majority);

- Image cleaning has a good efficiency for all scenarios, this indicates good physics performance of the trigger algorithms.

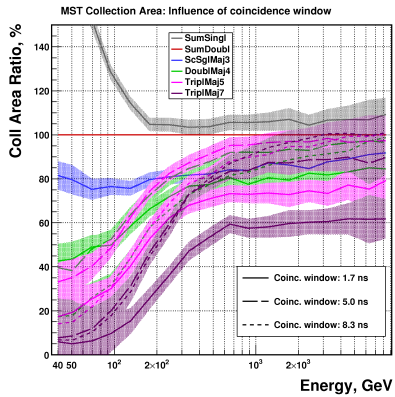

Figure 7: Relative MST collection area as function of coincidence window: 1.7, 5.0, 8.3ns (NSB/AP as in fig.**??**). At high energy, larger windows are favored.

fwhm (ie. lower bandwidth) results in worse sensitivities already for 5ns fwhm.

Next steps of this analysis aim at an improved sensitivity at high energies (≫1 TeV) by including larger time intervals and by improved image search topologies. For better low-energy performance, we plan to go beyond simple digital majority-schemes. With a second discriminator threshold, improved majority performance is expected. To realistically quantify the relative performance gain of sum- and majority-trigger concepts, we extend the analysis to the CTA-array trigger.

## 5 Conclusions

We presented the basic design ideas and a detailed MC-simulation for a FPGA based digital camera trigger, that could be used for CTA telescopes. The concept offers

- a simple and realistic hardware implementation for a camera trigger; detailed hardware tests underway;

- majority algorithms yield good collection area; FPGA resources allow for more complex and for several, parallel running algorithms;

- a natural implemenation of a 2nd-level trigger, and a baseline image analysis at trigger level ;

**Acknowledgements** We gratefully acknowledge financial support from the agencies and organisations listed in this page: http://www.cta-observatory.org/?q=node/22 .

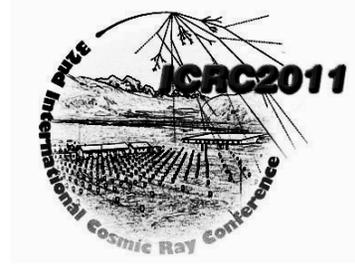

# Development of an ASIC for Dual Mirror Telescopes of the Cherenkov Telescope Array

JUSTIN VANDENBROUCKE[1], KEITH BECHTOL[1], STEFAN FUNK[1], AKIRA OKUMURA[2,1], HIRO TAJIMA[3],
GARY VARNER[4], FOR THE CTA CONSORTIUM

[1]*W. W. Hansen Experimental Physics Laboratory, Kavli Institute for Particle Astrophysics and Cosmology, Department of Physics and SLAC National Accelerator Laboratory, Stanford University, Stanford, CA 94305, USA*
[2]*Institute of Space and Astronautical Science, JAXA, Sagamihara, Kanagawa 252-5210, Japan*
[3]*Solar-Terrestrial Environment Laboratory, Nagoya University, Nagoya 464-8601, Japan*
[4]*Department of Physics and Astronomy, University of Hawaii, Honolulu HI 96822, USA*
*justinv@stanford.edu*

**Abstract:** We have developed an application-specific integrated circui (ASIC) for photomultiplier tube (PMT) waveform digitization which is well-suited for the Schwarzschild-Couder optical system under development for the Cherenkov Telescope Array (CTA) project. The key feature of the "TARGET" ASIC is the ability to read 16 channels in parallel at a sampling speed of 1 GSa/s or faster. In combination with a focal plane instrumented with 64-channel multi-anode PMTs (MAPMTs), TARGET digitizers will enable CTA to achieve a wider field of view than the current Cherenkov telescopes and significantly reduce the cost per channel of the camera and readout electronics. We have also developed a prototype camera module, consisting of 4 TARGET ASICs and a 64-channel MAPMT. We report results from performance testing of the camera module and of the TARGET ASIC itself.

**Keywords:** ASIC, Instrumentation, CTA, MAPMT, TARGET

## 1 Introduction

The Cherenkov Telescope Array (CTA) experiment is a next-generation verhy-high-energy gamma-ray observatory featureing an array of imaging atmospheric Cherenkov telescopes (IACTs) that will be an order of magnitude more sensitive than the current generation of instruments [1]. The energy band covered by CTA will range from a few tens of GeV to beyond 100 TeV. To achieve the highest gamma-ray sensitivity ever with this wide energy coverage, CTA will be an array of $\sim 100$ telescopes consisting of a mix of a few different telescope designs.

One candidate of telescope designs is Schwarzschild-Couder midium-size telescope (SC-MST) which is being developed to realize a wide field of view (FOV) ($\sim 8°$ in diameter) and high angular resolution ($< \sim 0.1°$) at the same time by using dual mirrors in the optical system [2]. The focal-plane camera of the SC optical system consists of an array of 64-channel multi-anode photomultiplier tubes (MAPMTs), because the f-ratio of the SC optics is a few times large than normal IACTs, and thus the pixel size of the camera is required to be smaller than regular PMTs with diamters of $\sim 25$ mm.

In order to read Cherenkov signals from an MAPMT array, a compact and modular readout system running at a sampling speed of $> \sim 1$ GSa/s (giga-samples per second) is required. In addition, the cost per channel of the readout system is required to be as low as possible because a large number of telescopes are to be built.

## 2 TARGET

We have developed an application-specific integrated circuit (ASIC) which was designed to match the requirements of the SC-MST. The first generation of this ASIC, TeV Array Readout with GSa/s sampling and Event Trigger (TARGET 1), has self-trigger functionality, 16-channel parallel input, and a 4096-sample buffer for each channel [3]. The total cost per channel including front-end and back-end electronics is expected to be $\sim$ \$20 not including photo detectors.

Figure 1 illustrates a schematic diagram of the TARGET 1 ASIC which has an array of 4096 capacitors divided into 256 blocks aligned in 32 columns by 8 rows, where each block consists of 16 capacitors. The sampling speed of the array can be adjusted between 0.7 GSa/s and 2.3 GSa/s by changing an external voltage input, and is typically driven at 1.0 GSa/s. Therefore, each capacitor and the total buffer depth correspond to 1 ns and 4096 ns, respectively. By reading three blocks upon each trigger, the waveform length becomes 48 ns, while the length can be changed through a field-programmable gate array (FPGA).

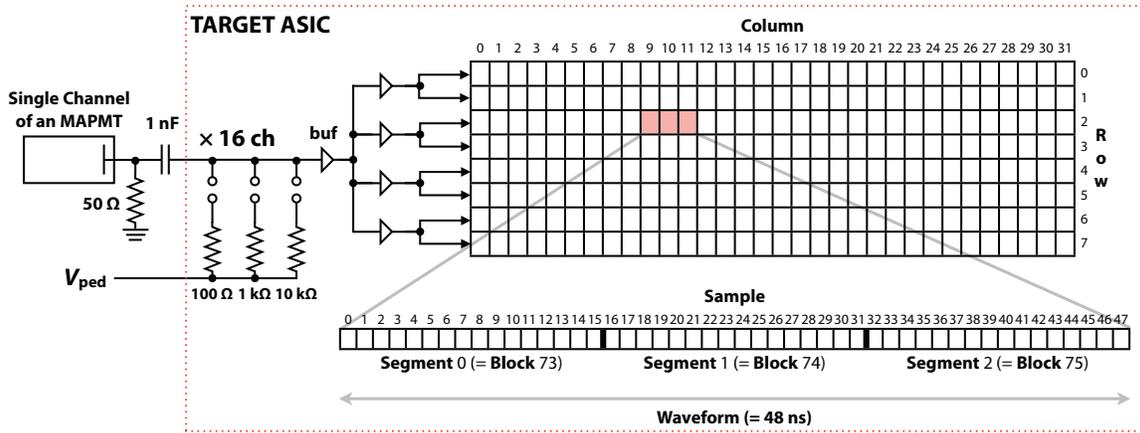

Figure 1: A schematic diagram of the TARGET 1 ASIC. A single AC-coupled channel of an MAPMT is connected to an array of 4096 capacitors. The 16 channels are digitized in pairs, with the two channels of each pair digitized simultaneously.. Three input impedances are selectable via the FPGA firmware. The pedestal level can also be changed by changing an external voltage, $V_{ped}$.

We typically set the waveform length to 48 or 64 ns in testing.

The storage capacitor voltage is digitized by Wilkinson-type analog-to-digital converters (ADC) equipped in TARGET 1. The voltage is measured by a gray-code counter which starts at the beginning of the Wilkinson ramp voltage, and stops when the ramp voltages equals the capacitor voltage. In the TARGET 1 evaluation board and camera module prototype to be explained in Section 3, the ADC resolution is 10 bits and 9 bits, respectively.

Table 2 is a summary of the specifications of TARGET 1. The same parameters for the 2nd version of TARGET (TARGET 2) are also listed. TARGET 2 chips have been designed and fabricated, and will be tested in 2011.

## 3 The TARGET 1 Camera Module Prototype

We have also developed the TARGET 1 evaluation board and the TARGET 1 camera module prototype as shown in Figure 2 and 3. Since the evaluation board was fabricated to study the basic characteristics of a TARGET 1 chip, it consists of a minimum set of components to operate a single ASIC. The camera module was designed to validate the concept of a combination of a 64-channel MAPMT and 4 TARGET 1 chips (16 channels by 4 ASICs). ~ 200 camera modules will be installed on the focal-plane camera of a single SC telescope in the future, where each ~ 36 modules will be controlled separately by a backplane board.

The camera module prototype has an MAPMT[1], a high-voltage (HV) power unit[2], a universal serial bus (USB) interface, a fiber optic interface, 4 separate ASIC boards, and an FPGA. The fiber optic interface enables us to acquire waveforms at a rate of > 3.3 kHz, while the speed of the USB interface is only ~ 40 Hz. The USB interface is used in initial testing at low trigger rate. When multiple camera modules are operated by a subfield board at higher trigger rate, the fiber interface is used instead.

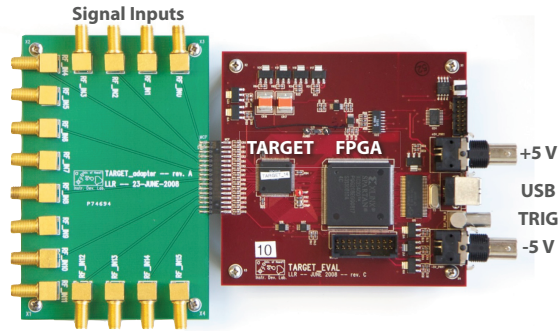

Figure 2: A photo of the TARGET 1 evaluation board (right) and 16-channel input board (left). Basic characteristics of TARGET 1 were measured with the evaluation board. Its main components are an ASIC, an FPGA, ±5 V DC input, and a USB interface to be connected to a computer.

## 4 Performance Tests

Many tests have been done using the evaluation board and the camera module, details of which are fully covered elsewhere [3] (hereafter TARGET 1 paper). Figure 4 is one of the measurements, showing the analog bandwidth of TARGET 1 against sinusoid input at various frequencies. The measured bandwidth, $-3$ dB at $\sim 150$ MHz, is somewhat low, but it is expected to be improved to $> 380$ MHz in TARGET 2 (see Table 1).

---

1. Hamamatsu Photonics H8500D-03 is currently used. It can be replaced with an array of multi-pixel photon counters (MPPCs) in the future.

2. Negative HV for an MAPMT and positive for an MPPC array

Table 1: Performance parameters of TARGET 1 and TARGET 2 [3]. The sampling frequency, bandwidth, and cross talk of TARGET 1 are based on actual laboratory measurements, while those of TARGET 2 are simulated.

| Parameter | TARGET 1 | TARGET 2 |
|---|---|---|
| Channels | 16 | 16 |
| Dynamic range (bits) | 9 or 10 | up to 12 |
| Sampling frequency (GSa/s) | $0.7 - 2.3$ | $0.2 - 1.8$ |
| 3 dB analog bandwidth (MHz) | 150 | $> 380$ |
| Cross talk at 3 dB frequency | $< 4\%$ | $1\%$ |
| Buffer depth (cells per channel) | $4,096$ | $16,384$ |
| Wilkinson ADC counter speed (MHz) | 445 | 700 |
| Samples per digitization (block size) | 16 | 32 |
| Digitization time per block ($\mu$s) | 1 (9-bit) or 2 (10-bit) | 0.7 (9-bit) or 1.5 (10-bit) |
| Number of cells digitized simultaneously | 16 cells $\times$ 2 channels | 32 cells $\times$ 16 channels |
| Clock speed for serial data transfer (Mbps) | – | 100 |
| Channels for simultaneous data transfer | – | 16 |
| Dead time for 48 samples $\times$ 16 ch ($\mu$s) (Digitization time + readout time ) | $24 + 0$ (9-bit) or $48 + 0$ (10-bit) | $1.5 + 7.2$ (9-bit) or $2.9 + 7.2$ (10-bit) |
| Trigger outputs | 1 (OR of 16 channels) | 4 (each is analog sum of 4 channels) |

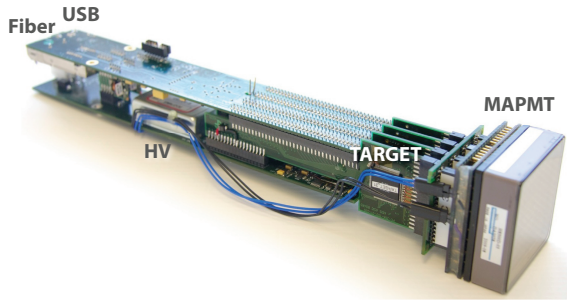

Figure 3: A photo of the TARGET 1 camera module. A TARGET 1 chip is located on each of four vertical boards connected to a 64-ch MAPMT. An FPGA is located on the backside of the top board.

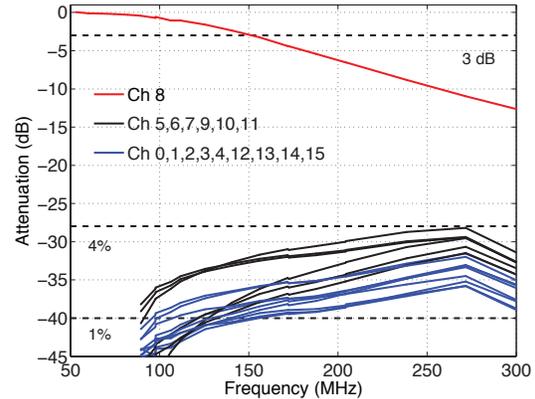

Figure 4: Analog bandwidth and cross talk of TARGET 1. A sinusoidal signal was input to only channel 8, and all 16 channels were read. Channel 8 data (red) shows that the attenuation is $-3$ dB at 150 MHz. Cross talk is 4% at maximum when $\sim 250$ MHz signal is given.

Figures 5 and 6 show an example of waveform and single photoelectron distribution of the MAPMT, respectively, which were digitized using the camera module. The self-trigger functionality, waveform digitization, and datastream chain from MAPMT to an external data-acquisition computer have been validated.

Since the design of the backplane board is not completed yet, we are testing the fiber interface using another project's board temporarily which uses the same data transfer protocol as the module. In the current configuration, the board is able to receive waveforms from two camera modules simultaneously. Using the temporary board, we achieved an event rate of 3.3 kHz. The speed can be improved to be two times faster with a faster optic interface and firmware upgrade.

Some more test results will be shown at the poster session in Beijing, and the present paper will be updated later.

## 5 Conclusion and Prospects

The TARGET 1 chip and the camera module prototype have been fabricated and their performance has been well evaluated. We found that most performance characteristics meet our requirements for a SC-MST. However, a couple of problems, such as low bandwidth and AC-linearity saturation against high-frequency signals ($> 50$ MHz), are known as reported in the TARGET 1 paper. Digitization noise in the camera module, which is not negligible, is also known, but it is not an intrinsic problem of TARGET 1. Those problems are to be removed in TARGET 2 and the second version of the camera module to be developed. The new system will be tested in 2011 and 2012.

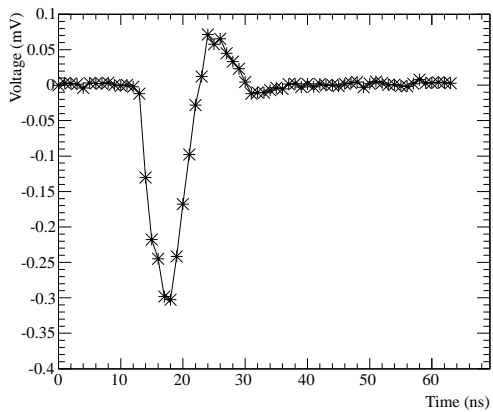

Figure 5: An example waveform of the MAPMT digitized by the camera module prototype. The waveform length is 64 ns (4 blocks) in this example.

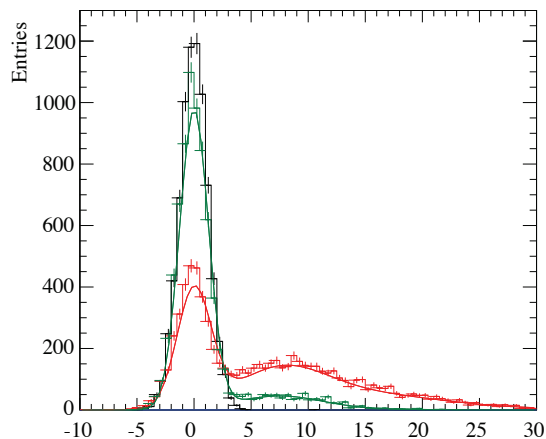

Figure 6: Single photoelectron distribution of the MAPMT obtained by the camera module prototype. Black line shows the pedestal distribution (no signal), green and red show the single photoelectron peak. Darker ND filter was used in the green data. The unit of horizontal axis is arbitrary.

In addition to the TARGET development, SC telescopes, subfield boards, back-end electronics, MAPMTs, and full Monte Carlo simulations are also being studied and developed in the CTA collaboration.

## 6 Acknowledgments

Testing was supported in part by Department of Energy Advanced Detector Research Award #DE–FG02–06ER41424. This work is supported by the Department of Energy, Laboratory Directed Research and Development funding, under contract #DE–AC02–76SF00515. J. V. is supported by a Kavli Fellowship from the Kavli Foundation. A. O. is supported by Grant-in-Aid for JSPS Fellows. S. F. acknowledges generous support by the KIPAC Enterprise funds and by the Mel Schwartz Award from Stanford University. K. B. is supported by a Stanford Graduate Fellowship. We gratefully acknowledge support from the agencies and organisations listed in this page: `http://www.cta-observatory.org/?q=node/22`

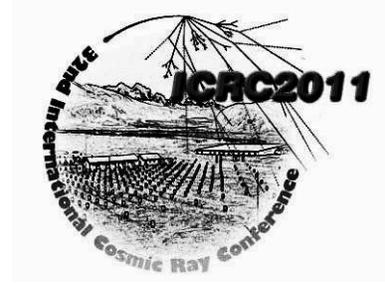

# Design Study of a CTA Large Size Telescope (LST)

MASAHIRO TESHIMA[1,2], OSCAR BLANCH[3], THOMAS SCHWEIZER[1], JUAN ABEL BARRIO[4], LAURENT BRUNETTI[5] EMILIANO CARMONA[6], PIERRE COLIN[1], JUAN CORTINA[3], FRANCESCO DAZZI[7], CARLOS DELGADO[6], GUILLAUME DELEGLISE[5], CARLOS DIAZ[6], DAVID FINK[1], NICOLAS GEFFROY[5] CHRISTOPHER JABLONSKI[1], HIDETOSHI KUBO[8], GIOVANNI LAMANNA[5], BRUNO LIEUNARD[5], ECKART LORENZ[1], GUSTAVO MARTINEZ[6], ABELARDO MORALEJO[3], REIKO ORITO[9], RICCARDO PAOLETTI[10], OLAF REIMANN[1], VICTOR STAMATESCU[3], HOLGER WETTESKIND[1], ON BEHALF OF THE CTA CONSORTIUM

[1] *Max-Planck-Institute for Physics, Munich, Germany*
[2] *Insistitute for Cosmic Ray Research, University of Tokyo, Chiba, Japan*
[3] *Institut de Fisica d'Altes Energies, Barcelona, Spain*
[4] *Universidad Compultense, Madrid, Spain*
[5] *Laboratoire d'Annecy-le-Vieux de Physique des Particules(LAPP), Universite' de Savoie, CNRS/IN2P3, Annecy,France*
[6] *Centro de Investigaciones Energeticas, Medioambientales y Tecnologicas, Madrid, Spain*
[7] *Dipartimento di Fisica dell'Universit di Udine and INFN sez. di Trieste, Italy*
[8] *Department of Science, Kyoto University, Kyoto, Japan*
[9] *Socio-Arts and Sciences, The University of Tokushima, Tokushima, Japan*
[10] *INFN Pisa, Pisa, Itally*

*masahiro.teshima@gmail.com*

**Abstract:** Following the great success of the current generation Imaging Atmospheric Cherenkov Telescopes, the preparation of the next generation VHE gamma ray observatory Cherenkov Telescope Array (CTA) is under way. It is designed to enhance the sensitivity to gamma ray sources, to enlarge the energy band and to improve the quality of data, i.e. angular and energy resolutions, and the gamma-hadron separation (low background and systematics). A few Large Size Telescopes (LSTs) of 23m diameter will be arranged at the centre of the array to lower the threshold energy and to improve the sensitivity of CTA below 200-300 GeV. The low threshold energy provided by the LSTs will be critical for CTA studies of pulsars, and high-redshift AGNs and GRBs. The status of the design study and prototyping of elements on CTA-LST will be presented.

**Keywords:** Gamma Ray, VHE Gamma Ray, Instruments

## 1 CTA Large Size Telescope

During the past few years, Very High Energy (VHE) gamma ray astronomy has made spectacular progress and has established itself as a vital branch of astrophysics. To advance this field even further, we propose the Cherenkov Telescope Array (CTA), the next generation VHE gamma ray observatory, in the framework of a worldwide, international collaboration. CTA is the ultimate VHE gamma ray observatory, whose sensitivity and broad energy coverage will attain an order of magnitude improvement above those of current Imaging Atmospheric Cherenkov Telescopes (IACTs). By observing the highest energy photons known, CTA will clarify many aspects of the extreme Universe, including the origin of the highest energy cosmic rays in our Galaxy and beyond, the physics of energetic particle generation in neutron stars and black holes, as well as the star formation history of the Universe. CTA will also address critical issues in fundamental physics, such as the identity of dark matter particles and the nature of quantum gravity.

CTA consists of three types of telescopes to cover a broader energy band, Large Size Telescopes (LSTs) of 23m diameters, Mid Size Telescopes of 12m meters, and Small Size Telescopes of 4-6m meters. The purpose of LST is to enhance the sensitivity below 200-300GeV and to lower the effective threshold down to 20-30GeV. The science case of LST is the observation of high redshift AGNs up to $z \leq 3$, GRBs up to $z \leq 10$, and pulsars and galactic transients. LST surely expands the domain of science to the cosmological distances and fainter sources with soft energy spectra.

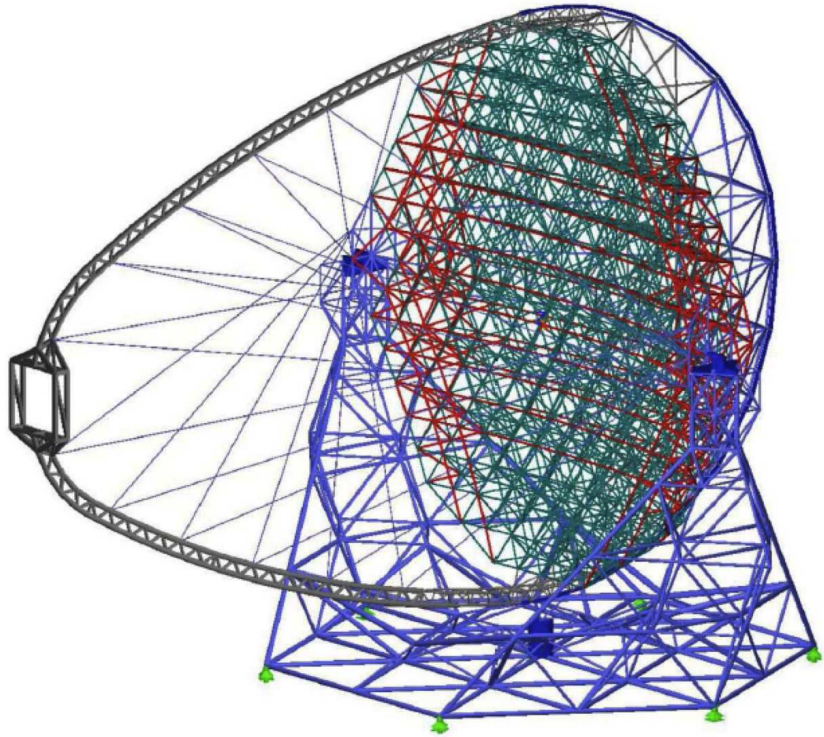

Figure 1: The telescope structure of LST designed by the MPI Munich group and MERO-TSK. The dish diameter is 23m and the total mirror area is about 400m². The focal length is 28m and F/D =1.2.

## 2 The Structure of the Large Size Telescope

### 2.1 Structure

The structure of the large size telescope (LST) as shown in figure 1 was designed by the MPI Munich group together with MERO-TSK. The major part of the telescope consists of the space frame structure with carbon fiber reinforced plastic (CFRP) tubes. The total weight of the telescope is designed to be about 50 tons and allows the fast rotation of the telescope, 180 degrees in 20 seconds, for the fast follow-up observations of gamma ray bursts using the location determined by gamma ray satellites.

The telescope geometry is optimized to maximize the cost performance by Monte Carlo simulations and toy models. The baseline parameters are defined with the dish size of 23m, the focal length of 28m and then F/D = 1.2, and the camera FoV of 4.5 degrees with a pixel size of 0.1 degrees.

### 2.2 Mirrors

The reflector with its diameter of 23m diameter consists of 198 units of hexagonal shape 1.5m flat to flat segmented mirrors of 2m². Total area of the reflector is about 400m². The individual segmented mirror is attached to the knots of the space frame structure with an universal joint, and two actuators. The segmented mirrors have a sandwich structure consisting of glass sheet of 2.7mm thickness - aluminum honeycomb of 60mm thickness - glass sheet. The weight of a segmented mirror is 45kg. The reflective layer of the mirror is coated with Cr and Al on the surface of the glass sheet with a protective multi-coat layer of $SiO_2$, $HfO_2$ and $SiO_2$. By adjusting the thickness of individual layers with $SiO_2$ and $HfO_2$, we can optimize the reflectivity to 95% due to the interference effect of multi-layers.

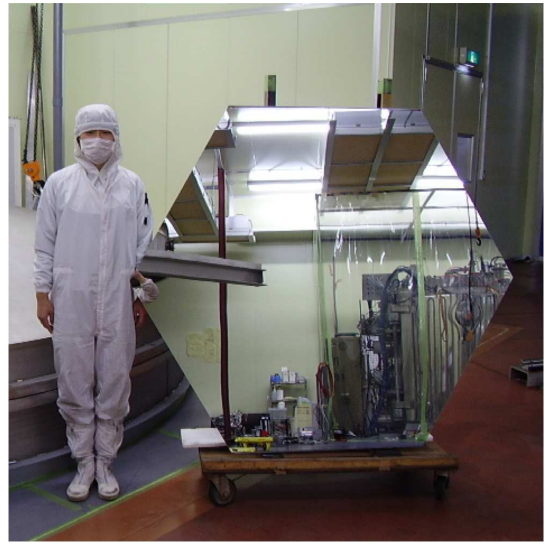

Figure 2: The prototype mirror of 1.5m flat to flat produced by CTA-Japan and the Company Sanko. The mirror with the sandwich structure of glass sheet of 2.7mm thickness - aluminum honeycomb of 60mm thickness - glass sheet of 2.7mm is produced with the cold slumping technique. The radius of curvature is about 56m. The weight is about 45kg.

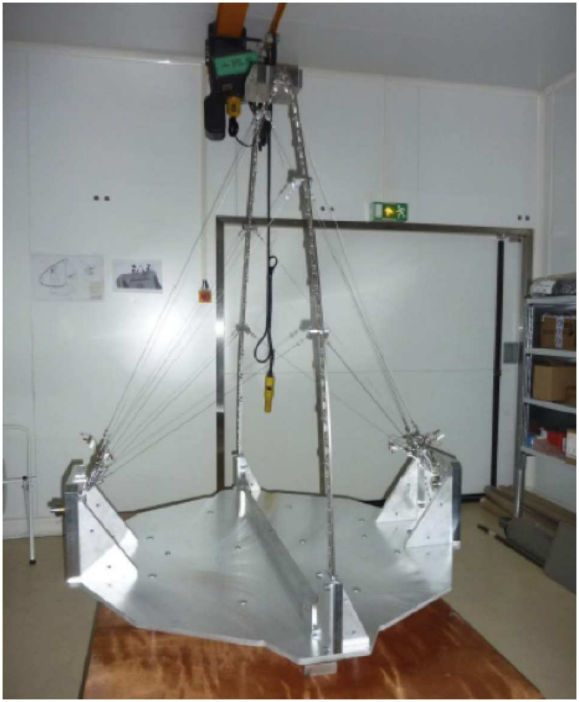

Figure 3: The 1/20 scaled telescope arch model for testing the arch oscillation damping by the LAPP group.

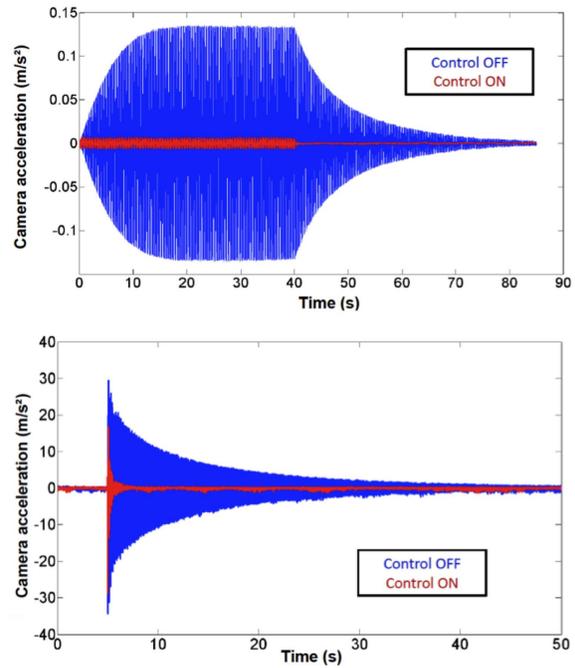

Figure 4: Measurements of the oscillations of the arch model with (Control On) and without (Control OFF) activation of the damping system. The significant improvements in the oscilation damping can be seen in sinusoidal and hammer-shock perturbations.

## 2.3 Active Mirror Control

We will define the optical axis (OA) of the LST optics with two infra-red lazers at the center of the dish constantly shining two targets left and right of the imaging camera. The individual segmented mirror will also have an infra-red laser at the edge of the mirror (MIR) which makes the spot at the target near the imaging camera confirm the direction of the mirror facet relative to the OA laser (optical axis). The directional offset of the mirror facets will be estimated by taking pictures of the MIR-laser and OA-laser spots on the target near the Camera with a high resolution IR CCD camera viewing from the center of the dish. If any significant offsets are found, the direction of the corresponding mirror facets will be corrected by actuators. The mirror directional calibrations over 198 mirrors will be done sequentially and performed within one minute. This calibration and control will be done continuously during the observation. After the first rotation for the GRB follow-up observation, or at the beginning of the observation of any source, we will use the look-up table corresponding to the zenith angle of the tar-get source as the initial values of actuators and then move to the mode of permanent/continuous active mirror control loop.

## 2.4 Arch Oscillation Damping System

The long structure of the arch (camera supporting mast), designed with CFRP tubes by the LAPP group will introduce non-negligible oscillations of the imaging camera under strong wind or after fast movement of the telescope. Such oscillations will be a source of temporal mispointings and also introduce mechnical instabilities in the long term. We will introduce the oscillation damping system, which actively changes the tension of the wires connecting the arch structure and two edges of the mirror supporting structure near the elevation axis. A demonstration of the oscillation damping system with a 1/20 scaled model was carried out by the LAPP group as shown in figure 4. It shows how the oscillation is suppressed / damped as a function of time with and without the damping system. We can observe the significant improvement in the oscillation damping in case of sinusoidal perturbation, like under a strong gust of wind, and also in case of hammer-shock perturbation, which may correspond to the fast rotation of the telescope.

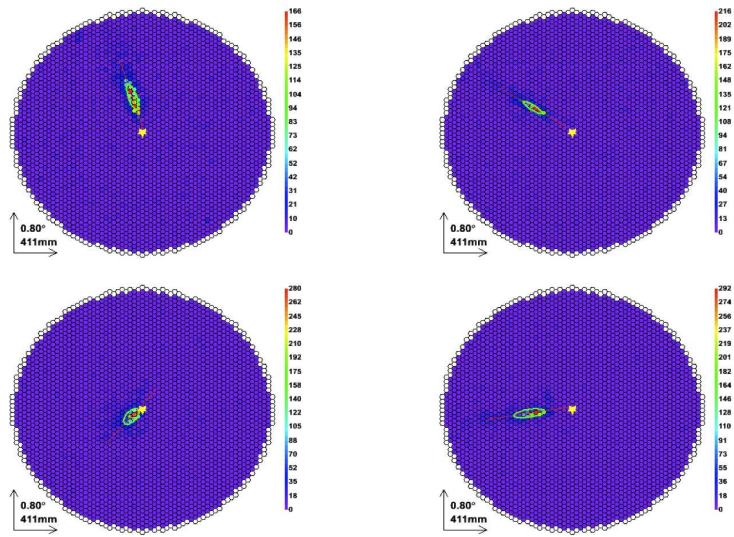

Figure 5: The typical image of the cherekov light images for a 50GeV gamma ray.

## 2.5 Imaging Camera

The imaging camera has a FOV of 4.5 degrees and a pixel size of 0.1 degrees. The actual size of the image plane will be about 2.2m in diameter. The signals from the photomultipliers will be read with 1G samples/sec speed and be stored in the ring capacitors of 4096 depth, which corresponds to 4 micro-seconds.

The camera should be sealed to resist the humidity and dust in the field. The front side (entrance window) of the imaging camera will be covered with uv-transparent plexiglass. Two water cooling plates are used to keep the temperature of the camera and the electronics constant. As a part of camera mechanical structure they will also serve as a support of PMT/electronics clusters. support. The readout electronics and the auxiliary electronics (HV, and amplifiers) will dissipate a heat of 2W/ch. 7-PMTs and readout electronics are mechanically bundled as a PMT/electronics cluster. The total number of pixels and clusters will become about 2000-2500 and 300-350, respectively. The total heat dissipation inside the camera will amount to 4-5kW.

## 2.6 The Sensitivity and Telescope Parameters

In order to optimize the cost performance of the LST array system, studies with toy models and Monte Carlo simulations are performed. We have assumed the following cost model: costs of the camera and the structure are proportional to the number of pixels, and to (mirror area)$^{1.35}$, respectively. We can formulate the sensitivity of the LST array system with parameters of the number of telescopes, diameter of the telescope dish, and FOV of the Camera using Monte Carlo simulations. Then with a fixed amount of budget, we can see the sensitivity of the LST array system as a function of the number of telescopes, FOV of camera, and telescope diameter. Figure 6 shows the sensitivity as a function of FOV for point source. The FOV of 4 degrees will give us the best sensitivity, be understood increasing the number of telescopes is more effective than increasing the FOV with a fixed budget. For extended sources of 1 degrees the best sensitivity is achieved with a FOV of about 4.5 degrees. The baseline design of the LST array system can be defined as the array of four LSTs with a dish diameter of 23m, FOV of 4.5 degrees.

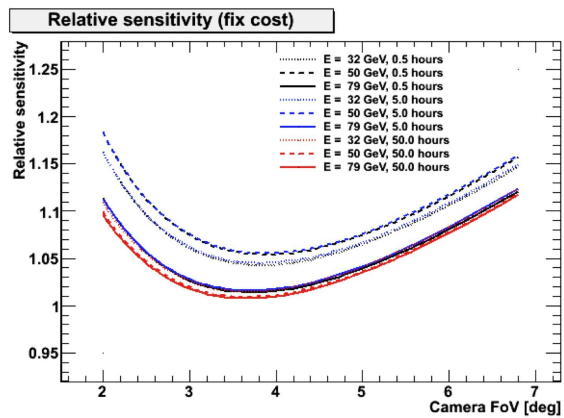

Figure 6: The sensitivity as a function of FOV with a constant budget. If we increase the FOV of the imaging camera, the cost of the camera will increase and the total number of telescopes will decrease and lose performance.

**Acknowledgement** We gratefully acknowledge financial support from the agencies and organisations listed in this page: http://www.cta-observatory.org/?q=node/22

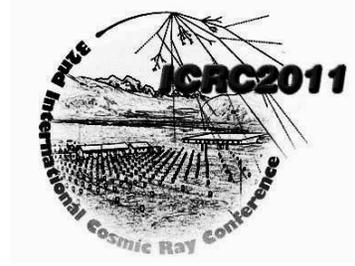

# Telescopes for the High Energy Section of the Cherenkov Telescope Array

R. WHITE[1], F. DI PIERRO[2], T. GREENSHAW[3], G. PARESCHI[4], R. CANESTRARI[4] FOR THE CTA CONSORTIUM

[1]*University of Leicester, Leicester, UK*
[2]*Istituto di Fisica dello Spazio Interplanetario dell'Istituto Nazionale di Astrofisica, Torino, Italy*
[3]*University of Liverpool, Liverpool, UK*
[4]*INAF Osservatorio Astronomico di Brera, Merate, Italy*
*richard.white@leicester.ac.uk*

**Abstract:** The Cherenkov Telescope Array (CTA) will provide an unprecedented level of sensitivity to very-high-energy (VHE) gamma rays across an energy range from below 50 GeV to above 100 TeV. To achieve this, large, medium and small size telescopes are required, covering the low, intermediate and high energy regimes, respectively. The small sizes telescope (SST) array will occupy an area on the ground of up to 10 km$^2$ and operate from around 1 TeV to several hundred TeV. It will allow CTA to probe the gamma-ray universe with the highest angular resolution yet obtained above the hard X-ray band, enabling morphological studies of PeV particle acceleration sites and discrimination between hadronic and leptonic cosmic ray acceleration models. Under consideration for the SST are a Davies-Cotton design with a mirror diameter of about 7 m and a dual mirror telescope with primary mirror diameter of about 4 m. The status of the optical and mechanical designs of each of these alternatives presented and the pros and cons of each approach discussed. Comments are also made on the cameras needed for each of these instruments.

**Keywords:** CTA, Imaging Atmospheric Cherenkov Telescope, gamma-rays, optics

## 1 Introduction

To extend the sensitivity of CTA to the highest energies, an array of small telescopes - the SST subsystem, will be utilised. The SST sub-system will operate from 1 TeV to 300 TeV and has a number of unique requirements:

- At the energies of interest to the SST sub-system astrophysical gamma rays are rare due to the power-law spectra of typical sources. Therefore the SST part of CTA must cover a large area (1 - 10 km$^2$).

- Images due to gamma rays above 1 TeV are bright out to many hundreds of meters from the shower impact point and so only relatively small dish diameters (3 - 7 m) are needed to collect enough light to trigger each telescope (see Fig. 1).

- Although images are visible out to large impact parameters, they also shift away from the centre of the camera with increasing distance from the shower impact point so each SST camera must have a field of view (FoV) >7$^o$ (see Fig. 1).

- An angular pixel size of 0.18 - 0.28$^o$ is needed to resolve the width of gamma-ray images above 1 TeV (see Fig. 1).

- The fundamental limit of angular resolution for the IACT technique decreases with energy (1 arc minute at several TeV), see Fig. 2 [6]. The SST array offers the unique opportunity of high angular resolution at high energies. Monte Carlo simulations show that the angular resolution improve with telescope multiplicity and therefore an SST geometry in which a fixed area (i.e. fixed cost) is covered with small dishes placed closer together is favoured over fewer larger dishes placed further apart.

A SST sub-system that achieves these requirements will allow CTA to probe the gamma-ray universe with the highest angular resolution yet obtained above the hard X-ray band. Sensitivity at arc-minute scales above several TeV will enable morphological studies of PeV particle acceleration sites essential for discrimination between hadronic and leptonic cosmic ray acceleration models.

The SST sub-system will be capable of operation in conjunction with the other sections of CTA, or independently, in the case of a strong source, or a source with a hard spectrum.

## 2 A Davies-Cotton SST

Traditionally IACTs with dish diameters of under ~12 m are constructed with a Davies-Cotton geometry. This op-tion forms the baseline solution for the SST, and has the advantage of being tried and tested.

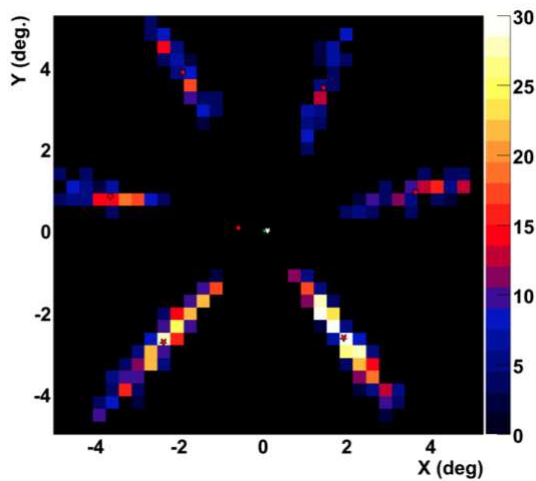

Figure 1: Images from a simulated 14 TeV gamma-ray in $10^o$ FoV cameras with $0.25^o$ pixels. Each dish has a diameter of 3 m diameter and the telescopes are ∼500 m from the shower core. The z-scale is in units of photoelectrons/pixel. Taken from [4].

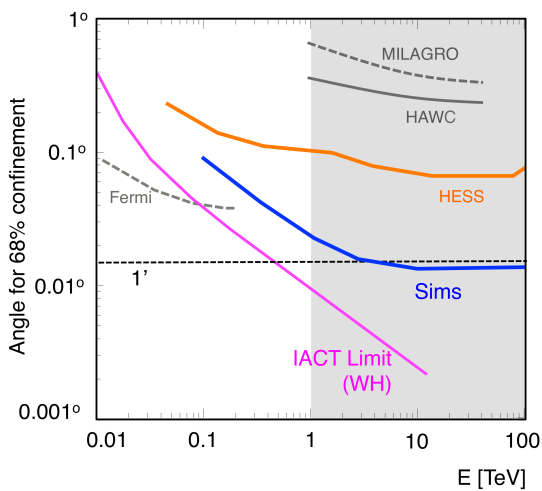

Figure 2: The angular resolution of gamma-ray instruments including a simulation of a high-energy IACT array, taken from [5] and the fundamental limit of angular resolution for IACTs from [6].

The CTA Medium Sized Telescope (MST) will use a 50 mm pixel pitch (including the PMT and associated light guide). It is desirable both in terms of cost and maintenance to use the same physical PMT size for all telescopes. Several Davies-Cotton designs for the SST based on this requirement are under investigation by Polish, Italian and Argentinian collaborators [8].

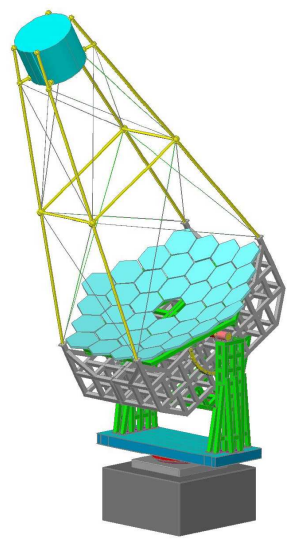

Figure 3: The Polish Davies-Cotton SST design with 11.46 m focal length and 7.64 m dish diameter.

A structural Davies-Cotton SST prototype is planned in Poland with a 11.46 m focal length and 7.64 m diameter dish as shown in Fig. 3. A $0.25^o$ pixel based on the 50 mm physical pixel size results in ∼1000 (1600) pixels for a $8^o(10^o)$ FoV and a 1.6 m (2.0 m) wide camera. Under the current cost model ∼35 such telescopes could be built.

A Davies-Cotton SST dish would consist of hexagonal segments tiled to approximate a circle. A segment size of ∼0.85 m (face-to-face) is under investigation by Italian groups. Larger sizes degrade the PSF, and are impossible to manufacture with the required radius of curvature using cold-slumping. A smaller segment size (and the additional mounts and actuators) increases the cost of the telescope.

## 3 Optimisation

The Davies-Cotton design for the SST is the baseline solution, but is not without challenges and may not be optimal. Mirror segments of the required radius and desired size are difficult to produce using traditional methods, supporting a 1000-1600 kg camera at 11.5 m is non-trivial, and the cost of the telescope is dominated by the large camera.

One method to optimise the SST is to consider alternative photodetectors. SiPMs (GaPDs) are more efficient than PMTs, weigh less, are impervious to day-light and require no high voltage. They have long since been considered for IACTs, but traditionally exhibit performance problems and have a high cost per unit area. Recently, however, the FACT collaboration proved that an IACT SiPM camera is realistic, and have overcome many technical issues [2]. In such a scenario DC optics could be used in conjunction with a SiPM camera. It may be possible to envisage a ∼4 m DC dish and a 0.7 m camera with 20 mm pixels constructed from SiPMs with solid light cones (assuming 5 mm SiPMs with the required performance and cost become available on the timescale of CTA).

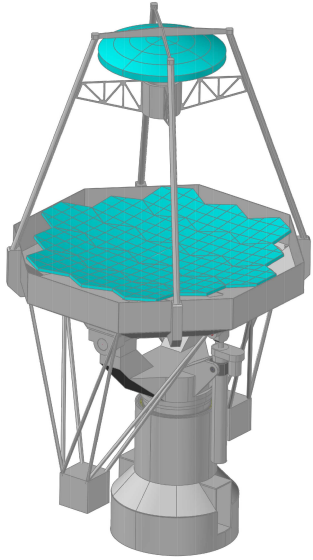

Figure 4: The Italian secondary optics SST design with a ∼4 m diameter primary and a ∼2 m secondary.

Another approach to optimise the SST is to address the cost imbalance between structure and camera in the DC design. The camera can be made drastically less expensive by reducing the plate scale to the point at which alternative photodetectors become affordable. This can be achieved with a secondary optics (SO) system. By reducing the camera cost to a similar level as the structure cost, many more telescopes can be built resulting in either a larger area on the ground (higher sensitivity) or tighter spacing (higher angular resolution).

## 4  A Secondary Optics SST

UK and Italian SO SST optical designs have been produced and independently converged to similar results. In the near future Italian and Anglo-French structural prototypes are planned, featuring aspherical mirrors with a segmented ∼4 m diameter primary and a (possibly monolithic) ∼2 m secondary with f/d∼0.5 resulting in a plate scale of ∼37.5 mm/°. The PSF (80% containment) is below the size of a single pixel across the entire FoV in both cases.

A 9.6° FoV can be implement with a 0.35 m camera. The focal plane is curved (radius 1 m) and must be approximated with photodetectors. The 64 channel Hamamatsu H10966 Multi Anode PMT MAPMT) can be used with a physical pixel of ∼6.5 x 6.5 mm² (∼0.18°). Alternatively, the Hamamatsu S11828-3344M SiPM array of 4 x 4 pixels, each of size 3 x 3 mm². In this case 4 SiPM pixels would be combined to form a 6 x 6 mm² (0.16°) camera pixel, and light collectors used to reduce dead space. The SO solution results in a curved focal plane and so the smaller SiPM array allows for a more accurate representation of the curvature. There are still questions over the performance of both the MAPMT and the SiPM array and these are under investigation within CTA.

Camera prototypes for the SO SST structures based on MAPMTs and SiPMs are planned by UK-US and Italian CTA groups. There are several front-end electronics solutions under development within CTA. The Target modules, based around the Target ASIC under development at SLAC [1, 9] offer a compact solution with a potentially low cost. Using the Target modules it would be feasible to construct a camera for the SO SST with electronics integrated into the compact camera. Alternatively the Flash-Cam project underdevelopment within CTA offers a potentially cost effective flash ADC based approach [7]. In this scenario signals would be amplified in the camera and transported in analogue form to the digitisation electronics located behind the primary reflector (potentially in the telescope counterweight as shown in 5). This approach would avoid a densely packed camera and the power dissipation problems associated with it, but would introduce a level of signal degradation and additional failure points at the cable interfaces.

Assuming a similar cost model to the Davies-Cotton SST, and that Target-based electronics are used, around ∼60 SO SSTs could be built.

## 5  Secondary Optics Technical Challenges

There are several technical challenges associated with the use of SO for the SST, not least of which is the novelty of the design. To date a SO IACTs has never been built. Beyond this:

- Mirrors: The unusual shape and size means that cold-slumping can not be used. Alternative techniques including hot slumping are under investigation.

- Photodetectors - MAPMTs: The 64 channel H10966A has a single HV supply. The device must be able to operate in star light otherwise a large fraction of the camera will have to be regularly disabled. The MAPMT must also be able to operate at large incidence angles of up to ∼60°.

- Photodetectors - SiPMs: These devices are typically over sensitive in the red and under sensitive in the blue. They also require compensation for temperature effects.

- Electronics: The small camera volume means reduced space for electronics presenting cabling, con-

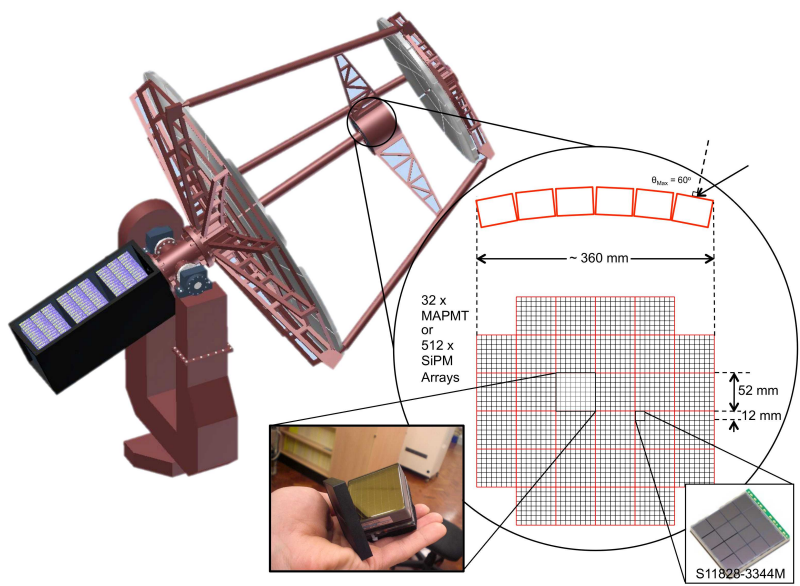

Figure 5: The Anglo-French secondary optics design with possible camera tiling and photodetectors.

nector and power dissipation problems. Gamma-ray images at the highest energies can take hundreds of nanoseconds to traverse a camera. This presents readout and dead-time challenges.

- Alignment: The successful operation of a secondary optics SST requires millimeter alignment precision between the mirror segments of the primary and secondary reflectors and the focal plane.

- Maintenance: The SST subsystem will be the largest array of IACTs ever built. The expected lifetime of CTA is the order of 20 years. Each telescope must be built with reliability and maintenance in mind to realise this goal.

## 6   Conclusion

CTA will consist of at least three types of telescope, each aimed at a particular energy range. Above 1 TeV the performance of the array will be dominated by the SST subsystem, consisting of 30 - 60 telescopes covering 1 - 10 km², each with a >7° FoV. Covering a larger area increases sensitivity whilst spacing the telescopes closer together improves angular resolution. A Davies-Cotton telescope design has the advantage of using proven technology, but the assumed cost is dominated heavily by the camera. One method to address this is to use secondary optics. In this way the plate scale is reduced and smaller, cheaper, photodetectors become available. Decreasing the SST cost allows more telescopes to be built and can be used to increase sensitivity and improve angular resolution.

Prototypes of both Davies-Cotton and SO SST designs are planned, and detailed designs are underway. Components for the SO camera are currently being tested, and two full camera prototypes using alternative photodetectors are planned.

An optimised SST array will provide CTA with unprecedented sensitivity and angular resolution in the energy range 1 TeV - 300 TeV.

## Acknowledgements

We gratefully acknowledge support from the agencies and organisations listed in this page: http://www.cta-observatory.org/?q=node/22. RW is supported by an STFC post-doctoral fellowship.

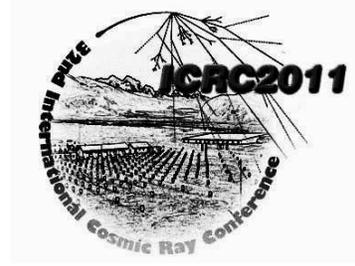

# Optical performance related to mechanical deformations of a Davies-Cotton mount for the high energy section of the Cherenkov Telescope Array

A.C. ROVERO[1], A.D. SUPANITSKY[1], M. ACTIS[2], P. RINGEGNI[2], F. ANTICO[2], A. BOTANI[2], G. VALLEJO[2], I. OCHOA[2], G. HUGHES[3], AND D. MARCONI[2] FOR THE CTA CONSORTIUM
[1]*Instituto de Astronomía y Física del Espacio (CONICET-UBA), Argentina*
[2]*Unidad de Investigación y Desarrollo GEMA, Ingeniería Aeronáutica (UNLP), Argentina*
[3]*Deutsches Elektronen-Synchrotron (DESY), Zeuthen, Germany*

*rovero@iafe.uba.ar*

**Abstract:** The Cherenkov Telescope Array is the next generation ground-based instrument for the observation of very-high-energy gamma-rays. It will provide an order of magnitude more sensitivity and better angular and energy resolution than present systems as well as an increased energy range. For the high energy portion of this range, the construction of $\sim 6\,m$ diameter Cherenkov telescopes is an option under study. We have proposed an innovative design of a Davies-Cotton mount for such a telescope, within Cherenkov Telescope Array specifications, and evaluated its mechanical and optical performance. The mount is a reticulated-type structure with steel tubes and tensioned wires. It consists of three main parts to be assembled on site. In this work we focus on the study of the point-spread-function of collected light for different mirror facet sizes and telescope positions, related to mount deformations.

**Keywords:** CTA, Imaging Atmospheric Cherenkov Telescope, Gamma-rays, Optics, Mechanics

## 1 Introduction

After the great success of very-high-energy gamma-ray observatories like HESS, MAGIC and VERITAS, the international scientific community is organizing the next generation of ground-based instruments, the *Cherenkov Telescope Array* (CTA). With sensitivity an order of magnitude better than present systems and improved angular and energy resolution, the CTA will cover the full sky by constructing two observatories, one in each hemisphere [1]. The southern observatory will cover most galactic sources, so it will be optimized to be fully sensitive in an extended energy range ($20\,GeV$ to $100\,TeV$). To span so large an energy range, three different sizes of Cherenkov telescope will be constructed. The small size telescopes (SSTs) will be used to cover a large area ($\approx 10\,km^2$) and detect the highest energy gamma rays. Two types of optical systems are being considered for the SST, Schwarzschild-Couder (SC) [2] and Davies-Cotton (DC) [3]. The first of these, with primary and secondary mirrors, has short focal length and hence a reduced plate scale which translates into a small camera. The DC design has a simpler structure and mirrors and is proven as it is the type telescope used by present experiments. Based on the past experience and Monte Carlo simulations, CTA has specified parameters for the SST. The DC option should have a $6-7\,m$ diameter tessellated reflecting surface (dish). The dish will be an array of hexag-onal facets with three mounting points, two of them with actuators to align the mirror. The detector at the focal plane is an array of 1400 photomultipliers, each of $0.25°$ angular diameter, at a focal distance $f \approx 11\,m$, covering a f.o.v. of $8-10°$. The Point Spread Function (PSF) for the SST has to be less than the angular size of the camera pixel. Here the definition of the PSF is taken to be the diameter of a circle containing $80\%$ of the collected light at the focal plane (herein denoted D80).

We have proposed an innovative design of Davies-Cotton mount for the SST of CTA and evaluated its mechanical and optical performance. In this work we focus on the study of D80 for different mirror facet sizes and telescope conditions, related to mount deformations.

## 2 Mechanical design

In addition to the optical parameters mentioned above, CTA has established mechanical specifications for the SST, some of which are summarized in Table 1.

The SST mirror facet size is not yet specified. To reduce maintenance, it is desired that all telescope types in CTA have the same facet size, which was originally set to $120\,cm$ (L in figure 2). With this condition, we have designed a DC mount for a $D = 6\,m$ dish, as shown in the same figure. The optical system has a focal length of $10.8\,m$, resulting in an $f/d \sim 1.8$. The mount is a reticulated-type structure with steel tubes and tensioned wires. It consists of three main parts, as shown in figure 1, to be assembled on site. For simplicity, truss tubes are reduced to 4 different cross section sizes. Shafts are designed to reduce costs and achieve a first natural frequency of $3.3\,Hz$ (second natural frequency $= 4\,Hz$). Buckling analysis was carried out and compared with analytical calculations, proving the mount to have a good safety margin. Elevation and azimuth double drive systems were adopted to improve structure movements and better support wind moments. Servomotors in all shafts give constant torque at low speed. The total weight of the mount is $\approx 15\,tons$, without the camera ($1600\,kg$).

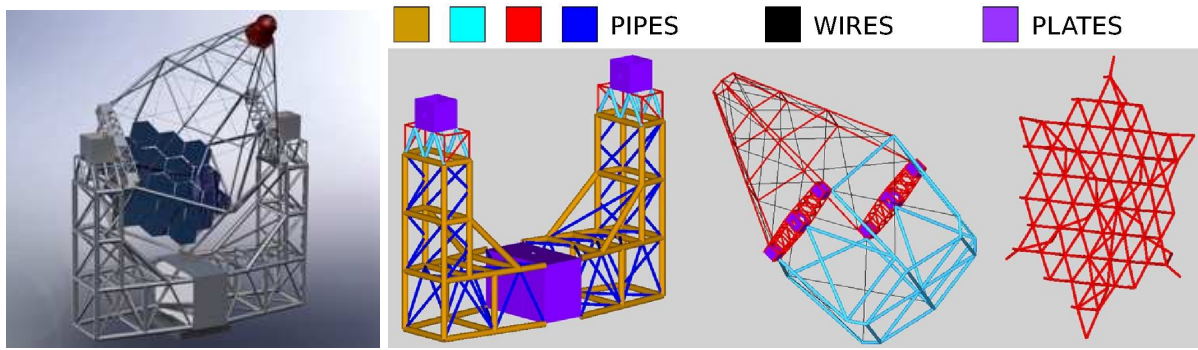

Figure 1: Telescope structure. *Left:* general artistic view. *Right:* Three main parts to be assembled in situ: the azimuth structure (left), the elevation structure (center) and the mirror structure (right).

| Parameter | Condition |
|---|---|
| Eigen frequencies | $> 2.5\,Hz$ |
| Positioning speed, azimuth | $180°/min$ |
| Positioning speed, elevation | $90°/min$ |
| Acceleration, azimuth | $> 1°/sec^2$ |
| Acceleration, elevation | $> 0.5°/sec^2$ |
| Tracking precision | $< 6\,arcmin$ |
| Pointing precision | $< 10\,arcsec$ |
| Temperature range, operational | $-10℃, 30℃$ |
| Temperature range, survival | $-15℃, 60℃$ |
| Wind speed, operational | $< 50\,km/h$ |
| Wind speed, emergency | $< 100\,km/h$ |
| Wind speed, survival | $< 180\,km/h$ |

Table 1: Some mechanical specifications for the SST. Under emergency wind speed, the telescope has to be able to go to the parking position. In this position, the structure should resist the survival wind speed.

## 3 Aberrations

Optical aberration is measured as the dispersion of photons collected at the focal plane. We have performed the analysis in three steps and studied aberrations for our standard design, then for two other alternative dishes with smaller facets, and finally adding mechanical deformations of the structure.

### 3.1 Standard design

The reflecting surface of our standard design has 18 hexagonal facet mirrors (the central mirror is not used) of $L = 120\,cm$, mounted on a spherical structure with radius of curvature $R = 10.8\,m$, *i.e.* equal to the focal length. Each facet is a spherical mirror with radius $2R$, following the DC design [3]. In our coordinate system the center of the dish is placed at the point $(0,0,-R)$, *i.e.* the origin is at the center of the focal plane (camera), $z$ is the optical axis, $x$ is parallel to the Earth's surface, and $y$ is perpendicular to both, as shown in figure 2. The direction of any incident photon is defined by two angles relatively to the $z$-axis: $\phi_x$ in the $x - z$ plane (azimuth) and $\phi_y$ in the $y - z$ plane (elevation). We describe the reflected photon using the angles $\xi$ and $\eta$. Thus, the incident and reflected photons are characterized by $(\phi_x, \phi_y)$ and $(\xi, \eta)$, respectively.

Given the incidence of a parallel photon beam on the dish, a non-pointlike distribution of reflected photons is formed at the focal plane due to system aberrations. To characterize this distribution we used MC ray-tracing code for $10^6$ photons homogeneously distributed over the dish. An example of such a distribution is shown in figure 3, for our standard dish and for a rather extreme angle of incidence, $(\phi_x, \phi_y) = (0°, 5°)$. In the same figure, the distribution on the elevation axis is shown, indicating the strong concentration of reflected photons at the specular position ($\eta = -5°$). The shape of the distribution is typical for a DC telescope.

The definition of D80 given above is not unique. The most intuitive interpretation would be to take the point defined by the mean values of the focal-plane photon-distribution ($< \xi >, < \eta >$), and increase the radius of a circle centered at that position until it encloses 80% of the photons ($D_{80}^\mu$). Another option would be to take the median as the circle center ($D_{80}^{med}$). The smallest possible values of D80 are obtained by choosing the position of the circle in the focal plane that minimizes its diameter. We call this the "optimized D80", or simply $D_{80}$. In figure 3 we show all three cases for a particular distribution, and in figure 4 their dependence on the incidence angle. We performed this for incidence angles on different planes, $\phi_x$ and $\phi_y$, which are not exactly the same as the dish shape is not symmetrical, being worst for $\phi_x = 0$ and varying $\phi_y$.

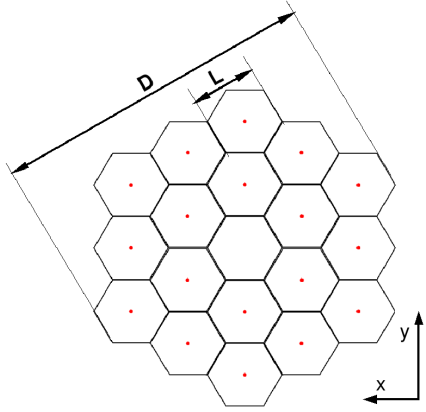

Figure 2: Front view of the standard mirror dish: 18 hexagonal 120 cm facets (the central facet is never used). Mirror facet is defined by its flat-to-flat edge distance ($L$). $D$ is defined as the "dish diameter". The plane $x - y$ of the coordinate system is indicated.

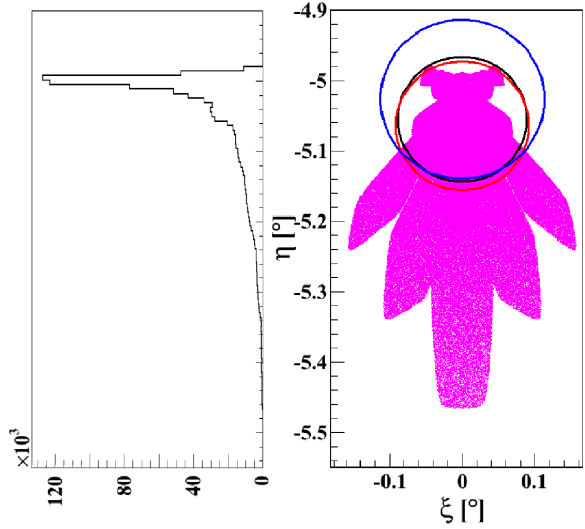

Figure 3: Simulated distribution of collected photons at the focal plane using the dish of figure 2, for an angle of incidence of $5°$ on the elevation axis. *Right:* bidimensional distribution with D80 computed for a circle centered at the median (blue-dashed), mean (red-doted) or optimized (black-solid). *Left:* projection on the elevation axis.

By inspecting the circles in figure 3 it is clear that $D_{80}^{med}$ is not the right way to define D80. The other two, however,

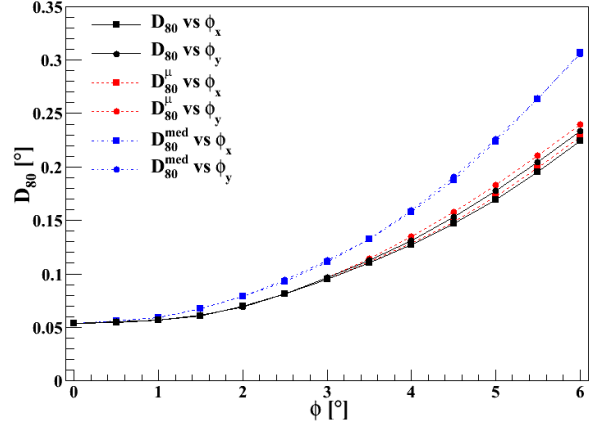

Figure 4: D80 as a function of incidence angle, $\phi$ ($\phi_x$ or $\phi_y$). $D_{80}^{med}$: centered at the median position; $D_{80}^\mu$: centered at the mean position; $D_{80}$: centered at the optimal position to minimize D80.

are very similar for all incidence angles, as shown in figure 4, $D_{80}$ being slightly better than $D_{80}^\mu$, although the latter could be taken as a good approximation. In this work we use $D_{80}$.

## 3.2 Dish with smaller facets

In principle, smaller facets produce smaller aberrations. To test how smaller facets perform we used two other sizes, $L = 90\,cm$ and $L = 60\,cm$. In order to get sizes of the dish similar to our standard case, we used 37 $90\,cm$ facets ($D = 6.3\,m$), and 90 $60\,cm$ facets ($D = 6.6\,m$). Computing D80 as defined in the previous section and using the worst incidence angle, *i.e.* setting $\phi_x = 0$ and varying $\phi_y$, we found the performance for these two cases, as shown in figure 5. The curves in this figure indicate that reducing the facet size improves the performance for incident angles $\phi_y < 2°$, and worsens it for larger incidence angles, while an improvement is expected for all incidence angles. Indeed, we note that this is not a direct effect but one caused by the fact that dish diameters increase for smaller facet sizes. To show this effect we recalculated the performance for a facet size of $L = 60\,cm$, removing the external ring of mirrors from the dish, resulting in $D = 5.4\,m$. This last curve is also shown in figure 5 and indicates a significant improvement relatively to the case $L = 60\,cm$ and $D = 6.6\,m$. In any case, the value of D80 is always below pixel size ($0.25°$) for the maximum required field of view ($10°$) and, consequently, within specifications.

To keep a similar mechanical design for all cases, we have considered only dishes with hexagonal shape as shown in figure 2. Other shapes will be considered in the future to ensure the dish is as circular as possible. This is rather difficult for big facets but possible for small ones.

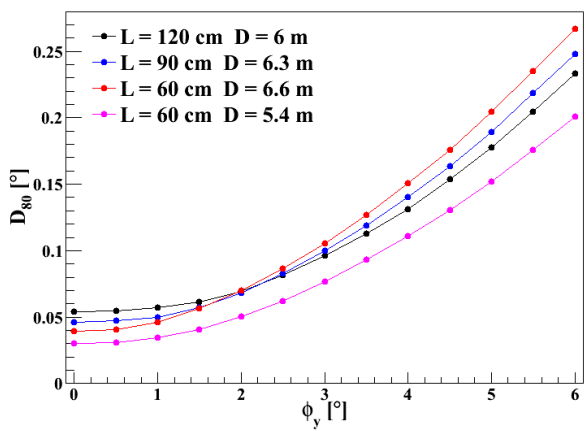

Figure 5: D80 as a function of the worst incidence angle, $\phi_y$, for different facet sizes (L) and dish diameters (D).

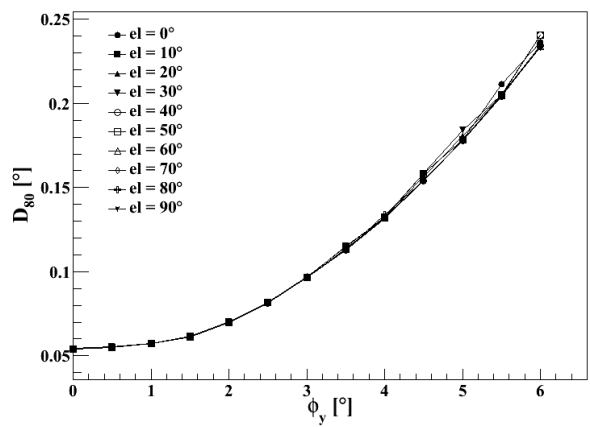

Figure 6: D80 as a function of the incidence angle, $\phi_y$, at different telescope elevation angles, considering structure deformations. It is done for our standard dish ($L = 120\,cm$ and $D = 6\,m$) under the worst operational conditions.

### 3.3 Including mechanical deformations

Telescope mount deformations cause facet mirrors to change their designed pointing direction and, consequently, to change the focal plane photon distribution. Here we do not consider mirror deformation or mirror roughness, only mechanical structure deformations. This translates into a calculation of displacements of the positions of the facet mounting points relatively to their ideal positions.

We have calculated the optical aberration of the worst possible case for our standard dish, by estimating D80 as described above. The worst possible case means under extreme wind speed conditions considered for telescope operation, *i.e.* $50\,km/h$, hitting from the most disadvantageous angle on the structure. We have also taken the worst case for photon incidence angles ($\phi_x = 0$, varying $\phi_y$). Gravity causes structure deformations to depend on elevation angle, so we have computed the system performance as a function of telescope elevation. In figure 6 we show D80 under these conditions for elevations $0°$ to $90°$, in steps of $10°$. No significant degradation of the optical performance is seen due to mechanical deformation of our mount design, for our standard dish.

Even for the worst case of telescope operation, figure 6 shows that elevation angle is not relevant for the telescope optical performance. For the largest field of view under consideration for the SST ($10°$), the deterioration of the optical performance due to mechanical deformations remains within 3.5%. This suggests that our mount design might be too rigid. Relaxing the mechanical tolerances could make the structure cheaper and still have resut in deformations that are within specifications.

### 4 Summary

We have proposed a design of a Davies-Cotton mount for the SST of CTA, and have evaluated the optical perfor-mance related to mount deformations. For our standard dish ($L = 120\,cm$, and $D = 6\,m$) mechanical deformations do not significantly change the optical performance, variations in the PSF are below 3.5%. This conclusion applies for the worst possible case of telescope operation and for any telescope elevation angle.

We have studied the influence of facet size on the optical performance of our SST design by evaluating D80, the diameter of the smallest circle containing 80% of the photons at the focal plane. For small facet sizes an improvement is seen for incidence angles $< 2°$. For angles larger than this, we found a degradation of D80 for small size facets, reflecting the fact that dish diameters are not the same for all cases we considered ($L = 120\,cm$; $90\,cm$; $60\,cm$). At these large incidence angles, the telescope's optical performance is determined more by the dish diameter than facet size.

We have only considered dishes with hexagonal shape. Evaluation of other facet arrays is under consideration to get a more circular dish, which is rather difficult for large facet sizes. For this, the mechanical design would have to be adapted.

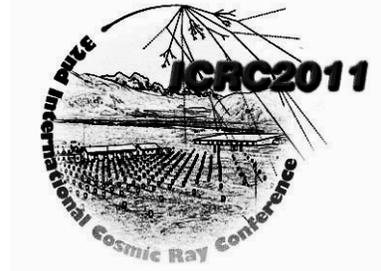

# Development of Raman LIDARs made with former CLUE telescopes for CTA

M. Barceló[1], O. Blanch[1], J. Boix[1], M. Bourgeat[2], M. Compin[2], M. Doro[3], M. Eizmendi[3], L. Font[3], D. Garrido[3], D. Glass[1], F. Grañena[1], A. López Oramas[1], M. Martinez[1], A. Moralejo[1], S. Rivoire[2], S. Royer[2] C. Sanchez[1], P. Valvin[4], G. Vasileiadis[2] FOR THE CTA CONSORTIUM

[1]Institut de Fisica d'Altes Energies (IFAE), Barcelona, Spain
[2]Laboratoire Univers et Particules de Montpellier (LUPM), Montpellier, France
[3]Universitat Autónoma Barcelona (UAB), Barcelona, Spain
[4]Laboratoire Charles Coulomb (L2C), Montpellier, France
alopez@ifae.es

**Abstract:** The Cherenkov Telescope Array (CTA) is the next generation high-energy gamma-ray observatory. A proper characterization of the atmosphere will be crucial for reducing systematic errors in any data collected and allowing an increase in effective observation time. In an elastic backscattering LIDAR, the systematic error in the derived extinction can be as large as 20%, whereas in a Raman LIDAR it can be constrained to 3-5%. The Barcelona IFAE-UAB and Montpellier LUPM groups are each building a Raman LIDAR based on former CLUE telescopes optimized for the future CTA observatory. In this work we present the basic design, adopted solutions and expected performance of the LIDARs.

**Keywords:** IACT, CTA, Raman LIDAR, Calibration

## 1 Introduction

Ground-based Cherenkov telescopes of the Imaging Atmospheric Cherenkov Telescope (IACT) class observe cosmic gamma rays in the GeV–TeV regime by collecting the Cherenkov light produced by electrons and positrons in electromagnetic showers initiated by primary gamma rays when interacting in the top Earth atmosphere.

The current generation of IACTs, and specially HESS[1], MAGIC[2] and VERITAS[3] have most of their systematic errors in the energy reconstruction and absolute scale of the gamma-ray measured fluxes due to systematics in the determination of atmospheric parameters. Of particular concern is the poorly known total extinction that Cherenkov photons undergo in their travel from the emission region, typically located between 20 and 10 km a.s.l., to the ground. The atmosphere in fact acts as a calorimeter for the development of the atmospheric showers, and therefore its characteristics (both the molecular and particle content and profile) affect the transmission of photons and eventually the reconstruction of the primary gamma-ray energy.

To date, moderate effort has been expended in characterizing and monitoring the atmosphere above the telescope and, in general, data are not corrected for the actual atmospheric conditions, but rather discarded when the atmosphere gets too opaque, as in the case of the presence of dense clouds. On the other hand, IACTs can operate in moderate hazy atmosphere, when the extinction of the

Cherenkov photons is not significant. In such conditions, the only thing that must be done is to check the effect on the data, e.g. the diminution of the effective area, the reduced efficiency, the possible variation in the shower intrinsic parameters, etc. and hopefully correct the data with this information. This can be done through Monte Carlo simulations and is currently under investigation in several groups of the Cherenkov Telescope Array (CTA) consortium.

Following another approach, the characterization of the atmosphere at night can be done through the use of Light Detection and Ranging (LIDAR) systems [1], which is the subject of this proceeding. A LIDAR is composed of a laser pointing at the atmosphere, a curved mirror which collects the light of the laser scattered back by atmospheric molecules and particles, and focusses it on a photon detection device. By accurate time measurements, considering the two-way path of the light in the atmosphere (upward and downward), one can also derive the altitude of the interaction region between the laser beam and the atmospheric constituents.

In a simplified approach, the laser photons can undergo two types of scattering (elastic and inelastic) and interact with two families of atmospheric constituents: the *molecules* (like $N_2$, $O_2$, $H_2O$, $CO_2$, etc.) and the *particles* or *aerosols*, mainly constituted by minerals, pollutants, sand, etc. The aerosols are characterized by bigger size than molecules, reaching up to several $\mu$m. Scatterings with molecules are called Rayleigh scattering and modeled with the well-known Rayleigh formula for the cross-section: $\sigma(\lambda) \sim \lambda^{-4}$. When the particle size is of the order of the impinging wavelength, as is the case of aerosols, the Rayleigh formula does not hold, and the more complex Mie interaction theory is used. In such situations, the cross-section scales with the Ångström coefficient, which typically goes from 0.5 to 1.5 depending on the aerosol type.

In the following, we will restrict ourselves to the description of a *Raman LIDAR*, which is a LIDAR that not only measures the elastic scattering, but also the rotation and vibrational inelastic scattering (which occurs basically only with the molecular component). Despite the fact that inelastic scatterings are two-three orders of magnitude less frequent than elastics ones, they are extremely useful because they allow to disentangle between the Rayleigh and Mie scattering, which is normally a large source of uncertainty in the estimation of atmospheric extinction from purely elastic LIDAR.

The development of these Raman LIDARs are conducted in the context of the ATAC working group of the CTA consortium. CTA is currently merging the worldwide effort for Cherenkov telescopes towards a new huge installation of several tens of dishes of several sizes to operate simultaneously[4] [4]. This will guarantee a boost in the performance compared to actual installation. The project is in the preparatory phase and will be completed around 2015–2020. The Raman LIDARs will be installed and operated at the CTA site, with the goal of reducing the systematic uncertainties of the imaging atmospheric Cherenkov technique of the telescopes and increasing the duty cycle thanks to a better knowledge of the atmosphere.

## 2 Overall Design

In the following we will describe the joint effort of three institutes: IFAE-UAB (Barcelona, Spain) and LUPM (Montpellier, France) for the construction of Raman LIDARs through the recycling of a CLUE telescope hosted in a foldable container. A visual picture of the container and the telescope is shown in Fig. 1.

### 2.1 The CLUE container and telescope

Three CLUE containers and telescopes were bought to be reassessed as LIDARs, two by IFAE and one by LUPM. They are standard 20-foot shipping containers, with the top and side walls modified as two shells that can be opened to let the telescope inside move. The telescope is composed of a parabolic solid-glass mirror, of 1.8 m diameter, produced at CERN [2, 3]. The focal plane is also at 1.8 m from the mirror (thus the $f/D$ ratio is 1) and was hosting a small (20×20 cm) MWPC (Multi Wire Proportional Cham-ber) that was removed. The container hosts two motors to open the shells and the telescope has one azimuthal and one zenithal motor for positioning, that can be remotely operated. The telescope has additional four motors to open four petals that protect the mirror.

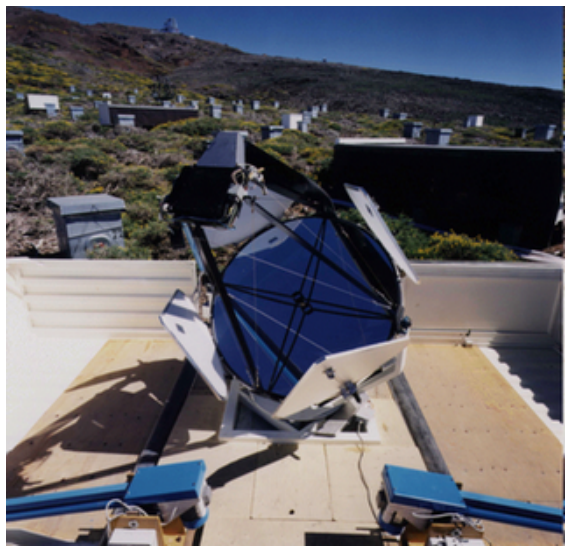

Figure 1: Picture of the CLUE telescope and the foldable container when it was operating in La Palma (Spain). The telescope will be re-designed as Raman LIDAR.

The primary mirror, produced in 1998, has maintained a good geometry, with 80% of the light contained in a circle of 6 mm diameter, almost as when produced. On the other hand, its reflectivity has dropped significantly, due to a missing protective coating. We are currently understanding whether refurbishment is feasible if not we could produce spare mirrors.

### 2.2 The LASER and its optics

IFAE-UAB and LUPM equipped the LIDARs with Nd:YAG lasers, that can provide simultaneously three wavelengths (1064 nm and its first two harmonics 532 and 355 nm). IFAE-UAB laser is a `Brilliant`[TM] compact Q-Switched laser, with a beam diameter of 6 mm and divergence of 0.5 $\mu$rad, with harmonic generator module including separation package in order to select the channels. At 355 nm, where our interests are concentrated[5], we can select 10, 20, 50 Hz as pulse repetition frequency, with 100, 70, 20 mJ energy per pulse respectively, with a pulse duration of ∼4 ns. LUPM will use a CFR-400 Q-Switched laser, with a beam divergence of less than 1.2 mrad and a beam diameter of 7 mm. The pulse repetition rate is fixed 20 Hz, while the beam energy is 400, 230 and 90 mJ for the first, second and third harmonic respectively with a pulse duration of ∼7 ns.

The laser is mounted at one edge of the mirror, at around 0.9 m from the optical axis. The bi-axial design in general does not allow a good sampling of the atmosphere in the proximity of the LIDAR (near range), because there is only a partial "overlap" between the backscattered light reflected onto the focal plane and the photon sensitive area. To maximize the overlap, the laser beam is guided to the optical axis through the use of two 45 deg high reflective mirrors mounted on special adjustable supports. IFAE-UAB bought two `MI1050-SBB` 1' fused-silica mirrors from `Precision Photonics`$^{TM}$ with surface figure of $\lambda/10$, damage threshold of $>1$ J cm$^{-2}$ at 355 nm and reflectivity above 99 %.

A high precision alignment between the laser and the optical axis is needed to be able to get information from the farthest distances. For the IFAE-UAB case a micrometer stepping motor moving the laser head ensures the requested precision.

## 2.3 Focal plane and light transmission

The scattered light that is collected by the primary mirror is focussed onto the focal plane. The parabolic shape ensures that light coming from infinity is focussed with the minimum point spread function (PSF), while light coming from the near range is focussed on a broader circle (but still centered in the optical axis in case of a mono-axial system). Given that 80 % of the light is focussed within 6 mm diameter, IFAE-UAB has chosen to transport the light with an optical fiber fixed at the center of the focal plane to the rear of the telescope where the optical bench and acquistion system are placed. The guide is a liquid optical guide from `Lumatec`$^{TM}$ of 8 mm diameter and 3.2 m length, guarantees an almost flat transmission of 75 % in the range from 300 to 650 nm (intense light with frequency above 650 nm would damage the guide) and can operate between -5 and 35°C. LUPM uses a *ThorLabs*$^{TM}$ 8 mm liquid fiber with a total length of 4 m. Transmission is better than 70 % from 400 to 750 nm.

IFAE-UAB tested the light guide, and confirmed the producer specifications. In general, the light guide transmission does not depend on the way the light guide is enrolled, does not change within temperature between 0 and 30°C, and does not depend significantly from the incident angle below 30 deg (given the mirror diameter and the focal distance, the maximum incident angle is 27 deg). At the exit, the light beam keeps the incident angle, although the direction is randomized forming a circumference from a single incident point.

For the IFAE-UAB LIDAR, the light guide needs to be protected with a low-pass optical filter placed at the focal plane before the entrance to cut above 650 nm. In addition, possibly a collimating lens will be placed at the entrance of the fiber to reduce beam divergence.

## 2.4 Optical bench

The optical bench will be placed at the exit of the optical fiber. It will be composed of a collimating system (still to be defined) and by a series of dychroic mirrors, filters and reflective mirrors, that are needed to select specific wavelengths of interest. There is still an important debate about how many and which are the wavelengths that one should measure to well determine the atmosphere for the purpose of IACTs. Typical ones are the elastic channels at 1064, 532, 355 nm and the Raman channels of the N$_2$ of the first and second harmonics at 607, 387 nm. On the other hand, different solutions can be applied that can use O$_2$ Raman channels, water vapor channels, CO$_2$ channels, the transverse and parallel polarization of the elastic channels, etc. The baseline design foresees the use of 355 (elastic) and 387 nm (N$_2$ Raman) for IFAE-UAB LIDAR and the additional 532 (elastic) for LUPM. On the other hand, in the case of IFAE-UAB LIDAR, the optical bench has been designed to have enough modularity to easily add or remove channels.

Once the wavelength is selected, the beam is focussed onto an hemispherical photomultiplier tube (PMT). We received offers from `Hamamatsu`$^{TM}$ for two PMTs, `R1924A` and `R329P` of 25 and 51 mm cathode diameter respectively. The choice of the PMT will be done after the optical bench will be simulated to understand at which level the light beam can be collimated and transported. The PMTs are optimized for LIDAR purposed, with background noise at 3 nA, gain at $\sim 10^6$, single photo-electron capability, and in general with good environmental ruggedness to be used in outside installations.

## 2.5 Readout

In general, each channel must read return power that spans more than six order of magnitude, from the mW received from the near range in the elastic case, to the nW received from 10-15 km in the Raman case. Typically, this problem is approached by using a duplex readout that provides both an analogue readout for the near range and a photon counting readout for the far range, with an overlap region at moderate altitudes (around 5 km).

Commercial modules are available and both groups contemplate to equip the LIDAR as a default option with standard `LICEL` modules. In parallel, the IFAE engineer team is developing a customized alternative with emphasis on getting the maximum information for the highest distances.

## 2.6 System control

The system control still must be developed. It can be divided in two sectors. The front-end part controls the LIDAR hardware: container motors, the telescope petals, the telescope pointing, the trigger of the laser, etc. The integration part exchanges information with the central control system of the CTA observatory. While the former system will be probably controlled by `c++` code programs, the latter one will be written in the standard CTA control software `ACLI`.

## 3 Outlook and conclusions

In this proceedings, we have presented the effort for the reassessment of three decommissioned CLUE telescopes to be used as Raman LIDARs for the atmospheric monitoring and characterization for the future CTA observatory. The goal is to increase the duty time by observing during moderate hazy atmosphere conditions and to reduce the systematic errors on the energy distribution through a deeper knowledge of the effect of atmospheric properties on the data.

Two institutes in Barcelona (Spain), IFAE and UAB, and LUPM in Montpellier (France) are currently collaborating for the construction of the LIDARs, despite different solutions are currently adopted. A fourth insitute, CEILAP in Buenos Aires (Argentina) is also developing a different design of Raman LIDAR, presented elsewhere in this conference.

Our primary goal is to have the complete design fixed by the end of the year, specially for the optical bench, and finish the construction by summer 2012. After a characterization campaign in Barcelona and Montpellier, the LIDAR may be shipped at the (yet non defined) CTA sites.

We gratefully acknowledge financial support from the agencies and organisations listed in this page: http://www.cta-observatory.org/?q=node/22

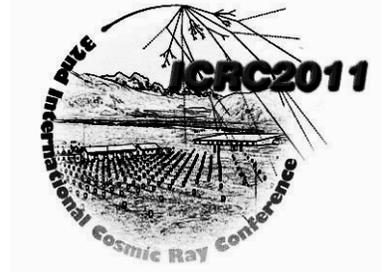

# Constraining the Extragalactic Background Light in the near-mid IR with the Cherenkov Telescope Array (CTA)

M. ORR[1] AND F. KRENNRICH[1] FOR THE CTA CONSORTIUM[2]

[1]*Department of Physics and Astronomy, Iowa State University, Ames, IA 50011*
[2] *see http://www.cta-observatory.org/?q=node/22*
*morr@iastate.edu*

**Abstract:** The next generation imaging atmospheric Cherenkov telescope (IACT), the Cherenkov Telescope Array (CTA), will have improved sensitivity, cover a broader energy range, and posses a wider field of view than the current generation of IACTs. CTA is therefore an ideal instrument for acquiring precise spectral measurements, at very high gamma-ray energies, on a large population of blazars. This is highly advantageous for studies of the extragalactic background light (EBL) since very-high-energy gamma rays from blazars interact with EBL photons as they propagate through the Universe. The signatures of EBL absorption are therefore imprinted on the observed spectra of blazars. One of the possible absorption features is that of a spectral hardening/softening in the observed emission between approximately 1 and 5 TeV. This spectral break is dependent on both the source redshift and EBL spectral energy distribution. One can therefore constrain the EBL by measuring the spectral break versus redshift distribution from a large sample of blazars. This analysis was originally developed using blazar measurements from the current generation of IACTs. Here we discuss how CTA can improve upon this technique by taking advantage of the instruments wide field of view and high sensitivity in the requisite energy range. In particular, we consider the exposures and source statistics CTA can reasonably obtain to improve current constraints on the EBL.

**Keywords:** blazars, CTA, extragalactic background light, gamma rays

## 1 Introduction

The extragalactic background light (EBL) is a diffuse photon field spanning wavelengths from $\sim 0.1\,\mu m$ to $1000\,\mu m$ and is second in intensity only to the cosmic microwave background. It has a bimodal distribution with peaks at $\sim 1\,\mu m$ (near-infrared) and $\sim 100\,\mu m$ (far-infrared). The near-infrared peak derives from the collective emission of nuclear (stars/galaxies) and gravitational (accretion onto active galactic nuclei) energy releases while the peak in the far-infrared is due to the absorption and re-radiation of this near-infrared emission. See [1] and [2] for reviews on the origins and cosmological implications of the EBL.

The EBL is difficult to measure directly due to overwhelming foreground emission from both zodiacal light within the solar system and radiation from the Galaxy. Limits on the EBL can be obtained using the integrated light from galaxy counts as well as gamma-ray observations of extragalactic objects at very-high-energies (VHEs). The latter approach exploits the fact that as VHE gamma rays propagate over cosmological distances they can interact with the diffuse infrared photons of the EBL via pair production (i.e., $\gamma_{VHE}\,\gamma_{EBL} \rightarrow e^+\,e^-$) [3]. The signature of this absorption is left imprinted on the VHE spectral energy distribution

(SED). The characteristics of this feature are dependent on the intensity and shape of the EBL SED.

The amount of absorption present in VHE spectra is dependent on, among other things, the distance of the emitting source. Blazars are therefore suitable sources for constraining the intensity of the EBL given their large redshifts and high VHE fluxes. The work discussed here investigates the potential for the Cherenkov Telescope Array (CTA), a next generation imaging atmospheric Cherenkov telescope (IACT), to place strong constraints on the EBL and potentially *detect* an EBL absorption signature.

## 2 Cherenkov Telescope Array

Two separate Cherenkov telescope arrays will constitute CTA, with the aim to: increase sensitivity by and order of magnitude for deep observations around 1 TeV, increase the overall detection area and therefore detection rates, increase angular resolution to improve capabilities for resolving the morphological properties of extended sources, provide uniform energy coverage for photons from tens of GeV to more than 100 TeV, and enhance sky survey and monitoring capabilities [6]. These two arrays will be located in the nothern and southern hemispheres. In addition, CTA will have the ability to operate sub-arrays in different

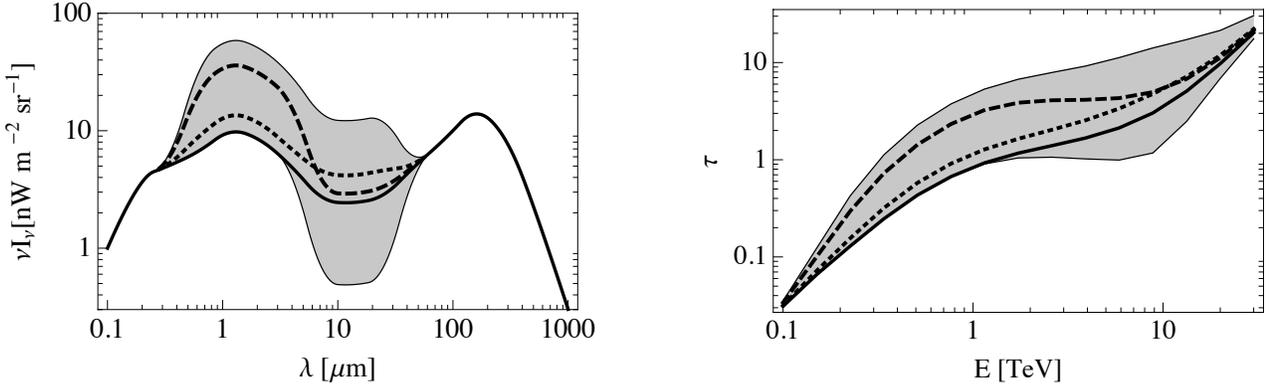

Figure 1: *Left*: EBL intensity versus photon wavelength. The shaded region indicates the range of scenarios tested. The thick solid line designates the baseline shape used, from which all other scaled shapes are generated. For clarity, two additional models are shown (dotted and dashed) illustrating the independent scaling of the near- and mid-IR regions. *Right*: Optical depth $\tau$ (at $z = 0.1$) versus gamma-ray energy in TeV for each EBL scenario tested. The optical depths for the baseline and two additional EBL models shown in the left panel are shown as well.

observing modes to maximize both source coverage and science return.

A significant change for CTA with respect to the current generation of IACTs is its open data access policy. A Science Data Center will provide public access to all CTA data as well as tools for their analysis and associated tutorials to facilitate the data processing. These changes will open the field of TeV gamma-ray astronomy to the astronomical community at large.

## 3 Detecting an EBL Absorption Signature with CTA

Limits on the intensity and shape of the EBL SED can be obtained by calculating the gamma-ray absorption resulting from different EBL scenarios and comparing this with VHE observations. Figure 1 shows a range of EBL scenarios and their calculated $\gamma_{\mathrm{VHE}}\,\gamma_{\mathrm{EBL}}$ optical depths. The increase in absorption with gamma-ray energy (right panel of Figure 1) produces a softening in the observed VHE spectra of extragalactic sources such as blazars. The flattening of the optical depth at 1 TeV, for some EBL SEDs, produces a break in observed spectra at approximately this energy. Figure 2 illustrates both of these effects. The magnitude of the spectral softening and break increases with source redshift.

Evidence for this EBL-induced, redshift-dependent, spectral break can be searched for using a large sample of VHE blazars. This was done using observations of 12 blazars from the current generation of IACTs [4]. Each source in the sample was fit with a broken power-law with a break energy of 1.3 TeV. This choice of break energy is a direct result of the location of the near-infrared peak of the EBL, which governs where the $\gamma_{\mathrm{VHE}}\,\gamma_{\mathrm{EBL}}$ optical depth significantly changes slope.

The peak of the photon-photon cross-section can be approximated using the relation,

$$\lambda_\epsilon \; [\mu\mathrm{m}] \approx 1.24 E_\gamma \; [\mathrm{TeV}] \,, \qquad (1)$$

where $\lambda_\epsilon$ is the EBL photon wavelength in $\mu$m and $E_\gamma$ is the gamma-ray energy in TeV [5]. A near-infrared EBL SED peak at $\sim 1.6\,\mu$m therefore corresponds to an interaction cross-section peak of $\sim 1.3$ TeV.

Figure 3 shows the spectral break versus redshift distribution for the 12 blazars used in the study by [4] (open symbols). It can clearly be seen that the size of the error bars on the spectral break measurements vary substantially from source to source. This is due to the different instrument sensitivities and exposure times associated with each source. Nonetheless, a fit to this data with a flat line located at $\Delta\Gamma(z) = 0$ can be excluded with a significance of 3.2 standard deviations.

To determine the expected spectral break versus redshift distribution for CTA, the set of analysis tools `ctatools`, provided by the CTA consortium, were used. The currently measured flux and spectral index for each source was used, along with the expected spectral break given the current IACT data shown in Figure 3, to simulate the expected result from 50 hours of CTA observations. A broken power-law was simulated between 100 GeV and 4 TeV following the form

$$\frac{\mathrm{d}N}{\mathrm{d}E} = \begin{cases} N_0 \left(\dfrac{E}{E_{\mathrm{break}}}\right)^{-(\Gamma + \Delta\Gamma/2)} & , \; E \le E_{\mathrm{break}} \\[2ex] N_0 \left(\dfrac{E}{E_{\mathrm{break}}}\right)^{-(\Gamma - \Delta\Gamma/2)} & , \; E > E_{\mathrm{break}} \end{cases} \,,$$
$$(2)$$

where $N_0$ is the normalization at the break energy $E_{\mathrm{break}}$, $\Gamma$ is the spectral index measured by current IACTs, $\Delta\Gamma$ is the expected spectral break given the current best linear fit to the data in Figure 3, and $E$ is the energy. The flux normalization at $E_{break}$ for the simulated spectrum was chosen based on existing measurements. These simulations were then processed through the CTA data selection and unbinned likelihood analysis tools. The resulting spectral break versus redshift distribution is shown in Figure 3 (filled symbols).

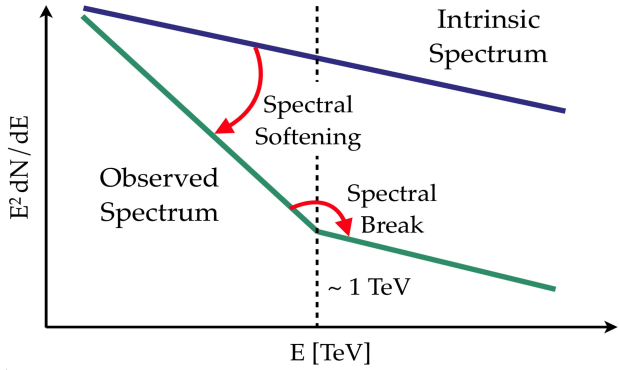

Figure 2: Illustration of two EBL absorption effects - a spectral softening due to increased absorption with increasing gamma-ray energy and a spectral break due to the slope of the EBL SED between near- and mid-infrared wavelengths.

There are a few important things to note about the CTA spectral break versus redshift distribution. First, while no spectral break measurement currently exists for the distant blazar 1ES 0414+009 ($z = 0.287$), CTA will be able to achieve this within 50 hours. The detection of EBL absorption signatures at such large redshifts will be critical to understanding how the EBL has evolved over cosmic time. Second, there are no CTA simulation results shown for the blazars PKS 0548-322 and PKS 2005-489. This is due to the fact that 50 hours of CTA observations will not yield a spectral break measurement for either of these sources that will contribute to the overall significance of the spectral break versus redshift distribution result.

There is still a large uncertainty, given current data, in the spectral break versus redshift distribution. What the CTA simulated observations show, however, is that the excellent sensitivity of CTA in the regime around 1 TeV will yield high precision measurements of blazar spectra. This in turn provides high precision measurements of blazar spectral breaks which yields valuable results for the spectral break versus redshift distribution, even for small breaks.

## 4 Constraining the EBL with CTA

The spectral break versus redshift distribution shown in Figure 3, in addition to providing evidence for an EBL absorption signature, can be used to constrain the ratio of near- to mid-infrared intensities of the EBL. Each EBL SED possessing a different near- to mid-infrared ratio predicts a unique spectral break versus redshift distribution.[1] The predicted redshift dependence can be calculated using a test spectrum for the intrinsic blazar emission, placed over a range of redshifts sufficient to sample the distribution, which then gets absorbed according to the calculated optical depth for the EBL scenario in question. The binning of the test spectrum can be chosen in a variety of ways, but the method best suited for CTA is to use the energy binning and redshift sampling representative of the actual blazar observations.

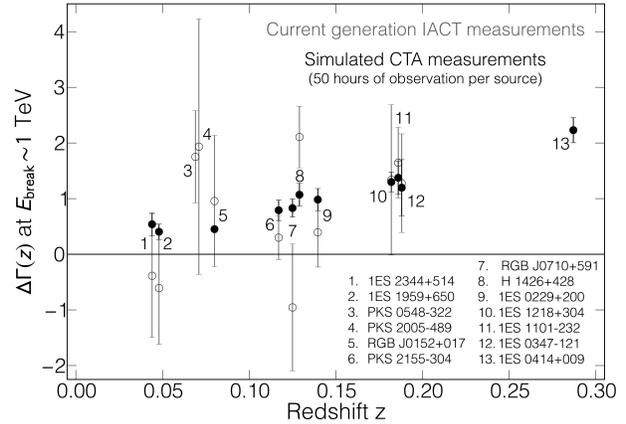

Figure 3: Spectral break versus redshift distribution for 12 blazars measured by the current generation of IACTs (open symbols). Also shown are the projected spectral break measurements for 50 hours on each source with CTA. While no spectral break measurement currently exists for the distant blazar 1ES 0414+009 ($z = 0.287$), CTA will be able to achieve this with high significance in 50 hours. However, no anticipated CTA results are shown for PKS 0548-322 or PKS 2005-489 as CTA will not obtain a significant spectral break measurement for either these sources with 50 hours of observing time.

The predicted spectral break versus redshift distribution can be calculated for a wide range of EBL SEDs and then compared with the observations (i.e., Figure 3). EBL scenarios can then be ruled out based on their level of disagreement with CTA observations. This method, combined with techniques along the lines of [7], can provide strong constraints on the EBL.

## 5 Advancing Studies of the EBL with CTA

Given the results shown in Figure 3, it is likely that CTA will have the capability to detect the signature of EBL absorption in blazar spectra and trace its trend with redshift. There is another aspect CTA's capabilities, not yet considered, that will further advance studies of the EBL well beyond anything achievable with the current generation of IACTs. This key component is the ability of CTA to monitor and survey the sky. Not only will the CTA telescopes have a wider field of view ($\sim 6-8^\circ$), but the northern and southern arrays will provide full sky coverage and the many tens of telescopes will allow the observation of multiple sources simultaneously using subsections of the overall array (if so desired).

---

1. To a lesser degree, this distribution also depends on the overall intensity of the EBL, which alternative techniques are better suited to constrain (see [4]).

It is anticipated that CTA will detect on the order of 1000 sources, roughly a factor of 10 more than are currently detected at TeV energies. The CTA blazar catalog will therefore likely contain hundreds of sources. Of these blazars, not all will be suitable for these spectral break studies, but it is reasonable to assume that CTA will populate the spectral break versus redshift distribution with several tens of objects.

Not only can the evolution of the EBL over redshift be studied using this potentially large sample of blazars, but also the intrinsic (dis)similarities between sources. If conflicting constraints on the EBL are obtained from different source types (e.g., low-, intermediate-, and high-frequency-peaked BL Lac objects), they can be used to asses the validity of certain assumptions regarding the intrinsic properties of these same sources.

## 6    Discussion & Conclusions

CTA will have a greatly improved performance as compared with the current generation of IACTs. It therefore stands to significantly progress studies of the EBL using VHE observations. If a redshift-dependent spectral break exists, it is likely that CTA will definitively detect this EBL absorption signature in the spectra of blazars. The detection of this absorption signature can also be used to place constraints on the shape of the EBL SED. In addition, CTA will significantly expand the known population of VHE blazars which will advance studies of the EBL performed with IACTs beyond what is currently possible. Namely, the evolution of the EBL with redshift can be studied and the intrinsic properties of sources can begin to be disentangled from the effects of EBL absorption. The prospects for investigating both blazars and the EBL with CTA are very promising.

## Acknowledgements

We gratefully acknowledge support from the agencies and organizations listed in this page: http://www.cta-observatory.org/?q=node/22.

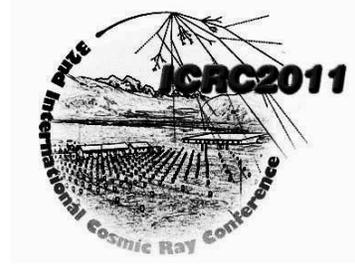

# On the Detectability of Dwarf Galaxies with the Cherenkov Telescope Array

NIETO D.[1], HASSAN T.[1] , MIRABAL N.[1], BARRIO J.A.[1], CONTRERAS J.L.[1] FOR THE CTA CONSORTIUM

[1]*Universidad Complutense de Madrid, SPAIN*

*nieto@gae.ucm.es,thassan@gae.ucm.es,mirabal@gae.ucm.es, barrio@gae.ucm.es,contrera@gae.ucm.es*

**Abstract:** The nature of dark matter is one of the greatest unsolved mysteries in science. Observations with the current generation of imaging atmospheric Cherenkov telescopes have taken the indirect search for dark matter to competitive levels. With a factor of 5-10 improvement in sensitivity, the Cherenkov Telescope Array (CTA) will improve the chances of detecting a dark matter signal. Assuming a hypothetical dark matter particle that generates observable $\gamma$-ray photons in the CTA energy range, we compute the detectability of dark matter subhalos as a function of the particle mass and its annihilation cross section. A wide range of subhalo astrophysical factors is considered, as well as different annihilation channels.

**Keywords:** HE 3.4 Direct and indirect dark matter searches, HE 3.6 New experiments and instrumentation.

## 1 Introduction

The existence of an additional non-baryonic particle component appears to be indispensable to accommodate a number of astrophysical observations, including gravitational lensing and galactic rotation curves [1]. An unequivocal detection of this so-called *dark matter* (DM) is one of the most important pursuits in astroparticle physics. Indirect searches for a DM signal from pair annihilation or particle decays are being performed through $\gamma$-ray observations of astrophysical objects with high dark matter density [2, 3, 4]. Unfortunately, no reliable DM smoking gun has been detected to date. Here, we explore the prospects for detecting a DM signal with observations of subhalos around the Milky Way with the upcoming Cherenkov Telescope Array (CTA).

### 1.1 The Cherenkov Telescope Array (CTA)

CTA is a global initiative to build the next generation ground-based $\gamma$-ray observatory [5]. When compared to current facilities such as H.E.S.S., MAGIC or VERITAS, a factor of 5-10 improvement in sensitivity is expected in the 100 GeV to some tens of TeV (or Very High Energy, VHE) energy domain. In addition CTA sensitivity will extend to both lower and higher energies. The observatory will consist of two arrays, one in each hemisphere. The Southern hemisphere array will be mainly dedicated to Galactic sources and the central part of our Galaxy, whereas the Northern one will complement the Southern, and be dedicated to northern extragalactic objects notably active galactic nuclei, gamma-ray bursts, and starburst galaxies.

Apart from this wide variety of astrophysical objects, CTA will allow a deep exploration of an energy window where DM signatures are expected, via DM particle annihilation or decay as $\gamma$-ray by-products. Therefore, a dedicated study of CTA response to possible DM signals is required.

Throughout this paper, we rely on the CTA design concepts summarized in [5], which has defined a number of possible array configurations and studied their sensitivities. The arrays considered thus far are composed principally of 3 types of telescopes: *large* (23 m diameter), *medium* (around 12 m) and *small* (6-7m). Apart from sizes, these telescope types also differ in other essential parameters, such as field of view and camera pixel diameter. Differences between arrays are due to the number of detectors of each kind that are used, and their spatial layout. Within the design study each considered array has been assigned a letter. Although in our study we have considered all the configurations, the results presented in this work concern only one of them (configuration E) that extends over a surface of $\approx$ 3 km$^2$ and it is characterized by a balanced sensitivity over the whole energy range.

### 1.2 Dark Matter targets and signatures

Despite bold efforts, no DM signal has been detected so far from any of the most promising DM targets including dwarf spheroidal (dSph) galaxies[6], the Galactic Center [4], and clusters of galaxies [7]. The null outcome is partially due to sensitivity limitations with the current generation of instruments. Yet, it also reflects the density and distance of the regions probed thus far.

High-resolution simulations indicate that DM halos must exhibit a wealth of substructure on all resolved mass scales [8]. Some of these subhalos would correspond to "classical" dwarf galaxies. Nevertheless, the low end of dark matter subhalos might not have attracted enough baryonic matter to ignite star-formation and would therefore be "invisible" to routine astronomical observations except for some ultra-faint dwarf galaxies. A fraction of these DM subhalos might become prominent at VHE due annihilating weakly interacting massive particles (WIMP) [9]. Such Galactic DM subhalos offer an intriguing sample where to search for a dark matter signature given that these might be out of the Galactic Plane and virtually free of astrophysical background.

A $\gamma$-ray signal from DM particle annihilation in the VHE domain would be characterized by a distinctive spectral shape that might include features such as lines, internal bremsstrahlung, as well as a characteristic cut-off at the DM particle mass. A genuine indicator of DM detection would be the detection of this distinct cut-off in several sources, given the universality of the DM spectrum. In the super-symmetric extension of the standard model (SUSY) the *neutralino* emerges as a natural DM particle candidate and the spectral cut-off is conveniently located within CTA energy range [10]. It is clear that even if one could capture a well measured spectrum, it might not be sufficient to reveal all the properties of the DM particle. However, a detection at VHE would provide valuable information for subsequent searches.

## 2 Detection prospects of dark matter subhalos with CTA

In order to quantify the detection prospects of DM subhalos with CTA one must compare the assumed annihilation $\gamma$-ray spectrum from DM annihilation with the expected CTA sensitivity.

### 2.1 $\gamma$-ray flux from dark matter subhalos

The hypothetical differential spectrum from DM annihilation in a galactic DM subhalo can be described as the product of two terms:

$$\phi(E, \Delta\Omega) = \phi^{PP}(E) \times J(\Delta\Omega), \quad (1)$$

The first term, so-called particle physics factor, depends on the particle physics model usually written as:

$$\phi^{PP}(E) = \frac{1}{4\pi} \frac{\langle \sigma_{\text{ann}} v \rangle}{2m_\chi^2} \sum_{i=1}^{n} B_i \frac{dN_i^\gamma}{dE}, \quad (2)$$

where $\langle \sigma_{\text{ann}} v \rangle$ is the thermally averaged annihilation cross section, $m_\chi$ is the DM particle mass, $\frac{dN_i^\gamma}{dE}$ is the photon spectrum per annihilation through channel $i$, and $B_i$ is the $i$ channel branching ratio. The second term, or the so-called astrophysical factor, consists on the integration of the DM density squared along the line of sight, considering a solid angle $\Delta\Omega$:

$$J(\Delta\Omega) = \int_{\Delta\Omega} \int_{l.o.s.} \rho^2(r(s, \Omega)) \, ds \, d\Omega. \quad (3)$$

There are many unknown variables in the former expressions. For simplicity, we make a number of assumptions to build a set of reasonable DM spectra. First, the DM particle is considered to be a weakly interacting massive particle (WIMP) whose freeze-out in standard cosmology implies an annihilation cross section of $\langle \sigma_{\text{ann}} v \rangle \sim 3 \times 10^{-26} cm^3 s^{-1}$ [11]. Specifically, we examine a *neutralino* in the minimal Super Gravity (mSUGRA) model that is the preferred DM particle attending to the simulation results from [6]. For the photon spectra per annihilation, the analytical expressions from [12] for the channels $b\bar{b}$, $W^+W^-$, $\tau^+\tau^-$, and $\mu^+\mu^-$ respectively were used. Throughout, branching ratios $B_i$ of 100% are fixed. For the sake of simplicity we assume that the annihilation proceeds entirely through each of the considered channel.

We have considered a range of possible masses for the DM particle $m_\chi$, that spans from 50 GeV to 10 TeV. The lower limit corresponds to a conservative energy threshold for CTA. It is also motivated by the current experimental lower limits from accelerator data [13]. The upper limit is justified by theoretical arguments (see [11]). Fig. 1 shows spectra for different annihilation channels for a potential 1 TeV WIMP mass and an astrophysical factor $J = 8 \times 10^{20}$ GeV$^2$ cm$^{-5}$.

In order to be considered statistically significant with CTA, we require a detection of a gamma-ray signal that exceed 5 standard deviations ($5\sigma$) over the background events [5]. In this work, statistical significances are calculated using Eq. 17 in [14]. The excess rate $R_{exc}$ over a certain energy threshold $E_{th}$ can be computed from the effective area of the instrument $A_{eff}(E)$ and the differential spectrum of the source $\phi(E)$ as:

$$R_{exc} = \int_{E_{th}}^{\infty} \phi(E) A_{eff}(E) \, dE \quad (4)$$

the estimated detection time for a certain source depends on the above formula and the background rate of the instrument. As it was previously mentioned, the considered effective area and background rate correspond to the CTA candidate configuration E.

## 3 Results

In order to survey the CTA capabilities to gamma-rays from DM annihilations at a specific subhalo object, we evaluate the statistical significance of the DM signal as a function of the DM particle mass $m_\chi$ and the astrophysical factor $J$, considering four different annihilation channels, namely $b\bar{b}$, $W^+W^-$, $\tau^+\tau^-$, and $\mu^+\mu^-$.

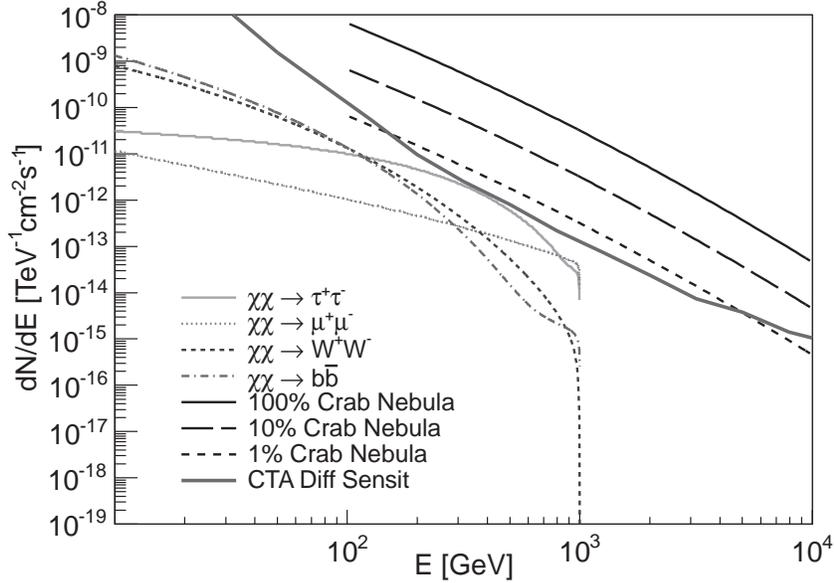

Figure 1: DM spectra for $b\bar{b}$, $W^+W^-$, $\tau^+\tau^-$, and $\mu^+\mu^-$ channels. We assume an astrophysical factor $J = 8 \times 10^{20}$ GeV$^2$ cm$^{-5}$ and $m_\chi = 1$ TeV. The lines indicate the 100%, 10% and 1% Crab Nebula differential flux. Also shown is the differential sensitivity expected for CTA.

We have chosen two ways to present our results. In the first one we calculate the astrophysical factor $J$ required to reach a statistical significance of $5\sigma$ assuming an effective observation time of 50 hours. In the second one we compute the so called *boost factor* for a set of well known dwarf spheroidal galaxies.

### 3.1 Astrophysical factors

Results for the first method are shown in Fig. 2 as a function of WIMP mass for each of the 4 channels considered. As illustrated, the lines mark to the minimum astrophysical factor $J_{min}$ necessary for a $5\sigma$ detection with CTA for each annihilation channel. In order to put these values in context, we note that known dwarf galaxies span a range between $4 \times 10^{17}$ GeV$^2$ cm$^{-5}$ for the Carina dSph and $1.8 \times 10^{19}$ GeV$^2$ cm$^{-5}$ for Segue 1 ultra-faint dwarf galaxy [15].

### 3.2 Boost Factors

A different way to evaluate the prospects of DM detection is by means of the *boost factor* $B_F$. The approach is justified by the fact that one current conjecture in this field is that the actual signal due to dark matter might be enhanced with respect to classical calculations. This intrinsic boost in the flux could be provided by the effect of substructures within the subhalos, which enhances the astrophysical factor [16] and/or by the *Sommerfeld effect* which enhances the particle physics term [17].

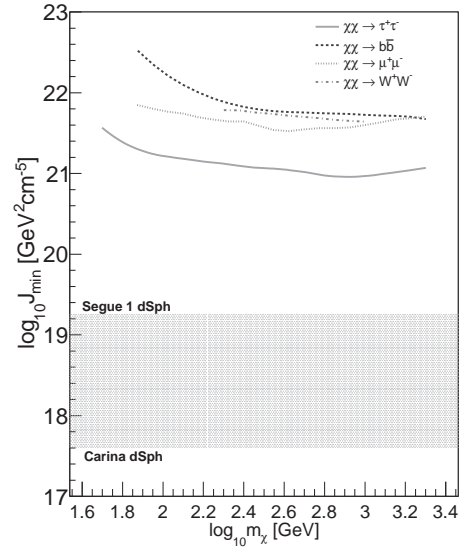

Figure 2: Astrophysical factor $J$ as a function of WIMP mass $m_\chi$. The lines depict the minimum value $J$ required for a $5\sigma$ detection with CTA for the respective annihilation channel. The shaded region illustrates the current estimated values from Carina to the ultra-faint Segue 1.

We simply estimate the minimum boost required by a set of well known dwarf galaxies in order to allow for a $5\sigma$ detection in 50 hours of observation time with CTA. We compute the value of this minimum $B_F$ as the ratio of the minimum astrophysical factor $J_{min}$ to the measured astrophysical factor $J_{obs}$ from observations/model ling, for a particular WIMP mass. Fig. 3 shows the minimum $B_F$ for a 1 TeV WIMP mass annihilating to $\tau^+\tau^-$ for a set of 8 dwarf galaxies.

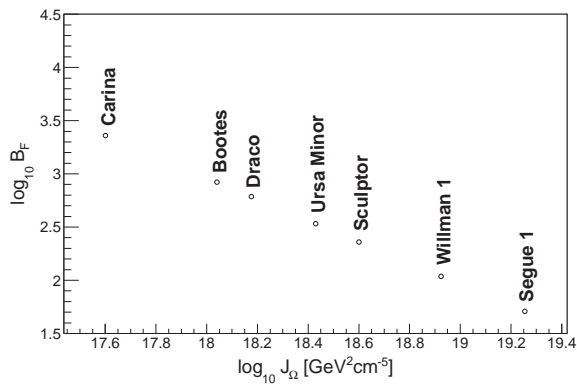

Figure 3: Minimum boost factor $B_F$ as a function of astrophysical factor $J$ for a set of 8 well-studied dSph galaxies, for a putative 1 TeV WIMP mass annihilating to $\tau^+\tau^-$. The minimum requirement is a $5\sigma$ detection in 50 hours by CTA. The least strict boost is for the ultra-fain Segue 1 that would only require a factor of 50.

## 4 Discussion and Conclusions

We have presented a preliminary study of the prospects of DM detection with CTA. In particular, DM annihilation could be realized for subhalos with astrophysical factors $J \geq 8 \times 10^{20}$ GeV$^2$ cm$^{-5}$ assuming a *neutralino* with a mass around 1 TeV. At present, Segue 1 appears to be one of the most intriguing DM targets especially given the $1\sigma$ error bars of its astrophysical factor [15]. However, there is consensus that further work is required to try to calculate $J$ directly from stellar dynamics [18]. From our results, it is obvious that the astrophysical community must continue to search for objects with extreme astrophysical factors *i.e.* either less distant or subhalos with enhanced density profiles. Our knowledge of the Galactic halo will certainly improve with upcoming Southern optical surveys that will scan for such subhalos prior or concurrent with CTA.

Potentially, boost factors $B_F \geq 50$ would alleviate the strict astrophysical factor requirements needed for a $5\sigma$ detection with CTA even down to some of the "classical" dwarf galaxies in our current sample. However, our unfamiliarity with the boost regime obliges us to be cautiously optimistic. It must be noted that longer observation times, larger annihilation cross sections, galaxy stacking and improved data analysis with CTA will allow us to improve these estimates even further.

We close by emphasizing that the results presented should be considered very conservative limits from the observational point of view, as further studies and design improvements are being introduced into the final CTA configuration. Specifically, the CTA dark matter prospects taking into account candidate arrays with augmented effective areas will be discussed in an upcoming publication. Nonetheless, albeit preliminary this study clearly illustrates the valuable impact that CTA could make on indirect dark matter searches over the upcoming years.

## 5 Acknowledgments

We thank Jan Conrad and Ulli Schwanke for useful comments. We are indebted to the CTA Montecarlo Work Package for their exceptional work during the design study. We gratefully acknowledge support from the agencies and organisations listed in this page: http://www.cta-observatory.org/?q=node/22 The authors acknowledge the support of the Spanish MICINN under project codes FPA2009-0838, FPA2010-22056-C06-06, and the Consolider-Ingenio MULTIDARK CSD2009-00064. N.M. gratefully acknowledges support from the Spanish MICINN through a Ramón y Cajal fellowship.

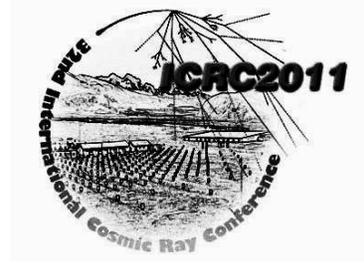

# Search for Lorentz Invariance Violation with flaring Active Galactic Nuclei: a prospect for the Cherenkov Telescope Array

J. Bolmont[1], D. Emmanoulopoulos[2], A. Jacholkowska[1], J.-P. Tavernet[1] for the CTA Consortium

[1]LPNHE, Université Pierre et Marie Curie Paris 6, Université Denis Diderot Paris 7, CNRS/IN2P3, 4 Place Jussieu, F-75252, Paris Cedex 5, France

[2]School of Physics and Astronomy, University of Southampton, SO17 1BJ Southampton, United Kingdom

bolmont@in2p3.fr

**Abstract:** In the recent years, many results have been published about a possible violation of Lorentz Invariance in the frame of Quantum Gravity (QG) models measuring time delays in the arrival times of very high energy (VHE, >100 GeV) gamma-ray photons from distant flaring active galaxies. These photons have been detected by the current ground-based VHE Cherenkov detectors (HESS,MAGIC, VERITAS) and so far, no deviations from the speed of light in vacuum have been seen either for linear or for quadratic scales. The new generation of ground based instruments "the Cherenkov Telescope Array" (CTA) will be able to probe deeper into this area due to its increased sensitivity (one order of magnitude better than the current detectors) and broader energy range (above 10 GeV). Based on a maximum likelihood technique, a quantitative study is presented with respect to the potential of CTA to detect possible QG effects. Based on simulations, predictions for linear and quadratic limits are given and discussed.

**Keywords:** CTA, active galaxies, Quantum Gravity, Lorentz Invariance Violation

## 1 Quantum Gravity and Lorentz Invariance Violation

The search for a quantum theory of gravitation is one of the outstanding tasks of modern physics [1]. As an important consequence of the time-space discretization, Lorentz Invariation Violation (LIV) may appear as predicted in some models of Loop Quantum Gravity [2] or String Theory [3]. The tiny effects in the photon propagation from distant astrophysical sources as Active Galactic Nuclei (AGNs) or Gamma-ray Bursts (GRBs) would add-up producing deviations in the value of the velocity of light [4]. These deviations could be represented by linear and quadratic terms in the so-called dispersion relation:

$$c^2 p^2 = E^2 \left(1 \pm \xi(E/M) \pm \zeta(E/M)^2 \pm ...\right), \quad (1)$$

where $M$ is the Quantum Gravity energy scale (in principle close to the Planck scale) and $\xi$ and $\zeta$ are positive parameters.

In this paper a search for LIV with photons emitted in flares of AGN is being presented as a prospect for the future Cherenkov Telescope Array (CTA) [5]. In particular, the effect of potentially improved statistical accuracy due to both increased sensitivity and better energy coverage is evaluated.

## 2 The analysis procedure

In this work a likelihood fit procedure, as described in detail in [6, 7], is used to measure the energy dependant time lags. This method makes use of individual photon information (energy and detection time) and requires a parameterization of both the light curve and the spectrum. The light curve is parameterized ($F_S$) at low energies where the time lags are supposed to be negligible and the measured spectrum ($\Lambda$) is parameterized in the full energy range of the instrument. Then the probability density function (pdf) is given by:

$$P(t, E) = \int_0^\infty \Lambda(E_S) G(E - E_S, \sigma(E_S))$$
$$F_S(t - \tau_l E_S) dE_S, \quad (2)$$

where $G$ takes into account the energy resolution of the detector, considered here to be 10%. After normalizing the pdf, the parameter $\tau_l$ (here for the linear correction to the dispersion relation of Eq. 1) is obtained by minimizing $-\ln(L)$ where

$$L = \prod_{\text{all photons}} P(t, E). \quad (3)$$

A toy Monte Carlo software developed for PKS 2155-304 analysis [6] is used to simulate sets of photons with given time and energy distributions. The injected lags range from -60 s TeV$^{-1}$ to 60 s TeV$^{-1}$ in steps of 20 s TeV$^{-1}$ for the linear case and from -60 s TeV$^{-2}$ to 60 s TeV$^{-2}$ in steps of 20 s TeV$^{-2}$ for the quadratic case. For each injected lag, 500 realizations of the lightcurve are simulated. The obtained distribution of reconstructed lags exhibits a Gaussian with standard deviation $\sigma_\tau$. The mean dispersion $\bar{\sigma}_\tau$ is the average of $\sigma_\tau$ for all injected lags.

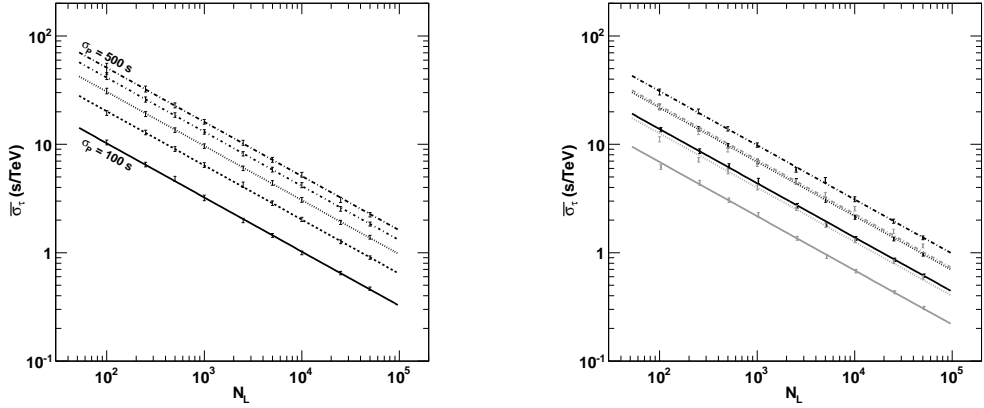

Figure 2: Average error on the reconstructed lag as a function of the number of photons used in the likelihood fit computation. Points are fitted with the function defined in Eq. 4. Left: the different set of points correspond to different light curve peak widths: from bottom to top, $\sigma_P = \{100, 200, 300, 400, 500\}$. Right: Black and gray lines correspond respectively to linear and quadratic effects. Dash-dotted, dotted and solid lines correspond respectively to $\Gamma = 3.4$, 2.8 and 2.2.

## 3 Parameterization of the light curve

The likelihood fit procedure requires the light curve to be parameterized at low energies. As the error $\sigma_\tau$ is smaller when the spikes of the light curve are narrower, the peak separation capability is essential: a single peak seen by present day experiments could be seen with sub-structures with an instrument such as CTA, which would in turn increase the performance of the analysis. In this section, this issue is studied using simple hypotheses.

A list of 3000 photons is generated randomly with energies following a power law distribution with index $\Gamma = 2.8$ and a time distribution chosen to be the sum of two Gaussian functions, which have the same standard deviation $\sigma_G$ and the separation between the peaks is varied from 0 to 1400 s. The time distribution is fitted with one or two Gaussian functions. The bin width of the light curve is set to 60 s for H.E.S.S./MAGIC and to 30 s for CTA. The minimal peak separation is then obtained when $\chi^2/$dof = 1.5.

Fig. 1 shows the minimal peak separation necessary to distinguish the two spikes as a function of $\sigma_G$. For example, for two spikes of width $\sigma_G = 300$ s, it will be possible to distinguish the peaks if they are separated by at least 900 s for H.E.S.S./MAGIC and by 500 s for CTA.

The impact of this effect is evaluated in the following.

## 4 Likelihood fit

The time distribution of photons is considered to be a Gaussian curve with a standard deviation $\sigma_P$. The energy distribution follows a power law $E^{-\Gamma}$ with index taking discrete values from $\Gamma = 2.2$ (as in Mkn 501 MAGIC data [8]) to $\Gamma = 3.4$ (PKS 2155-304 H.E.S.S. data [9]).

Fig. 2 (left) shows the average error $\bar{\sigma}_\tau$ on the reconstructed lag as a function of the number $N_L$ of photons included in the likelihood fit computation *ie* photons in the energy range 0.3–10 TeV. As expected, the error on reconstructed lags follows the relation:

$$\bar{\sigma}_\tau \sim 1/\sqrt{N}. \qquad (4)$$

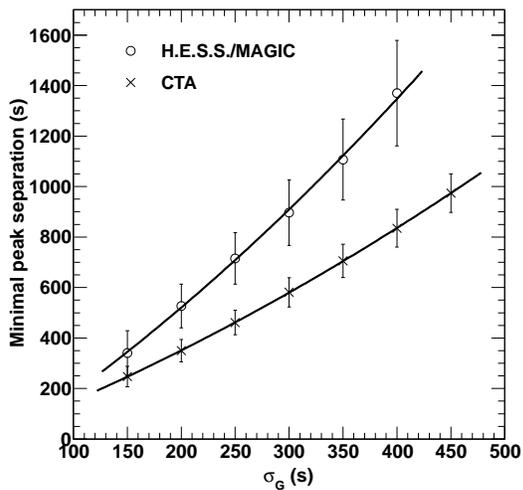

Figure 1: Minimal distinguishable separation of the two peaks as a function of their width, in case of a H.E.S.S.-like statistics (open circles) and a CTA-like statistics (crosses).

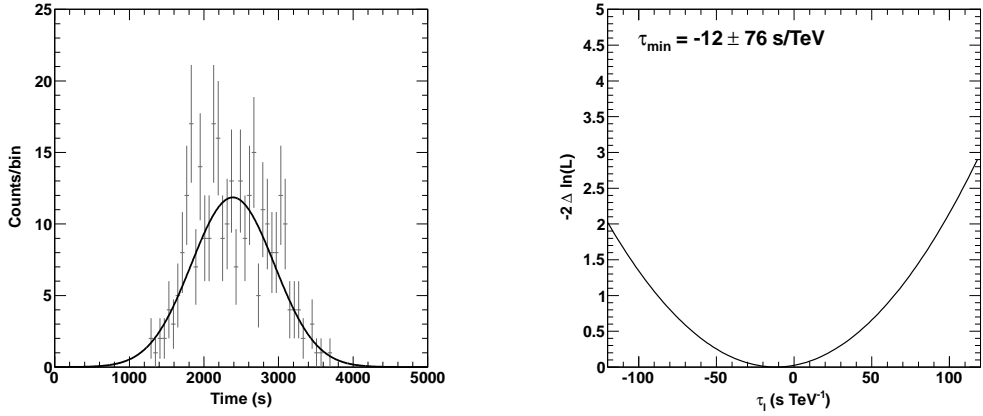

Figure 3: Left: realization of a light curve with 300 photons and a binning of 60 s (H.E.S.S. case). The fit with a Gaussian curve leads to $\chi^2/\text{dof} = 36.1/36$. Right: corresponding likelihood computed assuming a spectral index of $\Gamma = 2.8$. The position of the minimum is indicated on top left and the error is obtained requesting $-2\Delta \ln(L) = 1$.

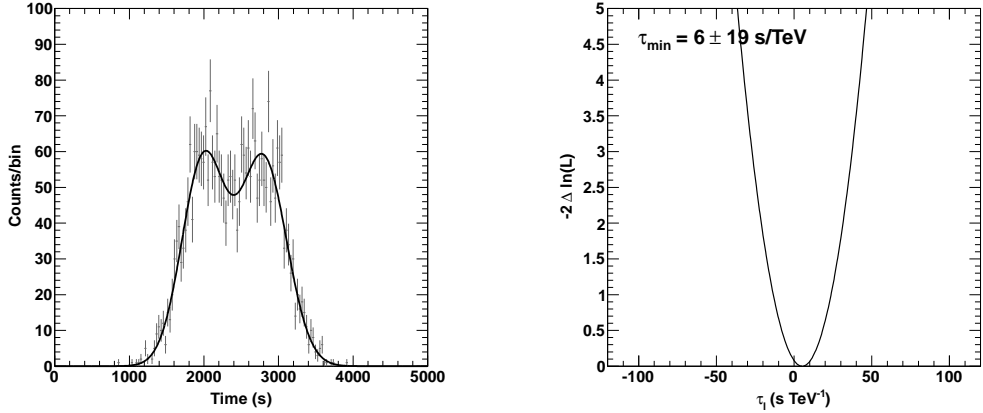

Figure 4: Left: the flare of Fig. 3 (left) with 3000 photons and a binning of 30 s (CTA case). The fit with a double Gaussian curve leads to $\chi^2/\text{dof} = 92.7/83$. Right: corresponding likelihood computed assuming a spectral index of $\Gamma = 2.8$. The position of the minimum is indicated on top left and the error is obtained requesting $-2\Delta \ln(L) = 1$.

When the width of the light curve peak $\sigma_P$ increases from 100 s to 500 s, $\bar{\sigma}_\tau$ increases as well. This reflects the fact that the error on the lag is strongly dependent on the variability amplitude of the source.

Fig. 2 (right) shows the average error $\bar{\sigma}_\tau$ on the reconstructed lag as a function of the number $N_L$ of photons included in the likelihood fit computation for three different values of the spectral index $\Gamma = 3.4$, 2.8 and 2.2 for linear and quadratic effects.

Here again, the error on reconstructed lags follows the relation of Eq. 4. The figure also illustrates the fact that harder spectra favour the detection of possible quadratic effects.

## 5 Results

### 5.1 Parameterization of the light curve

In order to quantify the effect of the result obtained in §3 on the limits on $M_{QG}$, the likelihood was computed for $\sigma_G = 300$ s and a peak separation of 700 s, using all the generated photons.

Fig. 3 shows the fit of a realization of the light curve (left) and the corresponding likelihood minimum (right) for an injected lag of 0 s TeV$^{-1}$ and for 300 photons. The fit with a single Gaussian curve gives a good value of $\chi^2/\text{dof} = 36.1/36$. The likelihood curve reaches a minimum at $-12 \pm 76$ s TeV$^{-1}$.

Fig. 4 shows the same plots for 3000 photons. The two peaks are clearly visible and the light curve is fitted with the sum of two Gaussian functions. The corresponding likelihood curve reaches a minimum at $6 \pm 19$ s TeV$^{-1}$. The error in this case is smaller by ∼20% including both effects of larger statistics and better peak separation power. Considering only the better peak separation leads to an improvement of ∼10%.

The effect of better peak separation comes in addition to the one discussed in §4 and in the next section and gives limits of $M_{QG}^l > 4 \times 10^{19}$ GeV and $M_{QG}^q > 1.2 \times 10^{11}$ GeV for the linear and quadratic case respectively for PKS flare and $M_{QG}^l > 10^{18}$ GeV and $M_{QG}^q > 0.5 \times 10^{11}$ GeV for Mkn 501 flare, at the 95% CL.

Here again, only the increase of the sensitivity by a factor of ten between H.E.S.S./MAGIC and CTA is taken into account. As the energy range will also be extended towards a couple tens of GeV with CTA, the increase of statistics will be even greater than the one considered here. The break in the spectrum due to the maximum of the Inverse Compton Peak will also tend to reduce the statistics to be included in the light curve parameterization. This point will be addressed in the final version of this proceeding.

### 5.2 Likelihood fit

As mentionned above, with the increase of the sensitivity for a given energy range, the error would be divided by a factor of three in case of CTA.

Then, observing the PKS 2155-304 flare of July 2006 in the energy range of H.E.S.S., the limits at 95% CL would be of the order of $M_{QG}^l > 1.7 \times 10^{19}$ GeV and $M_{QG}^q > 1.1 \times 10^{11}$ GeV for the linear and quadratic case respectively, as compared to the curent H.E.S.S. results of $M_{QG}^l > 2.1 \times 10^{18}$ GeV and $M_{QG}^q > 0.6 \times 10^{11}$ GeV [10].

For a flare like the one of Mkn 501 as detected by MAGIC [8], the limits would be of the order of $M_{QG}^l > 0.4 \times 10^{18}$ GeV and $M_{QG}^q > 0.4 \times 10^{11}$ GeV for the linear and quadratic case respectively, as compared to the curent MAGIC results of $M_{QG}^l > 0.2 \times 10^{18}$ GeV and $M_{QG}^q > 0.3 \times 10^{11}$ GeV [10].

The limits given here take only into account the increase in statistics in the high energy band, where the likelihood is computed.

In addition to the increase in statistics, the energy range will also be extended towards high energies to ∼100 TeV. However, for high energies, the effect of additional statistics will be moderated to some extent by the absorption of TeV photons by the Extra-galactic Background Light (e.g. [11]). This point will be addressed in the final version of this proceeding.

## 6 Discussion

The simulations performed for this work show that CTA will greatly improve the sensitivity of photon propagation studies with respect to LIV effects. Considering the fact that the sensitivity will be ten times higher than the one of present-day experiments and that the energy range covered will be much larger, the Planck scale will be easily reached for the "linear" models, conforting the present *Fermi* results [12, 13]. A flare as the one of Mkn 501 seen by MAGIC would reach Planck scale for the linear correction to the dispersion relations while a flare as the one seen by H.E.S.S. with PKS 2155-304 would largely exceed this value. The main increase in sensitivity will be reflected in the QG scale in case of the "quadratic" models where a new range of detection will emerge.

Another important question for future studies is how many AGN flares will be observed by CTA. Of all flares observed so far, only three had enough statistics and variability to be used to search for LIV. It is expected that CTA will also improve this number, especially in the so called "survey pointing mode" where each telescope aims at a different location of the sky.

## Acknowledgments

We gratefully acknowledge support from the agencies and organizations listed in this page: `http://www.cta-observatory.org/?q=node/22`. DE acknowledges the Science and Technology Facilities Council (STFC) for support under grant ST/G003084/1. AJ and JB acknowledges the support of GdR PCHE in France.